\documentclass[preprintnumbers,aps,prd,notitlepage,showpacs,nofootinbib,superscriptaddress]{revtex4}
\usepackage{eurosym}
\usepackage{lipsum}
\usepackage{CJK}
\usepackage{amsfonts}
\usepackage{amsmath}
\usepackage{amssymb}
\usepackage[normalem]{ulem}
\usepackage[english]{babel}
\usepackage{graphicx}
\usepackage{epsfig}
\usepackage{bm}
\usepackage{verbatim}
\usepackage[utf8]{inputenc}
\usepackage{booktabs}
\usepackage{multirow}
\usepackage{slashed}
\usepackage{xcolor}
\usepackage[colorlinks=true,urlcolor=red,citecolor=red]{hyperref}
\usepackage[font=small]{caption}
\usepackage{float}
\usepackage{blindtext}
\usepackage{placeins}
\usepackage[utf8]{inputenc}
\usepackage{soul}
\usepackage{bbm}
\usepackage[colorinlistoftodos]{todonotes}
\usepackage{subcaption}

\newenvironment{Eqnarray}{\arraycolsep 0.14em\begin{eqnarray}}{\end{eqnarray}}
\newcommand{\ba}{\begin{Eqnarray}}
\newcommand{\ea}{\end{Eqnarray}}
\newcommand{\be}{\begin{equation}}
\newcommand{\ee}{\end{equation}}
\newcommand{\bal}{\begin{aligned}}
\newcommand{\eal}{\end{aligned}}
\newcommand{\bea}{\begin{eqnarray}}
\newcommand{\eea}{\end{eqnarray}}
\newcommand{\ben}{\begin{enumerate}}
\newcommand{\een}{\end{enumerate}}
\newcommand{\bit}{\begin{itemize}}
\newcommand{\eit}{\end{itemize}}
\newcommand{\bde}{\begin{widetext}}
\newcommand{\ede}{\end{widetext}}

\renewcommand{\[}{\left[}

\def\lsim{\mathrel{\rlap{\lower4pt\hbox{\hskip1pt$\sim$}}
    \raise1pt\hbox{$<$}}}
\def\gsim{\mathrel{\rlap{\lower4pt\hbox{\hskip1pt$\sim$}}
    \raise1pt\hbox{$>$}}}

\def\3211{$\mathrm{SU(3) \otimes SU(2)_L \otimes U(1)_R \otimes U(1)_{B-L}}$ }
\def\321{$\mathrm{SU(3) \otimes SU(2) \otimes U(1)}$ }
\def\422{$\mathrm{SU(4) \otimes SU(2) \otimes SU(2)_R}$ }

\newcommand{\mathsym}[1]{{}}

\definecolor{bostonuniversityred}{rgb}{0.8, 0.0, 0.0}

\input{tcilatex}
\begin{document}

\title{Fermion mass hierarchy in an extended left-right symmetric model}
\author{Cesar Bonilla}
\email{cesar.bonilla@ucn.cl}
\affiliation{Departamento de F\'{\i}sica, Universidad Catolica del Norte, Avenida Angamos
0610, Casilla 1280, Antofagasta, Chile}
\author{A. E. C\'arcamo Hern\'andez}
\email{antonio.carcamo@usm.cl}
\affiliation{{Universidad T\'ecnica Federico Santa Mar\'{\i}%
a, Casilla 110-V, Valpara\'{\i}so, Chile}}
\affiliation{{Centro Cient\'{\i}fico-Tecnol\'ogico de Valpara\'{\i}so, Casilla 110-V,
Valpara\'{\i}so, Chile}}
\affiliation{{Millennium Institute for Subatomic Physics at High-Energy Frontier
(SAPHIR), Fern\'andez Concha 700, Santiago, Chile}}
\author{Sergey Kovalenko}
\email{sergey.kovalenko@unab.cl}
\affiliation{{Center for Theoretical and Experimental Particle Physics, Facultad de Ciencias Exactas,
Universidad Andr\'es Bello, Fernandez Concha 700, Santiago, Chile}}
\affiliation{{Millennium Institute for Subatomic Physics at High-Energy Frontier
(SAPHIR), Fern\'andez Concha 700, Santiago, Chile}}
\affiliation{{Centro Cient\'{\i}fico-Tecnol\'ogico de Valpara\'{\i}so, Casilla 110-V,
Valpara\'{\i}so, Chile}}
\author{H. Lee}
\email{huchan.lee@clermont.in2p3.fr}
\affiliation{Laboratoire de Physique de Clermont (UMR 6533), CNRS/IN2P3, Univ. Clermont Auvergne, 4 Av. Blaise Pascal, 63178 Aubiere Cedex, France}
\author{R. Pasechnik}
\email{roman.pasechnik@hep.lu.se}
\affiliation{Department of Physics, Lund University, SE-223 62 Lund, Sweden}
\author{Ivan Schmidt}
\email{ivan.schmidt@usm.cl}
\affiliation{{Departamento de F\'{\i}sica, Universidad T\'ecnica Federico Santa Mar\'{\i}%
a, Casilla 110-V, Valpara\'{\i}so, Chile}}
\affiliation{{Centro Cient\'{\i}fico-Tecnol\'ogico de Valpara\'{\i}so, Casilla 110-V,
Valpara\'{\i}so, Chile}}
\affiliation{{Millennium Institute for Subatomic Physics at High-Energy Frontier
(SAPHIR), Fern\'andez Concha 700, Santiago, Chile}}
\date{\today }

\begin{abstract}
We present a Left-Right symmetric model that provides an explanation for the
mass hierarchy of the charged fermions within the framework of the Standard
Model. This explanation is achieved through the utilization of both
tree-level and radiative seesaw mechanisms. In this model, the tiny masses
of the light active neutrinos are generated via a three-loop radiative
inverse seesaw mechanism, with Dirac and Majorana submatrices arising at
one-loop level. To the best of our knowledge, this is the first example of
the inverse seesaw mechanism being implemented with both submatrices
generated at one-loop level. The model contains a global $U(1)_{X}$ symmetry
which, after its spontaneous breaking, allows for the stabilization of the
Dark Matter (DM) candidates. We show that the electroweak precision
observables, the electron and muon anomalous magnetic moments as well as the
Charged Lepton Flavor Violating decays, $\mu \rightarrow e \gamma$, are
consistent with the current experimental limits. In addition, we analyze the
implications of the model for the $95$ GeV diphoton excess recently reported
by the CMS collaboration and demonstrate that such anomaly could be easily
accommodated. Finally, we discuss qualitative aspects of DM in the
considered model.
\end{abstract}

\maketitle



\section{Introduction}

\label{Sect:Intro} 

Despite the remarkable accomplishments of the Standard Model (SM) in
describing the fundamental interactions, it fails to address several
profound inquiries that remain unanswered. In the context of the present
work, we plan to address specific questions such as the SM flavor structure
(i.e. the observed pattern of SM fermion masses and mixing), the origin of
Dark Matter (DM) and the parity violation in weak interactions, whose
answers lie certainly beyond the SM. Even though the energy scale of New
Physics is still a mystery, current experimental searches keep restricting
the possibilities for new phenomena at experimentally accessible energy
scales and pushing potential New Physics signatures towards high energies.
For this reason, it is interesting to look into well motivated SM
extensions, whose signatures dwells at energies much higher than the
electroweak scale, such as for instance the Left-Right (LR) symmetric
models. These are particularly interesting since they provide a robust
explanation of parity violation in weak interactions, as a low-energy effect
of the spontaneously broken at high scales LR-symmetry.

In this study, we present a LR-symmetric extension of the SM, which offers a
comprehensive explanation for the fermion mass hierarchy. To start, the
masses of the third generation of SM charged fermions as well as the charm
quark mass are generated by means of a seesaw-like mechanism. This implies
the existence of a mixing between the SM charged fermions and the heavy
charged vector-like (with respect to the gauge symmetry) fermions\footnote{%
This is a variant of the original idea developed in Ref.~\cite%
{Davidson:1987mh} where all the SM charged fermion and light active neutrino
masses are generated through a tree-level seesaw mechanism mediated by heavy
fermions.}. In addition, the masses of the light (up, down and strange)
quarks, as well as those of electron and muon, arise from a radiative seesaw
mechanism at one-loop level. The complete picture of this setup considers
the tiny masses of the light active neutrinos generated via the inverse
seesaw mechanism at the three-loop level, with the Dirac as well as the
Majorana mass submatrices induced at the one-loop level. So, this is an
alternative to other radiative models where the Majorana mass submatrices
arise at the loop level~\cite%
{Ma:2009gu,Law:2012mj,Ahriche:2016acx,CarcamoHernandez:2017kra,CarcamoHernandez:2017cwi,CarcamoHernandez:2018hst,Mandal:2019oth,CarcamoHernandez:2019lhv,Abada:2021yot,Hernandez:2021kju,Hernandez:2021xet,Hernandez:2021uxx}
while the Dirac mass submatrix is generated at tree level. To the best of
our knowledge, the scenario proposed in this work is the first example of
the inverse seesaw mechanism with both Dirac and Majorana submatrices
generated at one-loop level. Other variants of the LR symmetric model have
been recently considered in Refs.~\cite%
{CarcamoHernandez:2018hst,Dekens:2014ina,Nomura:2016run,Brdar:2018sbk,
Ma:2020lnm,Babu:2020bgz,Hernandez:2021uxx}. Let us note that the model
studied here introduces the minimal ingredients for a LR-symmetric theory to
account for the fermion mass hierarchy, which means that the proposed set-up
turns out to be minimalistic in comparison with other existing proposals,
such as the one presented in \cite{Hernandez:2021uxx}.

For example, the scalar sector of our model has two $SU\left( 2\right)_{L}$
scalar doublets (8 degrees of freedom), two $SU\left( 2\right)_{R}$ scalar
doublets (8 degrees of freedom) and three electrically neutral gauge-singlet
complex scalar fields (6 degrees of freedom), thus, amounting to a total of
16 physical scalar degrees of freedom (after subtracting the number of
Goldstone bosons). The charged exotic fermion sector of the model has 7
charged exotic vector-like fermions. Thus, in total the model has 23
physical degrees of freedom (compared to 32 degrees of freedom of the model
of Ref.~\cite{Hernandez:2021uxx}).

It is worth noticing that the charged exotic vector-like leptons, which
mediate the seesaw mechanisms yielding the SM charged lepton masses,
contribute to the muon and electron anomalous magnetic moments. Therefore, a
scenario of this type offers a natural link between the fermion mass
generation mechanism and the $g-2$ anomalies. Another feature of the model
is the rich phenomenology at colliders related to the scalar and gauge boson
sectors.

The paper is organized as follows. In sec.~\ref{Sect:model} we introduce the
model theoretical setup, describing its field content and symmetries. The
scalar potential and scalar particles of the model are studied in sec.~\ref%
{scalarpotential}. In sec.~\ref{fermionmasses} we explore the quark, charged
lepton and neutrino mass spectra and mixing, as well as study the role of
heavy exotic fermions. Phenomenological implications of the model are
examined in secs.~\ref{TnS}-\ref{LFV}. Finally, in sec.~\ref{sec:DM} we
briefly discuss possible DM particle candidates provided by the model.


\section{Model setup}

\label{Sect:model} 
We start this section by explaining the reasoning that justifies the
inclusion of extra scalars, fermions and symmetries needed for implementing
an interplay of tree-level and radiative seesaw mechanisms, enabling us to
explain the SM charged fermion mass hierarchies, as well as the three-loop
level inverse seesaw mechanism, with one loop induced Dirac and Majorana
submatrices, in order to generate tiny masses of the light active neutrinos.
In our theoretical construction, detailed below, 
the basis $\left( \nu _{L},\nu _{R}^{C},N_{R}^{C}\right) $, has the
following structure: 
\begin{equation}
M_{\nu }=\left( 
\begin{array}{ccc}
0_{3\times 3} & m_{\nu D} & 0_{3\times 3} \\ 
m_{\nu D}^{T} & 0_{2\times 2} & M \\ 
0_{3\times 3} & M^{T} & \mu%
\end{array}%
\right) ,  \label{Mnufull}
\end{equation}%
where $\nu _{iL}$ ($i=1,2,3$) correspond to the active neutrinos, whereas $%
\nu _{iR}$ and $N_{iR}$ ($i=1,2$) are the sterile neutrinos. Furthermore,
the entries of the full neutrino mass matrix of Eq. (\ref{Mnufull}) should
obey the hierarchy $\mu _{ij}<<m_{ij}<<M_{ij}$ ($i=1,2,3$, $n,p=1,2$), where
the submatrices $\mu $ and $m_{\nu D}$ are radiatively generated at one loop
level, whereas the submatrix $M$ arises at tree level. Here we consider a
theory based on the $SU\left( 3\right) _{C}\times SU\left( 2\right)
_{L}\times SU\left( 2\right) _{R}\times U\left( 1\right) _{B-L}$ symmetry,
which is supplemented by the inclusion of the global $U\left( 1\right) _{X}$
symmetry. The $U\left( 1\right) _{X}$ global symmetry is assumed to be
spontaneously broken down to a preserved matter parity symmetry, which is
crucial for ensuring the radiative nature of the inverse seesaw mechanism
that produces the tiny active neutrino masses, as well as the seesaw-like
mechanism that generates the masses of the light (up, down and strange)
quarks, as well as those of electron and muon. To generate the Majorana
submatrix $\mu $ at one loop level, we require the following operators: 
\begin{equation}
\overline{N_{R_{i}}^{C}}\widetilde{\chi }_{R}^{\dagger }L_{R_{j}},\hspace{%
1.5cm}\overline{N}_{R_{i}}\Omega _{kR}^{C}\varphi ,\hspace{1.5cm}\overline{%
\Omega }_{R_{n}}\Omega _{R_{n}}^{C}\rho ,  \label{opmu}
\end{equation}
where $L_{R_{j}}$ ($j=1,2,3$) are $\chi _{R}$ are $SU\left(
2\right) _{R}$ lepton and scalar doublets, respectively. Besides that, $%
N_{R_{i}}$ and $\Omega _{R_{n}}$ ($n=1,2$) are right handed Majorana
neutrinos, whereas $\varphi $ and $\rho $ are electrically neutral gauge
singlet scalars.

Notice that the second and third operators in Eq.~(\ref{opmu}) generate a
one-loop level Majorana neutrino mass submatrix $\mu $, associated with the
breaking of lepton number, while the first operator in this equation
generates the mass scale of the heavy sterile neutrinos which mediate the
inverse seesaw mechanism.

On the other hand, a radiative generation of the Dirac neutrino mass
submatrix $m_{\nu D}$ at one-loop level requires the inclusion of the
following operators: 
\begin{equation}
\overline{L}_{L_{i}}\phi _{L}E_{R_{n}}^{\prime },\hspace{1.5cm}\bar{E}%
_{L_{n}}^{\prime }\phi _{R}^{\dagger }L_{R_{j}},\hspace{1.5cm}\bar{E}%
_{L_{n}}^{\prime }\sigma E_{R_{m}}^{\prime },\hspace{1.5cm}n,m=1,2\,,
\label{opmnuD}
\end{equation}%
which are also crucial to generate one loop level muon and
electron masses. Here $L_{L_{i}}$ ($\phi _{L}$)\ and $L_{R_{i}}$ ($\phi _{R}$%
)\ are $SU\left( 2\right) _{L}$ and $SU\left( 2\right) _{R}$ lepton (scalar)
doublets, $E_{n}$ ($n=1,2$)\ are heavy vector-like charged leptons and $%
\sigma $ is a electrically neutral scalar singlet.

It is worth mentioning that the operators presented in Eqs. (\ref{opmu}) and
(\ref{opmnuD}) allow for a successful implementation of the three-loop level
inverse seesaw mechanism that produces the tiny neutrino masses. This is due
to the fact that the light active neutrino mass matrix arising from the
inverse seesaw mechanism will have a quadratic and linear dependence with
the one loop induced Dirac $m_{\nu D}$ and Majorana $\mu $ submatrices,
respectively. Given that our proposed model does not consider a bidoublet
scalar in the particle spectrum, the invariance of the Yukawa interactions
with respect to the symmetries of the model forbids charged fermions Yukawa
terms involving a SM fermion-anti fermion pair. Consequently, heavy charged
vector-like (with respect to the gauge symmetry) fermions have to be
included in the particle spectrum in order to trigger a seesaw-like
mechanism that generates the masses of the SM charged fermions. The
symmetries of the model imply that the masses of the third generation of SM
charged fermions as well as the charm quark mass will be generated through
the mixing with heavy vector-like (with respect to the gauge symmetries)
fermions, arising from the following operators: 
\begin{eqnarray}
&&\overline{Q}_{L_{3}}\widetilde{\chi }_{L}T_{R_{1}},\hspace{1.5cm}\overline{%
T}_{L_{1}}\widetilde{\chi }_{R}^{\dagger }Q_{R_{i}},\hspace{1.5cm}m_{T}%
\overline{T}_{L_{1}}T_{R_{1}},\hspace{1.5cm}i,j=1,2,3,  \notag \\
&&\overline{Q}_{L_{n}}\widetilde{\chi }_{L}T_{R_{2}},\hspace{1.5cm}\overline{%
T}_{L_{2}}\widetilde{\chi }_{R}^{\dagger }Q_{R_{i}},\hspace{1.5cm}\overline{T%
}_{L_{2}}\sigma T_{R_{2}}  \notag \\
&&\overline{Q}_{L_{3}}\chi _{L}B_{R},\hspace{1.5cm}\overline{B}_{L}\chi
_{R}^{\dagger }Q_{R_{i}},\hspace{1.5cm}m_{B}\overline{B}_{L}B_{R}  \notag \\
&&\overline{L}_{L_{i}}\chi _{L}E_{R},\hspace{1.5cm}\overline{E}_{L}\chi
_{R}^{\dagger }L_{R_{j}},\hspace{1.5cm}\overline{E}_{L}\sigma E_{R},
\label{op0}
\end{eqnarray}%
where $L_{L_{i}}$ ($Q_{L_{i}}$)\ and $L_{R_{i}}$ ($Q_{R_{i}}$)\ are $%
SU\left( 2\right) _{L}$ and $SU\left( 2\right) _{R}$ lepton (quark)
doublets, respectively, and $\chi _{L}$ and $\chi _{R}$\ are $SU\left(
2\right) _{L}$ and $SU\left( 2\right) _{R}$ scalar doublets, respectively.
Furthermore, $T_{n}$ ($n=1,2$), $B$ and $E$ are heavy vector-like up, down
type quarks and charged leptons, respectively. The heavy quarks $T_{1}$, $%
T_{2}$, $B$ as well as the heavy lepton $E$ get their masses from mass terms
and Yukawa interactions involving the gauge-singlet scalar field $\sigma $.
We assume that the singlet scalar field $\sigma $ is charged under a global $%
U\left( 1\right) _{X}$ symmetry introduced in the model.

Furthermore, in order to radiatively generate the masses of the light up,
down and strange quarks, as well as the electron and muon masses, the
following operators are required: 
\begin{eqnarray}
&&\overline{Q}_{L_n}\widetilde{\phi }_{L}T_{R}^{\prime },\hspace{1.5cm}\bar{T%
}_{L}^{\prime }\widetilde{\phi }_{R}^{\dagger }Q_{R_i},\hspace{1.5cm}\bar{T}%
_{L}^{\prime }\sigma T_{R}^{\prime },\hspace{1.5cm}i,j=1,2,3,  \notag \\
&&\overline{Q}_{L_n}\phi _{L}B_{R_m}^{\prime },\hspace{1.5cm}\bar{B}%
_{L_n}^{\prime }\phi _{R}^{\dagger }Q_{R_i},\hspace{1.5cm}\bar{B}%
_{L_n}^{\prime }\sigma B_{R_m}^{\prime }  \notag \\
&&\overline{L}_{L_i}\phi _{L}E_{R_n}^{\prime },\hspace{1.5cm}\bar{E}%
_{L_n}^{\prime }\phi _{R}^{\dagger }L_{R_j},\hspace{1.5cm}\bar{E}%
_{L_n}^{\prime }\sigma E_{R_m}^{\prime },\hspace{1.5cm}n,m=1,2 \,,
\label{op2}
\end{eqnarray}%
where $\phi_{L}$ and $\phi_{R}$\ are $SU\left( 2\right)_{L}$ and $SU\left(
2\right)_{R}$ scalar doublets, respectively. These scalar doublets will be
assumed to have odd $U\left( 1\right)_{X}$ charge. The $U\left( 1\right)_{X}$
global symmetry is assumed to be spontaneously broken down to a preserved
matter parity symmetry, which will imply that the scalar doublets $\phi_{L}$
and $\phi_{R}$ will not acquire vacuum expectation values (VEVs).
Consequently, the radiative generation of masses for the light (up, down and
strange) quarks, as well as those of electron and muon can be achieved at
the one-loop level. The implementation of such a radiative seesaw mechanism
also requires the inclusion of the heavy vector-like up-type quarks $%
T^{\prime}$, down-type quarks $B_{n}^{\prime }$ and charged leptons $%
E_{n}^{\prime }$ ($n=1,2$) in the fermion spectrum of the model under
consideration.

The model under consideration is based on the gauge symmetry $%
SU(3)_{C}\times SU\left( 2\right) _{L}\times SU\left( 2\right) _{R}\times
U\left( 1\right) _{B-L}$ supplemented by the global $U\left( 1\right) _{X}$
symmetry, where the full symmetry $\mathcal{G}$ exhibits the following
breaking scheme: 
\begin{eqnarray}
&&\mathcal{G}=SU(3)_{C}\times SU\left( 2\right) _{L}\times SU\left( 2\right)
_{R}\times U\left( 1\right) _{B-L}\times U\left( 1\right) _{X}  \notag \\
&&\hspace{35mm}\Downarrow v_{\sigma },v_{R},v_{\rho }  \notag \\[0.12in]
&&\hspace{15mm}SU(3)_{C}\times SU\left( 2\right) _{L}\times U\left( 1\right)
_{Y}\times Z_{2}  \notag \\[0.12in]
&&\hspace{35mm}\Downarrow v_{L}  \notag \\[0.12in]
&&\hspace{15mm}SU(3)_{C}\otimes U\left( 1\right) _{Q}\times Z_{2}\,.
\end{eqnarray}%
The spontaneous breaking of the global $U\left( 1\right) _{X}$ symmetry is
assumed to occur together with the LR symmetry breaking at the same energy
scale. We further assume that this scale is about $v_{R}\sim \mathcal{O}(10)$
TeV. With the scalar field assignments given in Table~\ref{scalars}, the
global $U\left( 1\right) _{X}$ symmetry is spontaneously broken down to a
preserved $Z_{2}$ discrete symmetry. The latter allows a successful
implementation of a radiative seesaw mechanism at one-loop level, that
generates masses of the light (up, down and strange) quarks, as well as
those of electron and muon. Furthermore, thanks to the preserved $Z_{2}$
symmetry, the masses of the light active neutrinos arise from a three-loop
inverse seesaw mechanism, while the masses of the third family of SM charged
fermions as well the charm quark mass are generated by means of a seesaw
mechanism responsible for their tree-level mixing with heavy charged
vector-like fermions. The fermion assignments under the $SU(3)_{C}\times
SU\left( 2\right) _{L}\times SU\left( 2\right) _{R}\times U\left( 1\right)
_{B-L}\times U\left( 1\right) _{X}$ group are displayed in Table \ref%
{fermions}. 
\begin{table}[tbp]
\begin{tabular}{|c|c|c|c|c|c|c|c|c|c|c|c|c|c|c|c|c|c|c|c|c|}
\hline
& $Q_{L_n}$ & $Q_{L_3}$ & $Q_{R_i}$ & $L_{L_i}$ & $L_{R_i}$ & $T_{L_n}$ & $%
T_{R_1}$ & $T_{R_2}$ & $B_{L}$ & $B_{R}$ & $T_{L}^{\prime }$ & $%
T_{R}^{\prime }$ & $B_{L_n}^{\prime }$ & $B_{R_n}^{\prime }$ & $E_{L}$ & $%
E_{R}$ & $E_{L_n}^{\prime }$ & $E_{R_n}^{\prime }$ & $N_{R_i}$ & $\Omega
_{R_n}$ \\ \hline
$SU(3)_{C}$ & $\mathbf{3}$ & $\mathbf{3}$ & $\mathbf{3}$ & $\mathbf{1}$ & $%
\mathbf{1}$ & $\mathbf{3}$ & $\mathbf{3}$ & $3$ & $\mathbf{3}$ & $\mathbf{3}$
& $\mathbf{3}$ & $\mathbf{3}$ & $\mathbf{3}$ & $\mathbf{3}$ & $\mathbf{1}$ & 
$\mathbf{1}$ & $\mathbf{1}$ & $\mathbf{1}$ & $\mathbf{1}$ & $\mathbf{1}$ \\ 
\hline
$SU\left( 2\right) _{L}$ & $\mathbf{2}$ & $\mathbf{2}$ & $\mathbf{1}$ & $%
\mathbf{2}$ & $\mathbf{1}$ & $\mathbf{1}$ & $\mathbf{1}$ & $1$ & $\mathbf{1}$
& $\mathbf{1}$ & $\mathbf{1}$ & $\mathbf{1}$ & $\mathbf{1}$ & $\mathbf{1}$ & 
$\mathbf{1}$ & $\mathbf{1}$ & $\mathbf{1}$ & $\mathbf{1}$ & $\mathbf{1}$ & $%
\mathbf{1}$ \\ \hline
$SU\left( 2\right) _{R}$ & $\mathbf{1}$ & $\mathbf{1}$ & $\mathbf{2}$ & $%
\mathbf{1}$ & $\mathbf{2}$ & $\mathbf{1}$ & $\mathbf{1}$ & $1$ & $\mathbf{1}$
& $\mathbf{1}$ & $1$ & $1$ & $\mathbf{1}$ & $\mathbf{1}$ & $\mathbf{1}$ & $%
\mathbf{1}$ & $\mathbf{1}$ & $\mathbf{1}$ & $\mathbf{1}$ & $\mathbf{1}$ \\ 
\hline
$U\left( 1\right) _{B-L}$ & $\frac{1}{3}$ & $\frac{1}{3}$ & $\frac{1}{3}$ & $%
-1$ & $-1$ & $\frac{4}{3}$ & $\frac{4}{3}$ & $\frac{4}{3}$ & $-\frac{2}{3}$
& $-\frac{2}{3}$ & $\frac{4}{3}$ & $\frac{4}{3}$ & $-\frac{2}{3}$ & $-\frac{2%
}{3}$ & $-2$ & $-2$ & $-2$ & $-2$ & $0$ & $0$ \\ \hline
$U\left( 1\right) _{X}$ & $2$ & $0$ & $0$ & $0$ & $-2$ & $0$ & $0$ & $2$ & $%
0 $ & $0$ & $-1$ & $1$ & $1$ & $3$ & $-2$ & $0$ & $-1$ & $1$ & $2$ & $-3$ \\ 
\hline
\end{tabular}%
\caption{Fermion charge assignments under the $SU\left( 3\right) _{C}\times
SU\left( 2\right) _{L}\times SU\left( 2\right) _{R}\times U\left( 1\right)
_{B-L}\times U(1)_{X}$ symmetry. Here, $i=1,2,3$ and $%
n=1,2$.}
\label{fermions}
\end{table}

Notice that the fermion sector of the original LR symmetric model has
been enlarged by introducing only gauge group 
singlet Majorana neutrinos $N_{R_{i}}$ ($i=1,2,3$)\ and $\Omega _{R_{n}}$ ($%
n=1,2$) as well as vector-like exotic fermions: three exotic up-type quarks $T_{n}$(%
$n=1,2$), $T^{\prime }$, one exotic down-type quark $B$, three charged
leptons $E$, $E_{n}^{\prime }$. Since gauge singlet and vector-like fermions do not contribute to the gauge group chiral anomaly, our model, following the original LR-symmetric model, is also anomaly free.
 


Moreover the exotic fermions of our model are assigned to singlet
representations of the $SU\left( 2\right) _{L}\times SU\left( 2\right) _{R}$
group. The above mentioned exotic fermion content is the minimal one
required to generate tree-level masses via a seesaw mechanism for the third
generation of SM charged fermions and for the charm quark, one-loop level
masses for the light (up, down and strange) quarks and for the electron and
muon, as well as Dirac neutrino submatrix at one-loop level and the one-loop
Majorana submatrix $\mu $ of the inverse seesaw mechanism. The exotic lepton
spectrum of the model allows to generate the light active neutrino masses
via an inverse seesaw mechanism at three-loop level.

The scalar boson assignments under the $SU(3)_{C}\times SU\left( 2\right)
_{L}\times SU\left( 2\right) _{R}\times U\left( 1\right) _{B-L}\times
U\left( 1\right) _{X}$ group are: 
\begin{eqnarray}
\chi _{L} &=&\left( 
\begin{array}{c}
\chi _{L}^{+} \\ 
\frac{1}{\sqrt{2}}\left( v_{L}+\func{Re}\chi _{L}^{0}+i\func{Im}\chi
_{L}^{0}\right)%
\end{array}%
\right) \sim \left( \mathbf{1},\mathbf{2,1},1,0\right) ,\hspace{1cm}\chi
_{R}=\left( 
\begin{array}{c}
\chi _{R}^{+} \\ 
\frac{1}{\sqrt{2}}\left( v_{R}+\func{Re}\chi _{R}^{0}+i\func{Im}\chi
_{R}^{0}\right)%
\end{array}%
\right) \sim \left( \mathbf{1},\mathbf{1,2},1,0\right) ,  \notag \\
\phi _{L} &=&\left( 
\begin{array}{c}
\phi _{L}^{+} \\ 
\frac{1}{\sqrt{2}}\left( \func{Re}\phi _{L}^{0}+i\func{Im}\phi
_{L}^{0}\right)%
\end{array}%
\right) \sim \left( \mathbf{1},\mathbf{2,1},1,-1\right) ,\hspace{1cm}\phi
_{R}=\left( 
\begin{array}{c}
\phi _{R}^{+} \\ 
\frac{1}{\sqrt{2}}\left( \func{Re}\phi _{R}^{0}+i\func{Im}\phi
_{R}^{0}\right)%
\end{array}%
\right) \sim \left( \mathbf{1},\mathbf{1,2},1,-1\right) ,  \notag \\
\sigma &=&\frac{1}{\sqrt{2}}\left( v_{\sigma }+\func{Re}\sigma +i\func{Im}%
\sigma \right) \sim \left( \mathbf{1},\mathbf{1,1},0,-2\right) ,  \notag \\
\varphi &=&\frac{1}{\sqrt{2}}\left( \func{Re}\varphi +i\func{Im}\varphi
\right) \sim \left( \mathbf{1},\mathbf{1,1},0,-1\right) ,  \notag \\
\rho &=&\frac{1}{\sqrt{2}}\left( v_{\rho }+\func{Re}\rho +i\func{Im}\rho
\right) \sim \left( \mathbf{1},\mathbf{1,1},0,-6\right) .\hspace{1cm}.
\end{eqnarray}%
The fermion and scalar boson charge assignments under the $SU\left( 3\right)
_{C}\times SU\left( 2\right) _{L}\times SU\left( 2\right) _{R}\times U\left(
1\right) _{B-L}\times U(1)_{X}$ symmetry are shown in Tables \ref{fermions}
and \ref{scalars}, respectively. 
\begin{table}[tp]
\begin{tabular}{|c|c|c|c|c|c|c|c|}
\hline
& $\chi _{L}$ & $\chi _{R}$ & $\phi _{L}$ & $\phi _{R}$ & $\sigma $ & $\rho $
& $\varphi $ \\ \hline
$SU(3)_{C}$ & $\mathbf{1}$ & $\mathbf{1}$ & $\mathbf{1}$ & $\mathbf{1}$ & $%
\mathbf{1}$ & $\mathbf{1}$ & $\mathbf{1}$ \\ \hline
$SU\left( 2\right) _{L}$ & $\mathbf{2}$ & $\mathbf{1}$ & $\mathbf{2}$ & $%
\mathbf{1}$ & $\mathbf{1}$ & $\mathbf{1}$ & $\mathbf{1}$ \\ \hline
$SU\left( 2\right) _{R}$ & $\mathbf{1}$ & $\mathbf{2}$ & $\mathbf{1}$ & $%
\mathbf{2}$ & $\mathbf{1}$ & $\mathbf{1}$ & $\mathbf{1}$ \\ \hline
$U\left( 1\right) _{B-L}$ & $1$ & $1$ & $1$ & $1$ & $0$ & $0$ & $0$ \\ \hline
$U\left( 1\right) _{X}$ & $0$ & $0$ & $-1$ & $-1$ & $-2$ & $-6$ & $-1$ \\ 
\hline
\end{tabular}%
\caption{Scalar boson charge assignments under the $SU\left( 3\right)
_{C}\times SU\left( 2\right) _{L}\times SU\left( 2\right) _{R}\times U\left(
1\right) _{B-L}\times U(1)_{X}$ symmetry.}
\label{scalars}
\end{table}

Notice that the scalars $\chi _{L}$ and $\chi _{R}$ have been introduced to
implement the seesaw mechanism that produce tree-level masses for the third
generation of SM charged fermions. Furthermore, the $SU\left( 2\right) _{R}$
scalar doublet $\chi _{R}$ is required to trigger the spontaneous breaking
of the $SU\left( 2\right) _{R}\times U\left( 1\right) _{B-L}\times U\left(
1\right) _{X}$ symmetry, whereas the $SU\left( 2\right) _{L}$ scalar doublet 
$\chi _{L}$ spontaneously breaks the SM electroweak gauge symmetry. Besides,
the scalar fields $\sigma $ and $\rho $ spontaneously break the $U\left(
1\right) _{X}$ global symmetry down to a preserved $Z_{2}$ matter parity
symmetry defined as 
\begin{eqnarray}  \label{eq:M-P}
M_P &=& \left( -1\right)^{X+2s} \,.
\end{eqnarray}
The gauge singlets $\sigma $ and $\rho $ are included to generate tree-level
masses for the charged exotic vector-like fermions and heavy neutral leptons 
$\Omega _{R_n}$ ($n=1,2$), respectively. In addition, a successful
implementation of the loop-level radiative seesaw mechanism that produces
the masses for the first and second family of SM charged fermions, requires
the inclusion of the $SU\left( 2\right) _{R}$ and $SU\left( 2\right) _{L}$
inert scalar doublets $\phi _{R}$ and $\phi _{L} $, respectively. Moreover,
in order to generate the $\mu $ submatrix of the inverse seesaw at one-loop
level, the inert gauge singlet scalar field $\varphi $ is also required in
the scalar spectrum. The residual $M_P$ matter parity symmetry implies that
the scalar fields having odd $U\left( 1\right) _{X}$ charges do not acquire
VEVs, enabling an appropriate implementation of the radiative seesaw
mechanisms.

The VEVs of the scalars $\chi _{L}$ and $\chi _{R}$ read: 
\begin{equation}
\left\langle \chi _{L}\right\rangle =\left( 
\begin{array}{c}
0 \\ 
\frac{v_{L}}{\sqrt{2}}%
\end{array}%
\right) ,\hspace{1.5cm}\left\langle \chi _{R}\right\rangle =\left( 
\begin{array}{c}
0 \\ 
\frac{v_{R}}{\sqrt{2}}%
\end{array}%
\right) \,.
\end{equation}
The scalar fields having odd $U\left( 1\right) _{X}$ charges do not acquire
VEVs, due to the preserved matter parity symmetry defined as $\left(
-1\right) ^{X+2s}$.

With the above particle content, the following relevant Yukawa terms arise: 
\begin{eqnarray}
-\mathcal{L}_{Y} &=&x_{3}^{\left( T\right) }\overline{Q}_{L_3}\widetilde{%
\chi }_{L}T_{R_1}+\dsum\limits_{i=1}^{3}z_{1i}^{\left( T\right) }\overline{T}%
_{L_1}\widetilde{\chi }_{R}^{\dagger }Q_{R_i}+x_{3}^{\left( B\right) }%
\overline{Q}_{L_3}\chi _{L}B_{R}+\dsum\limits_{i=1}^{3}z_{i}^{\left(
B\right) }\overline{B}_{L}\chi _{R}^{\dagger }Q_{R_i}  \notag \\
&&+\dsum\limits_{n=1}^{2}x_{n}^{\left( T\right) }\overline{Q}_{L_n}%
\widetilde{\chi }_{L}T_{R_2}+\dsum\limits_{i=1}^{3}z_{2i}^{\left( T\right) }%
\overline{T}_{L_2}\widetilde{\chi }_{R}^{\dagger
}Q_{R_i}+\dsum\limits_{n=1}^{2}w_{n}^{\left( T^{\prime }\right) }\overline{Q}%
_{L_n}\widetilde{\phi }_{L}T_{R}^{\prime
}+\dsum\limits_{i=1}^{3}r_{i}^{\left( T^{\prime }\right) }\bar{T}%
_{L}^{\prime }\widetilde{\phi }_{R}^{\dagger }Q_{R_i}  \notag \\
&&+\dsum\limits_{n=1}^{2}\dsum\limits_{m=1}^{2}w_{nm}^{\left( B^{\prime
}\right) }\overline{Q}_{L_n}\phi _{L}B_{R_m}^{\prime
}+\dsum\limits_{n=1}^{2}\dsum\limits_{i=1}^{3}r_{ni}^{\left( B^{\prime
}\right) }\bar{B}_{L_n}^{\prime }\phi _{R}^{\dagger }Q_{R_i}  \notag \\
&&+m_{T}\overline{T}_{L_1}T_{R_1}+y_{T}\overline{T}_{L_2}\sigma
T_{R_2}+y_{T^{\prime }}\bar{T}_{L}^{\prime }\sigma T_{R}^{\prime
}+\dsum\limits_{n=1}^{2}\dsum\limits_{m=1}^{2}\left( y_{B^{\prime }}\right)
_{nm}\bar{B}_{L_n}^{\prime }\sigma B_{R_m}^{\prime }+m_{B}\overline{B}%
_{L}B_{R}  \notag \\
&&+y_{E}\overline{E}_{L}\sigma
E_{R}+\dsum\limits_{n=1}^{2}\dsum\limits_{m=1}^{2}\left( y_{E^{\prime
}}\right) _{nm}\bar{E}_{L_n}^{\prime }\sigma E_{R_m}^{\prime }  \notag \\
&&+\dsum\limits_{i=1}^{3}x_{i}^{\left( E\right) }\overline{L}_{L_i}\chi
_{L}E_{R}+\dsum\limits_{j=1}^{3}z_{j}^{\left( E\right) }\overline{E}_{L}\chi
_{R}^{\dagger }L_{R_j}  \notag \\
&&+\dsum\limits_{i=1}^{3}\dsum\limits_{n=1}^{2}w_{in}^{\left( E^{\prime
}\right) }\overline{L}_{L_i}\phi _{L}E_{R_n}^{\prime
}+\dsum\limits_{n=1}^{2}\dsum\limits_{j=1}^{3}r_{nj}^{\left( E^{\prime
}\right) }\bar{E}_{L_n}^{\prime }\phi _{R}^{\dagger }L_{R_j}  \notag \\
&&+\dsum\limits_{i=1}^{3}\dsum\limits_{j=1}^{3}x_{ij}^{\left( N\right) }%
\overline{N_{R_i}^{C}}\widetilde{\chi }_{R}^{\dagger
}L_{R_j}+\dsum\limits_{n=1}^{2}\left( y_{\Omega }\right) _{n}\overline{%
\Omega }_{R_n}\Omega _{R_n}^{C}\rho
+\dsum\limits_{i=1}^{3}\dsum\limits_{k=1}^{2}x_{ik}^{\left( \Omega \right) }%
\overline{N}_{R_i}\Omega _{kR}^{C}\varphi +H.c.  \label{Ly}
\end{eqnarray}%
\begin{figure}[H]
\includegraphics[width=7cm, height=4cm]{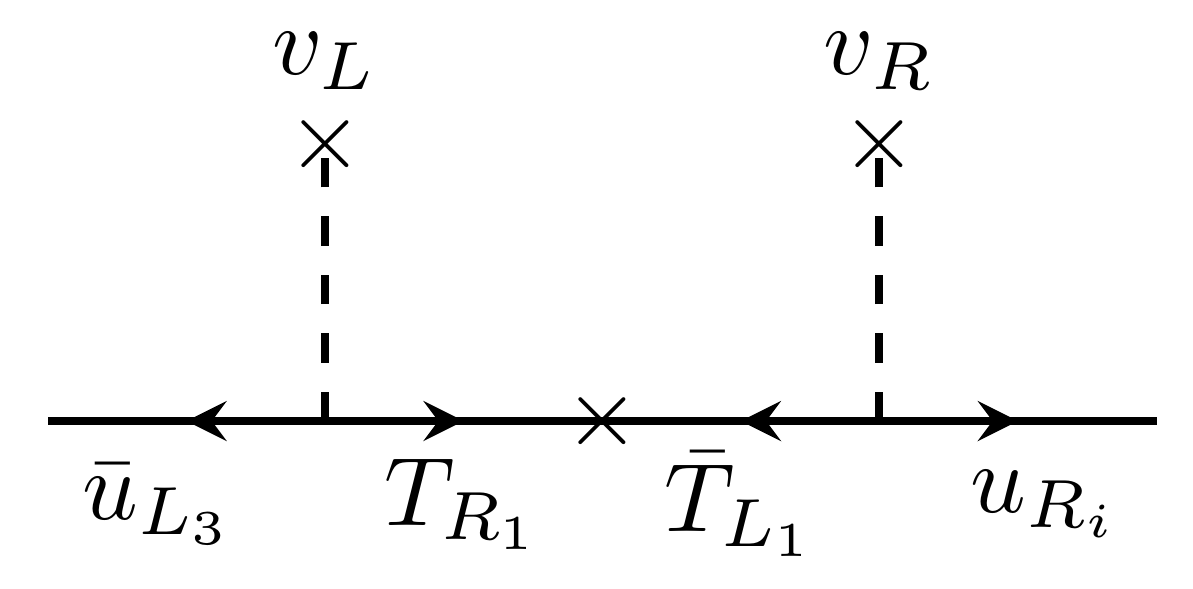} %
\includegraphics[width=7cm, height=5cm]{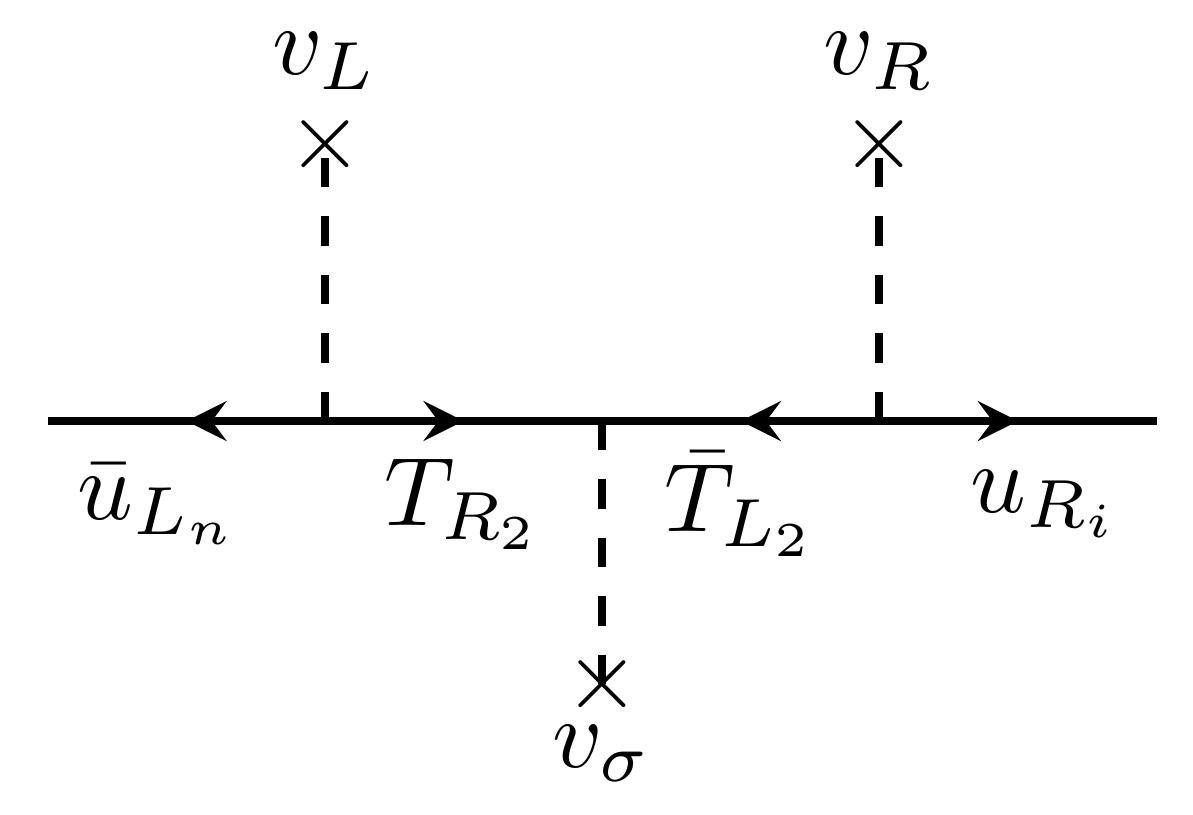}\newline
\includegraphics[width=8cm, height=4cm]{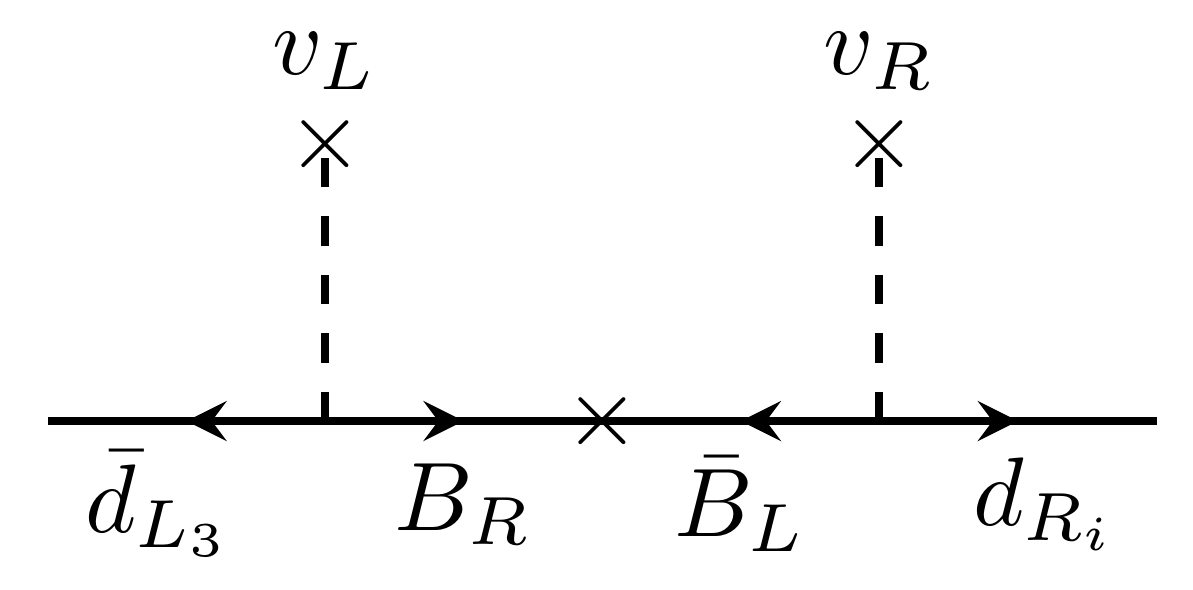} %
\includegraphics[width=8cm, height=5cm]{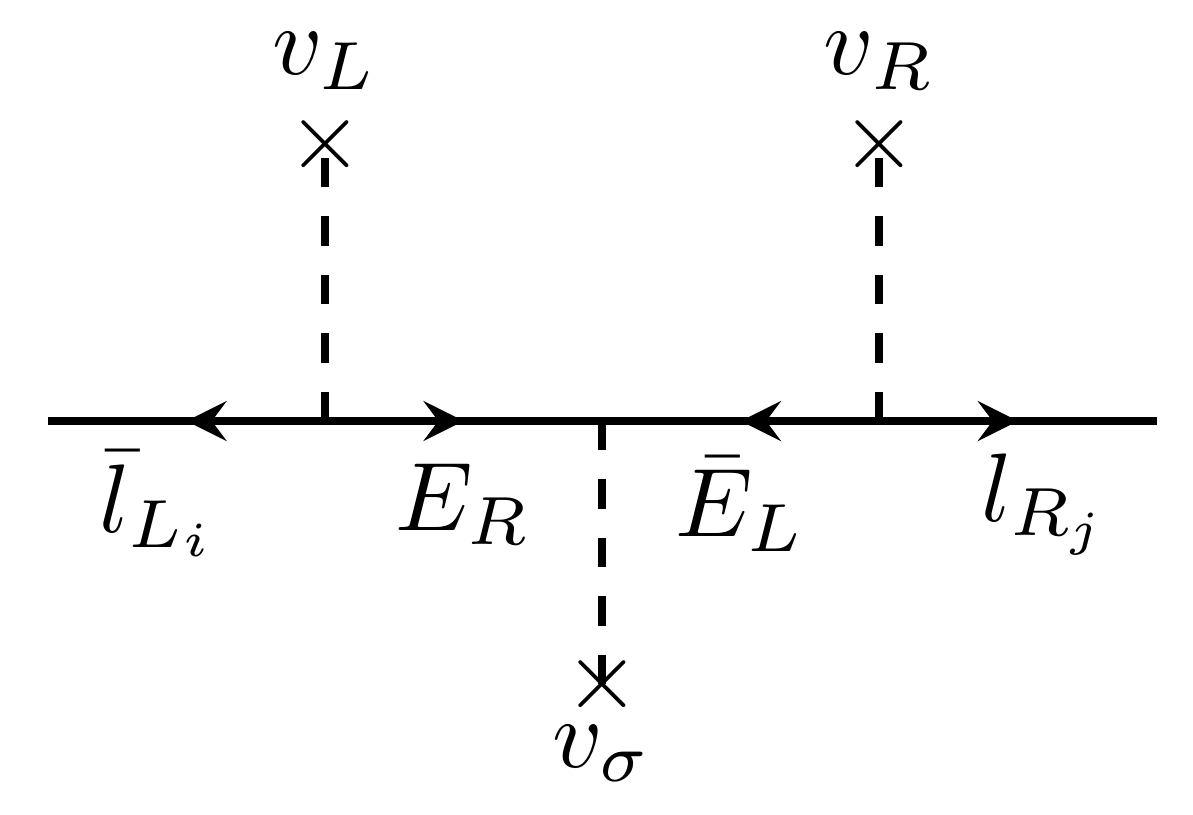}\newline
\includegraphics[width=8cm, height=4cm]{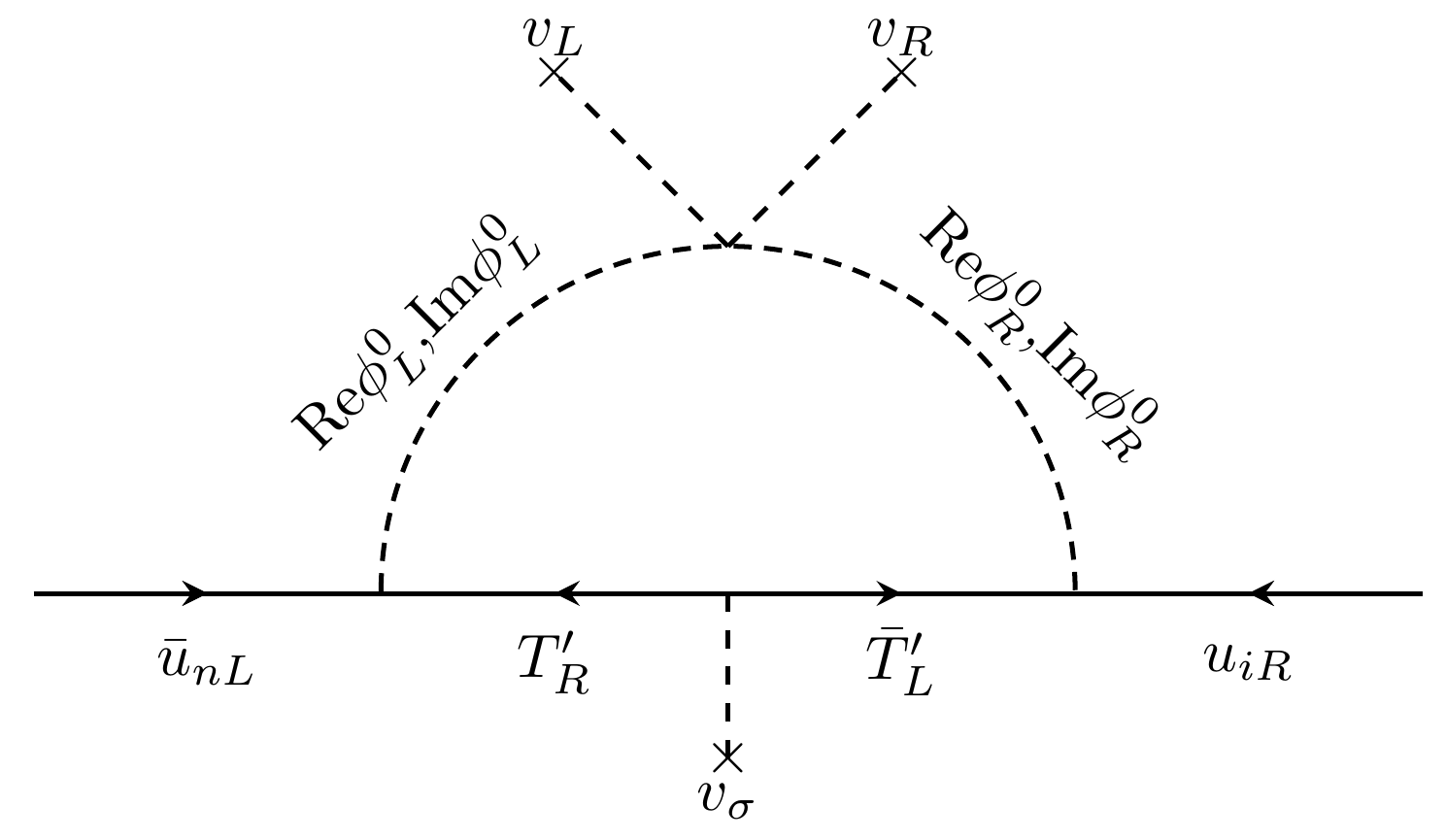} %
\includegraphics[width=8cm, height=4cm]{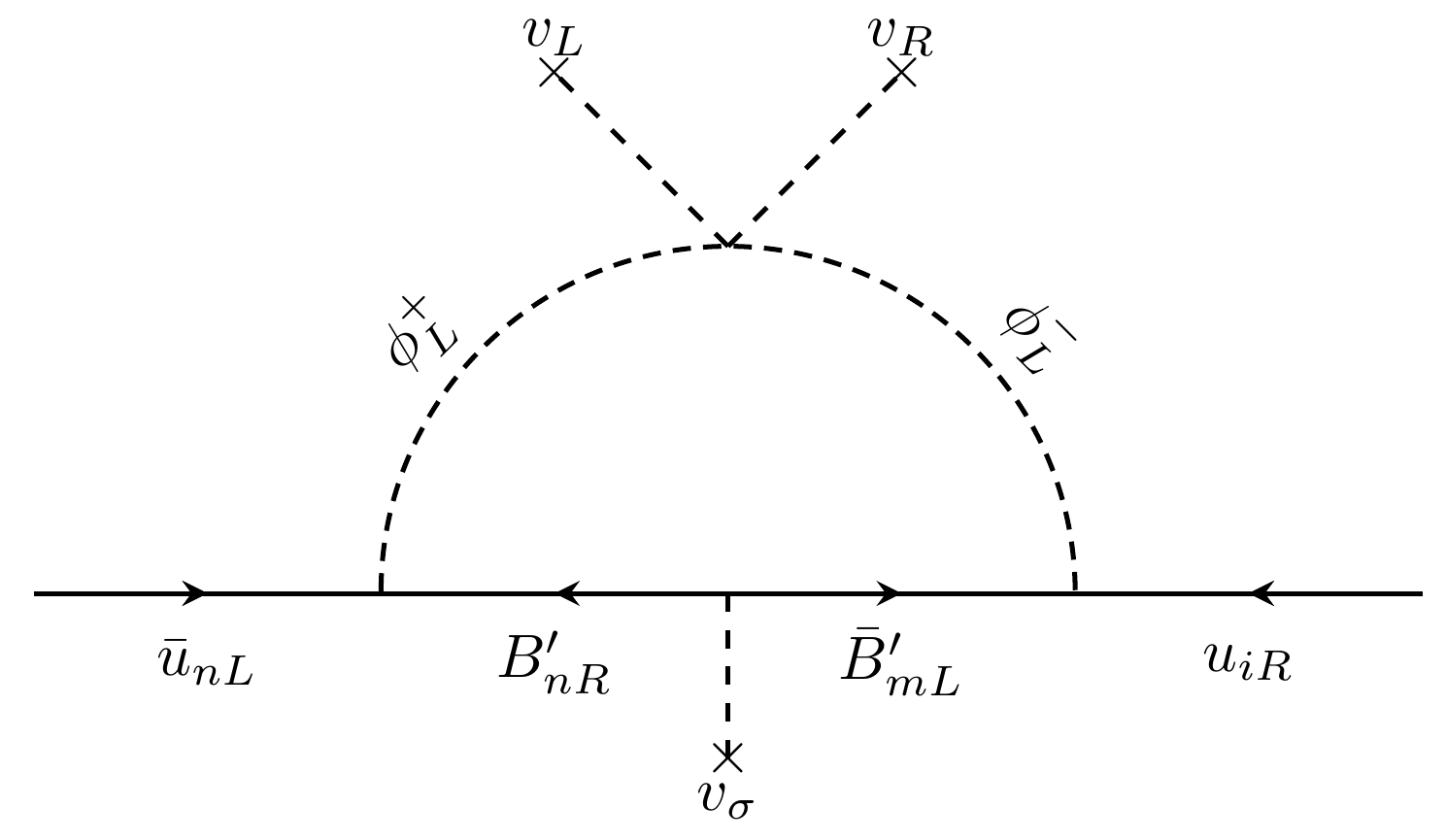}\newline
\includegraphics[width=8cm, height=4cm]{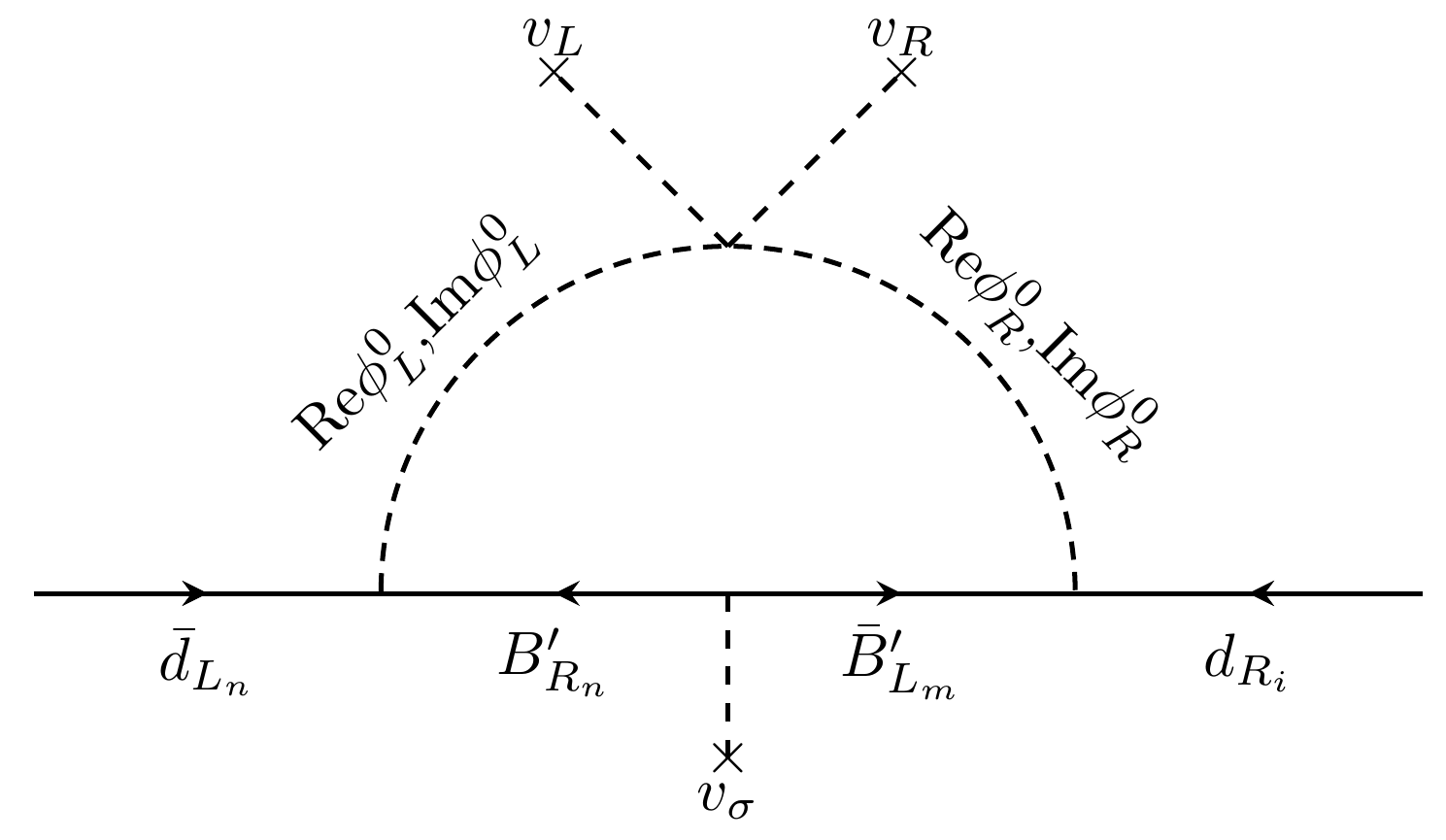} %
\includegraphics[width=8cm, height=4cm]{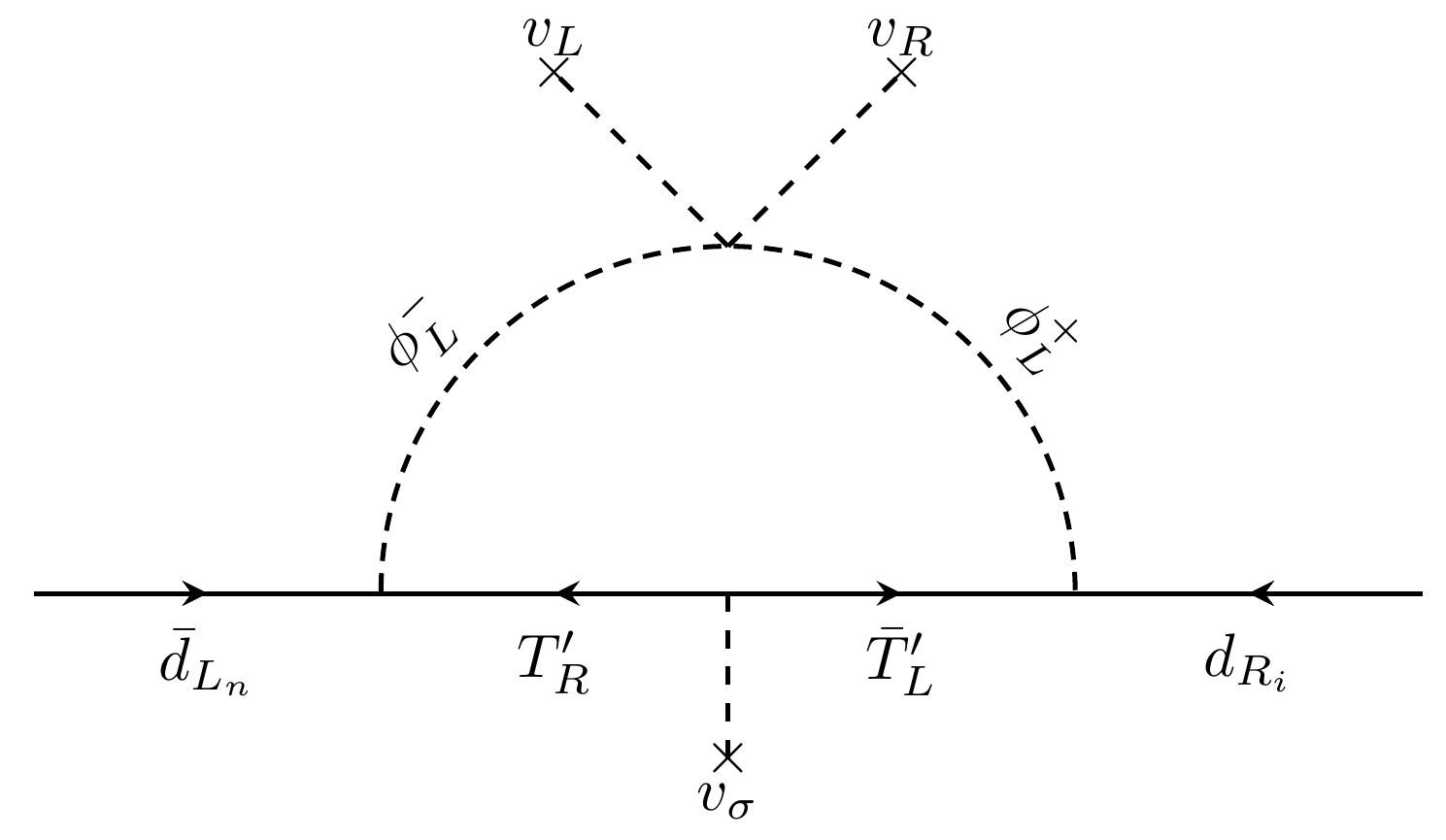}\newline
\includegraphics[width=8cm, height=4cm]{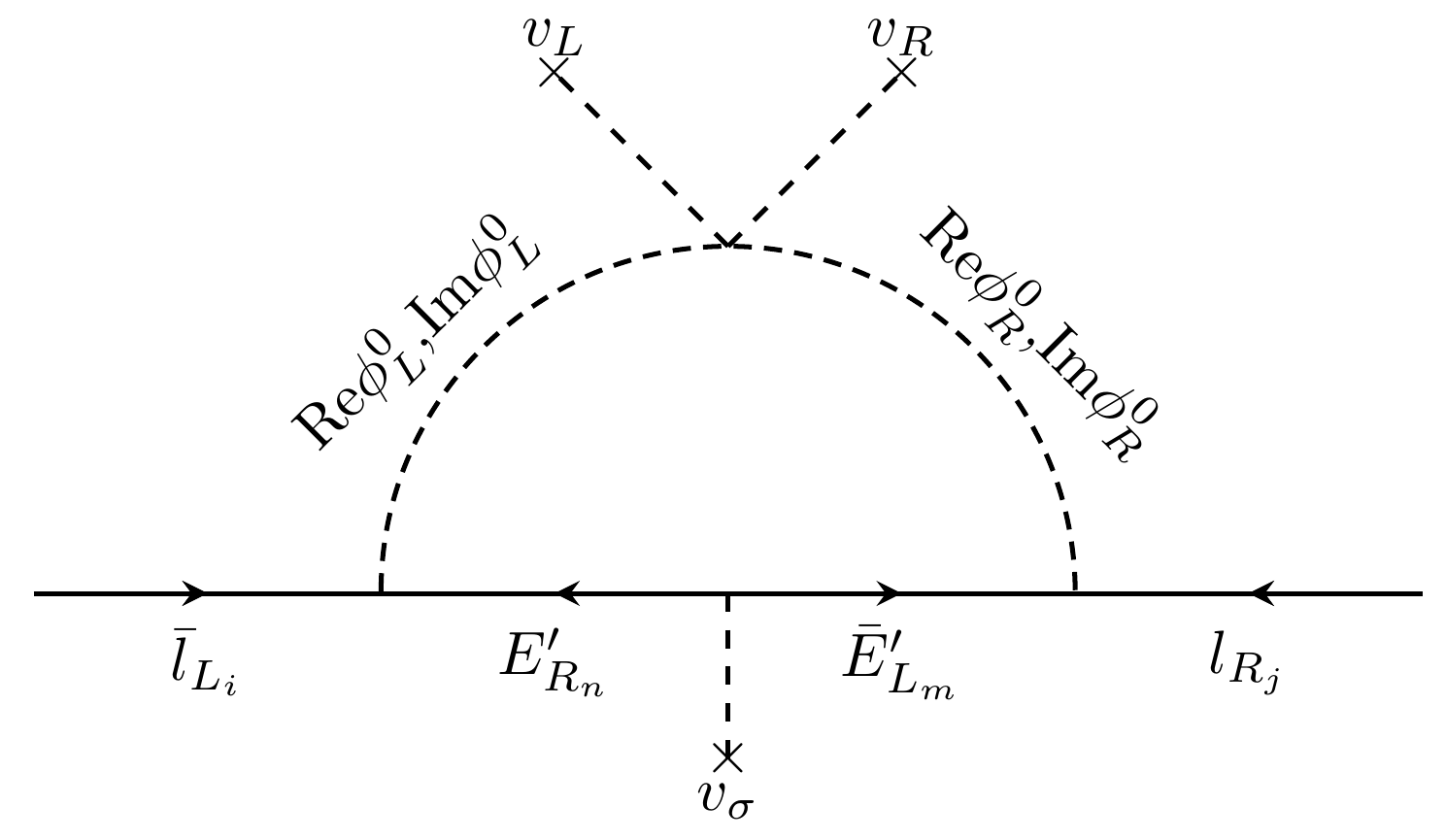}
\caption{Feynman diagrams contributing to the entries of the SM charged
fermion mass matrices. Here, $n,m=1,2$ and $i,j=1,2,3$.}
\label{Diagramschargedfermions}
\end{figure}
\begin{figure}[H]
\includegraphics[width=0.45\textwidth]{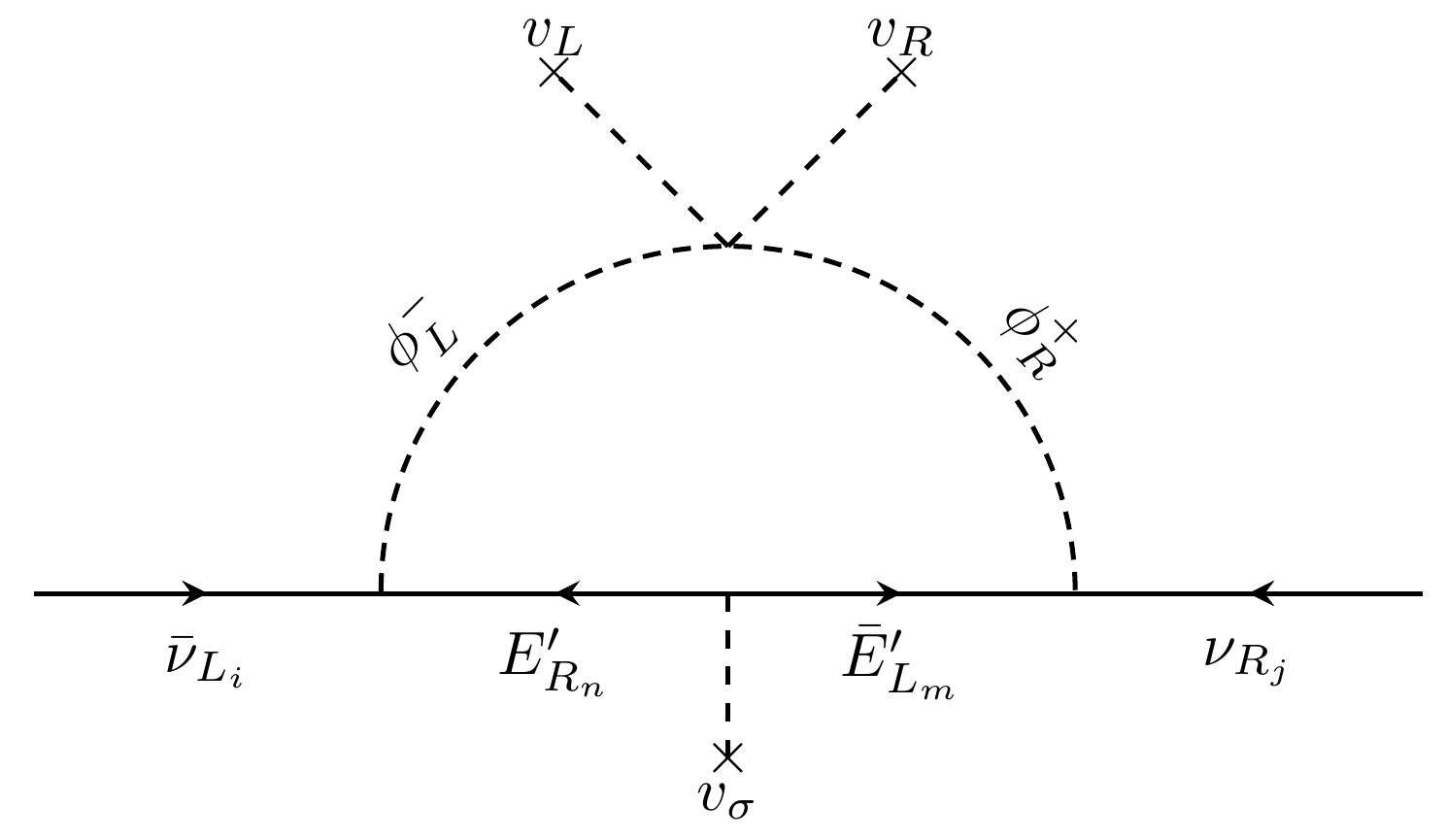} 
\includegraphics[width=0.45\textwidth]{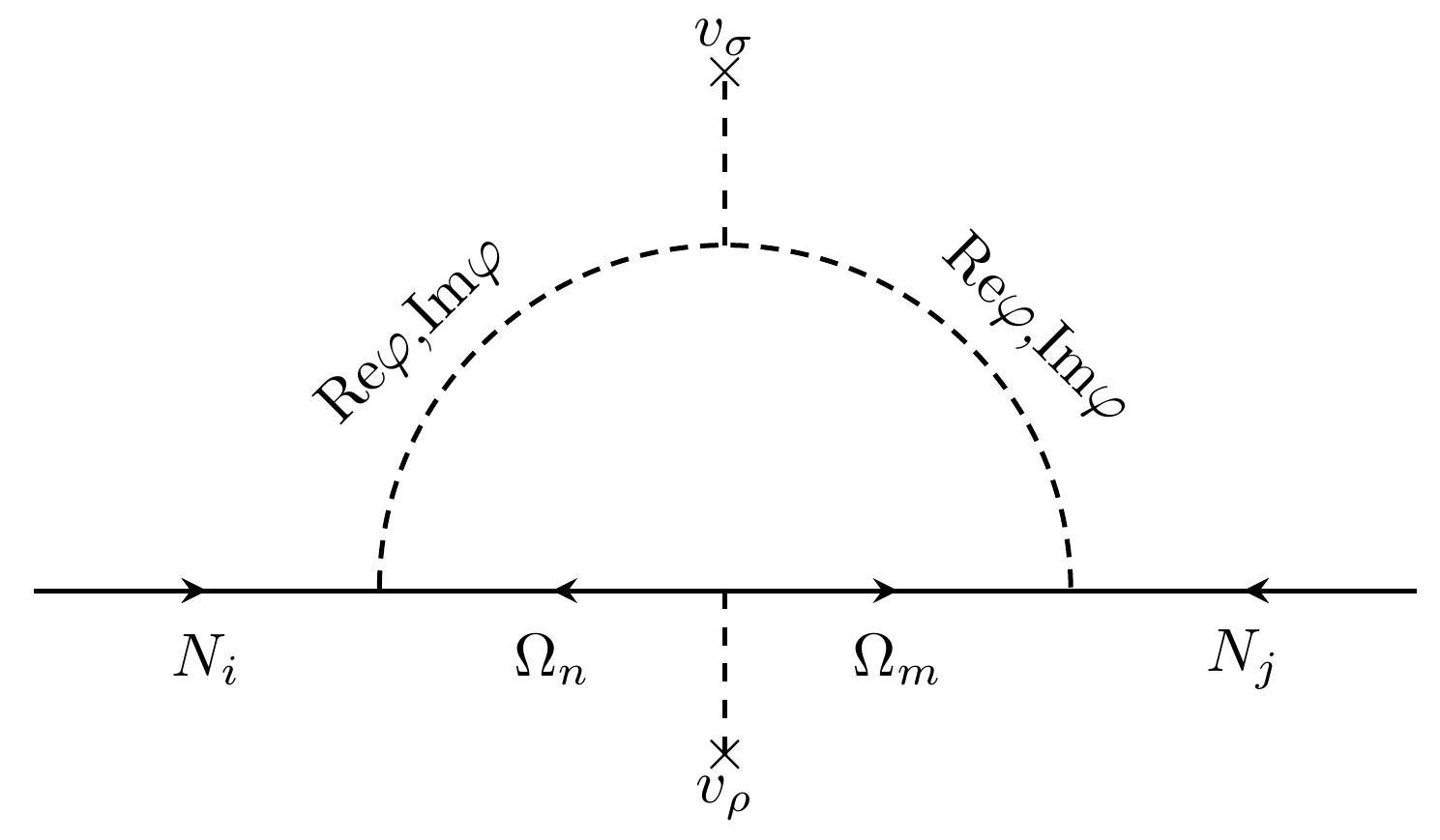}
\caption{Feynman diagrams contributing to the Dirac neutrino and Majorana
neutrino mass submatrices $m_{\protect\nu D}$ and $\protect\mu $. Here, $%
n,m=1,2$ and $i,j=1,2,3$.}
\label{Diagramsneutrinos}
\end{figure}

Finally, to close this section we provide a discussion about collider
signatures of exotic fermions of our model. From the Yukawa interactions it
follows that the charged exotic fermions have mixing mass terms with the SM
charged fermions, which allows the former to decay into any of the scalars
of the model and SM charged fermions. These heavy charged exotic fermions
can be produced in association with the charged fermions and can be pair
produced at the LHC via gluon fusion (for the exotic vector-like quarks
only) and also through the Drell-Yan mechanism. Consequently, observing an
excess of events in the multijet and multilepton final states can be a
signal in support of this model. Regarding the sterile neutrino sector, it
is worth mentioning that the sterile neutrinos can be produced at the LHC in
association with a SM charged lepton, via quark-antiquark annihilation
mediated by a $W^{\prime }$ gauge boson. 8

From the experimental point of view, vector-like quarks are among the most
searched for New Physics states at the LHC. Being colored particles, their
production rates in gluon-gluon fusion is expected to be rather high,
pushing the lower bounds on the the mass scale beyond a TeV scale, typically
landing between 1.4 and 2.0 TeV depending on the underlined assumptions. The
majority of existing searches assume dominant couplings to the
third-generation of SM quarks (see e.g.~Refs.~\cite%
{ATLAS:2018cjd,ATLAS:2018alq,ATLAS:2018uky,ATLAS:2017nap,ATLAS:2016scx,ATLAS:2015ktd}%
). There are a few searches focusing on couplings to light SM quarks, such
as Ref.~\cite{ATLAS:2015lpr} for a pair-production channel, with subsequent
decays into light jets via neutral and charged currents, and in Ref.~\cite%
{ATLAS:2011tvb} for a single-production channel, with the same decay modes.
In the case of charged-current decays, for instance, a down-type vector-like
quark can decay into a $W$-boson and an up-type SM quark, induced by a
possible mixing with down-type SM quarks.

In a more recent analysis of Ref.~\cite{Freitas:2022cno}, Machine learning
techniques have been effectively utilized to obtain robust exclusion bounds
and compute the statistical significance for potential discoveries at future
upgrades of the Large Hadron Collider (LHC). These techniques primarily rely
on analyzing the pair-production of vector-like quarks and their subsequent
charge-current decays. In a particular promising channel, when one of the $%
W^{\pm }$`s decays leptonically into a charged lepton $l=e,\mu $ and the
corresponding neutrino, while the other one decays hadronically into light
jets, it has been shown that the down-type vector-like quarks can be
excluded both at the run-III and the HL phase of the LHC for masses of up to
800 GeV.

Typical production channels of exotic vector-like leptons concern the
vector-boson fusion topologies, as well as via $\gamma/Z$ decays in a
Drell-Yan type reaction -- for pair production -- and via charged-channel $%
W^\pm$ decays -- for single production. The latter process can be
accompanied either by missing energy (in case of associated production of SM
neutrino in the final state) or a charged lepton plus $W$ (in case of
association production of the neutral doublet partner of a doublet-type
exotic lepton). The potential for possible discoveries of exotic vector-like
leptons at both the LHC and future electron and muon colliders has been
recently explored in Refs.~\cite{Freitas:2020ttd,Morais:2021ead} utilising
sophisticated deep learning methods. It has been demonstrated that
weak-doublet vector-like leptons can be probed at the LHC up to a TeV mass
scale, while the weak-singlet vector-like leptons can be barely probed
beyond several hundreds of GeV. In the considered model, only weak-singlet
exotic leptons are present, for which the existing exclusion limits are
weaker than for the doublet ones. This motivates future searches at leptonic
colliders, which can exclude larger parameter-space regions, probing larger
mass scales beyond those accessible at the high-luminosity LHC upgrade.


\section{The scalar potential}

\label{scalarpotential} 

The scalar potential of the model under consideration takes the form: 
\begin{eqnarray}
V &=&\mu _{1}^{2}(\chi _{L}^{\dagger }\chi _{L})+\mu _{2}^{2}(\chi
_{R}^{\dagger }\chi _{R})+\mu _{3}^{2}(\phi _{L}^{\dagger }\phi _{L})+\mu
_{4}^{2}(\phi _{R}^{\dagger }\phi _{R})+\mu _{5}^{2}(\sigma ^{\dagger
}\sigma )+\mu _{6}^{2}(\varphi ^{\dagger }\varphi )+\mu _{7}^{2}(\rho
^{\dagger }\rho )  \notag \\
&&+A_{1}\left[ \left( \phi _{L}^{\dagger }\chi _{L}\right) \varphi +h.c%
\right] +A_{2}\left[ \left( \phi _{R}^{\dagger }\chi _{R}\right) \varphi +h.c%
\right] +A_{3}\left( \varphi ^{2}\sigma ^{\ast }+h.c\right)  \notag \\
&&+\lambda _{1}(\chi _{L}^{\dagger }\chi _{L})^{2}+\lambda _{2}(\chi
_{R}^{\dagger }\chi _{R})^{2}+\lambda _{3}(\chi _{L}^{\dagger }\chi
_{L})(\chi _{R}^{\dagger }\chi _{R})+\lambda _{4}(\phi _{L}^{\dagger }\phi
_{L})^{2}  \notag \\
&&+\lambda _{5}(\phi _{R}^{\dagger }\phi _{R})^{2}+\lambda _{6}(\phi
_{L}^{\dagger }\phi _{L})(\phi _{R}^{\dagger }\phi _{R})+\lambda _{7}\left[
(\phi _{L}^{\dagger }\chi _{L})(\chi _{R}^{\dagger }\phi _{R})+(\chi
_{L}^{\dagger }\phi _{L})(\phi _{R}^{\dagger }\chi _{R})\right]  \notag \\
&&+\lambda _{8}\left[ \left( \varepsilon _{ij}\chi _{L}^{i}\phi
_{L}^{j}\right) \left( \varepsilon _{kl}(\chi _{R}^{\dagger })^{k}\left(
\phi _{R}^{\dagger }\right) ^{l}\right) +h.c\right] +\lambda _{9}(\chi
_{L}^{\dagger }\chi _{L})(\phi _{L}^{\dagger }\phi _{L})+\lambda _{10}(\chi
_{R}^{\dagger }\chi _{R})(\phi _{R}^{\dagger }\phi _{R})  \notag \\
&&+\lambda _{11}(\chi _{L}^{\dagger }\phi _{L})(\phi _{L}^{\dagger }\chi
_{L})+\lambda _{12}(\chi _{R}^{\dagger }\phi _{R})(\phi _{R}^{\dagger }\chi
_{R})+\lambda _{13}(\sigma ^{\dagger }\sigma )^{2}+\lambda _{14}(\varphi
^{\dagger }\varphi )^{2}  \notag \\
&&+\lambda _{15}(\rho ^{\dagger }\rho )^{2}+\lambda _{16}(\sigma ^{\dagger
}\sigma )(\varphi ^{\dagger }\varphi )+\lambda _{17}(\sigma ^{\dagger
}\sigma )(\rho ^{\dagger }\rho )+\lambda _{18}(\varphi ^{\dagger }\varphi
)(\rho ^{\dagger }\rho )  \notag \\
&&+\lambda _{19}\left( \sigma ^{3}\rho ^{\ast }+h.c\right) +\lambda
_{20}(\sigma ^{\dagger }\sigma )(\chi _{L}^{\dagger }\chi _{L})+\lambda
_{21}(\sigma ^{\dagger }\sigma )(\chi _{R}^{\dagger }\chi _{R})+\lambda
_{22}(\sigma ^{\dagger }\sigma )(\phi _{L}^{\dagger }\phi _{L})  \notag \\
&&+\lambda _{23}(\sigma ^{\dagger }\sigma )(\phi _{R}^{\dagger }\phi
_{R})+\lambda _{24}(\varphi ^{\dagger }\varphi )(\chi _{L}^{\dagger }\chi
_{L})+\lambda _{25}(\varphi ^{\dagger }\varphi )(\chi _{R}^{\dagger }\chi
_{R})+\lambda _{26}(\varphi ^{\dagger }\varphi )(\phi _{L}^{\dagger }\phi
_{L})  \notag \\
&&+\lambda _{27}(\varphi ^{\dagger }\varphi )(\phi _{R}^{\dagger }\phi
_{R})+\lambda _{28}(\rho ^{\dagger }\rho )(\chi _{L}^{\dagger }\chi
_{L})+\lambda _{29}(\rho ^{\dagger }\rho )(\chi _{R}^{\dagger }\chi _{R}) 
\notag \\
&&+\lambda _{30}(\rho ^{\dagger }\rho )(\phi _{L}^{\dagger }\phi
_{L})+\lambda _{31}(\rho ^{\dagger }\rho )(\phi _{R}^{\dagger }\phi
_{R})+\lambda _{32}(\chi _{L}^{\dagger }\chi _{L})(\phi _{R}^{\dagger }\phi
_{R})+\lambda _{33}(\phi _{L}^{\dagger }\phi _{L})(\chi _{R}^{\dagger }\chi
_{R})
\end{eqnarray}%
In order to simplify our analysis of the scalar sector, instead of
considering all the terms in the scalar potential, it is enough to consider
up to the $\lambda _{15}$ term, since the rest of the terms would be
important only in the high energy regime. Furthermore, we neglect the terms
of the scalar potential involving a mixing between the singlet scalar fields
and the $SU(2)_{L,R}$ doublet scalars. For the purpose of focusing on the
low-energy phenomenology, we consider the following low-energy scalar
potential: 
\begin{eqnarray}
\tilde{V} &=&\mu _{1}^{2}(\chi _{L}^{\dagger }\chi _{L})+\mu _{2}^{2}(\chi
_{R}^{\dagger }\chi _{R})+\mu _{3}^{2}(\phi _{L}^{\dagger }\phi _{L})+\mu
_{4}^{2}(\phi _{R}^{\dagger }\phi _{R})+\mu _{5}^{2}(\sigma ^{\dagger
}\sigma )+\mu _{6}^{2}(\varphi ^{\dagger }\varphi )+\mu _{7}^{2}(\rho
^{\dagger }\rho )  \notag \\
&&+\lambda _{1}(\chi _{L}^{\dagger }\chi _{L})^{2}+\lambda _{2}(\chi
_{R}^{\dagger }\chi _{R})^{2}+\lambda _{3}(\chi _{L}^{\dagger }\chi
_{L})(\chi _{R}^{\dagger }\chi _{R})+\lambda _{4}(\phi _{L}^{\dagger }\phi
_{L})^{2}+\lambda _{5}(\phi _{R}^{\dagger }\phi _{R})^{2}  \notag \\
&&+\lambda _{6}(\phi _{L}^{\dagger }\phi _{L})(\phi _{R}^{\dagger }\phi
_{R})+\lambda _{7}\left[ (\phi _{L}^{\dagger }\chi _{L})(\chi _{R}^{\dagger
}\phi _{R})+(\chi _{L}^{\dagger }\phi _{L})(\phi _{R}^{\dagger }\chi _{R})%
\right]  \notag \\
&&+\lambda _{8}\left[ \left( \varepsilon _{ij}\chi _{L}^{i}\phi
_{L}^{j}\right) \left( \varepsilon _{kl}(\chi _{R}^{\dagger })^{k}\left(
\phi _{R}^{\dagger }\right) ^{l}\right) +h.c\right] +\lambda _{9}(\chi
_{L}^{\dagger }\chi _{L})(\phi _{L}^{\dagger }\phi _{L})  \notag \\
&&+\lambda _{10}(\chi _{R}^{\dagger }\chi _{R})(\phi _{R}^{\dagger }\phi
_{R})+\lambda _{11}(\chi _{L}^{\dagger }\phi _{L})(\phi _{L}^{\dagger }\chi
_{L})+\lambda _{12}(\chi _{R}^{\dagger }\phi _{R})(\phi _{R}^{\dagger }\chi
_{R})  \notag \\
&&+\lambda _{13}(\sigma ^{\dagger }\sigma )^{2}+\lambda _{14}(\varphi
^{\dagger }\varphi )^{2}+\lambda _{15}(\rho ^{\dagger }\rho
)^{2}+A_{3}\left( \varphi ^{2}\sigma ^{\ast }+h.c\right)  \notag \\
&&+\lambda _{32}(\chi _{L}^{\dagger }\chi _{L})(\phi _{R}^{\dagger }\phi
_{R})+\lambda _{33}(\phi _{L}^{\dagger }\phi _{L})(\chi _{R}^{\dagger }\chi
_{R})\,.
\end{eqnarray}%
The minimization of this scalar potential gives rise to the effective mass
terms: 
\begin{equation}
\begin{split}
\mu _{1}^{2}& =-\lambda _{1}v_{L}^{2}-\frac{1}{2}\lambda _{3}v_{R}^{2}, \\
\mu _{2}^{2}& =-\lambda _{2}v_{R}^{2}-\frac{1}{2}\lambda _{3}v_{L}^{2}, \\
\mu _{5}^{2}& =-\lambda _{13}v_{\sigma }^{2}, \\
\mu _{7}^{2}& =-\lambda _{15}v_{\rho }^{2}\,.
\end{split}%
\end{equation}

The squared mass matrix for the CP-even neutral fields, in the basis of $%
\left( \func{Re}\chi _{L}^{0},\func{Re}\chi _{R}^{0},\func{Re}\phi _{L}^{0},%
\func{Re}\phi _{R}^{0},\func{Re}\sigma ,\func{Re}\varphi ,\func{Re}\rho
_{R}\right) $, reads: 
\begin{eqnarray}  \label{M2CPeven}
&&\mathcal{M}_{\func{CP} even}^{2}=  \notag \\
&&\left( 
\begin{array}{ccccccc}
2\lambda _{1}v_{L}^{2} & \lambda _{3}v_{L}v_{R} & 0 & 0 & 0 & 0 & 0 \\ 
\lambda _{3}v_{L}v_{R} & 2\lambda _{2}v_{R}^{2} & 0 & 0 & 0 & 0 & 0 \\ 
0 & 0 & \frac{1}{2}\lambda _{33}v_{R}^{2}+\frac{1}{2}\left( \lambda
_{9}+\lambda _{11}\right) v_{L}^{2}+\mu _{3}^{2} & \frac{1}{2}\lambda
_{7}v_{L}v_{R} & 0 & 0 & 0 \\ 
0 & 0 & \frac{1}{2}\lambda _{7}v_{L}v_{R} & \frac{1}{2}\lambda
_{32}v_{L}^{2}+\frac{1}{2}\left( \lambda _{10}+\lambda _{12}\right)
v_{R}^{2}+\mu _{4}^{2} & 0 & 0 & 0 \\ 
0 & 0 & 0 & 0 & 2\lambda _{12}v_{\sigma }^{2} & 0 & 0 \\ 
0 & 0 & 0 & 0 & 0 & \mu _{6}^{2}+\sqrt{2}A_{3}v_{\sigma } & 0 \\ 
0 & 0 & 0 & 0 & 0 & 0 & 2\lambda _{16}v_{\rho }^{2} \\ 
&  &  &  &  &  & 
\end{array}%
\right) \,.  \notag \\
\end{eqnarray}

The mixing between the $\chi _{L}$ and $\chi _{R}$ fields can be
approximated by $\frac{v_{L}}{v_{R}}$ and this value is suppressed in the
case where the VEV of $v_{L}$ is the SM Higgs VEV of $246\func{GeV}$ and
that of $v_{R}$ is equal to $10\func{TeV}$, 
\begin{equation}
\theta _{\chi }\approx \frac{v_{L}}{v_{R}}=\frac{246}{10000}=0.0246 \,.
\end{equation}
This shows that the CP-even neutral component of the $SU\left( 2\right) _{L}$
scalar doublet corresponds to the $126$ GeV SM like Higgs boson, which in
our model has couplings very close to the SM expectation, which is
consistent with the experimental data. On the other hand, the mixing between 
$\func{Re}\phi _{L}^{0}$ and $\func{Re}\phi _{R}^{0}$ depends on the free
parameters $\lambda _{7}$, $\mu _{3}^{2}$ and $\mu _{4}^{2}$, so it can be
sizeable. Furthermore, this mixing gives rise to the heavy CP-even scalars $%
H_{1}$ and $H_{2}$ defined as follows: 
\begin{eqnarray}
\left( 
\begin{array}{c}
H_{1} \\ 
H_{2}%
\end{array}%
\right) &=&\left( 
\begin{array}{cc}
\cos \theta _{H} & \sin \theta _{H} \\ 
-\sin \theta _{H} & \cos \theta _{H}%
\end{array}%
\right) \left( 
\begin{array}{c}
\func{Re}\phi _{L}^{0} \\ 
\func{Re}\phi _{R}^{0}%
\end{array}%
\right) ,\hspace{1cm}  \notag \\
\tan 2\theta _{H} &=&\frac{\lambda _{7}v_{L}v_{R}}{\frac{1}{2}\lambda
_{33}v_{R}^{2}+\frac{1}{2}\left( \lambda _{9}+\lambda _{11}\right)
v_{L}^{2}+\mu _{3}^{2}-\frac{1}{2}\left( \lambda _{10}+\lambda _{12}\right)
v_{R}^{2}-\frac{1}{2}\lambda _{32}v_{L}^{2}-\mu _{4}^{2}}.
\end{eqnarray}

The squared mass matrix for the CP-odd neutral fields in the basis of $%
\left( \func{Im}\chi _{L}^{0},\func{Im}\chi _{R}^{0},\func{Im}\phi _{L}^{0},%
\func{Im}\phi _{R}^{0},\func{Im}\sigma ,\func{Im}\varphi ,\func{Im}\rho
\right) $ is given by: 
\begin{equation}
\mathcal{M}_{\func{CP}odd}^{2}=\left( 
\begin{array}{ccccccc}
0 & 0 & 0 & 0 & 0 & 0 & 0 \\ 
0 & 0 & 0 & 0 & 0 & 0 & 0 \\ 
0 & 0 & \frac{1}{2}\lambda _{33}v_{R}^{2}+\frac{1}{2}\left( \lambda
_{9}+\lambda _{11}\right) v_{L}^{2}+\mu _{3}^{2} & \frac{1}{2}\lambda
_{7}v_{L}v_{R} & 0 & 0 & 0 \\ 
0 & 0 & \frac{1}{2}\lambda _{7}v_{L}v_{R} & \frac{1}{2}\lambda
_{32}v_{L}^{2}+\frac{1}{2}\left( \lambda _{10}+\lambda _{12}\right)
v_{R}^{2}+\mu _{4}^{2} & 0 & 0 & 0 \\ 
0 & 0 & 0 & 0 & 0 & 0 & 0 \\ 
0 & 0 & 0 & 0 & 0 & \mu _{6}^{2}-\sqrt{2}A_{3}v_{\sigma } & 0 \\ 
0 & 0 & 0 & 0 & 0 & 0 & 0 \\ 
&  &  &  &  &  & 
\end{array}%
\right) \,.  \label{M2CPodd}
\end{equation}%
Then, in the CP-odd neutral scalar sector there are four massless neutral
fields, from which two of them, namely, $\func{Im}\chi _{L}^{0}$ and $\chi
_{R}^{0}$, correspond to the Goldstone bosons associated with the
longitudinal components of the $Z$ and $Z^{\prime }$ gauge bosons. The
remaining massless CP-odd neutral scalars, i.e., $\func{Im}\sigma $ and $%
\func{Im}\rho $, correspond to Majorons, massless scalars arising from the
spontaneous breaking of the $U\left( 1\right) _{X}$ global symmetry. These
Majorons are harmless since they are gauge singlets, annd can acquire
non-vanishing masses by including the soft-breaking mass terms $\mu
_{sb}^{\left( 1\right) }\left( \sigma ^{2}+h.c\right) $ and $\mu
_{sb}^{\left( 2\right) }\left( \rho ^{2}+h.c\right) $ in the scalar
potential. As in the CP-even scalar sector, the mixing between $\func{Im}%
\phi _{L}^{0}$ and $\func{Im}\phi _{R}$ depends on the free parameters $%
\lambda _{7}$, $\mu _{3}^{2}$ and $\mu _{4}^{2}$. Furthermore, there are
three massive scalars in the CP-odd scalar sector, i.e, $\func{Im}\varphi $
as well as $A_{1}$ and $A_{2}$, which are defined as follows: 
\begin{eqnarray}
\left( 
\begin{array}{c}
\ A_{1} \\ 
A_{2}%
\end{array}%
\right) &=&\left( 
\begin{array}{cc}
\cos \theta _{A} & \sin \theta _{A} \\ 
-\sin \theta _{A} & \cos \theta _{A}%
\end{array}%
\right) \left( 
\begin{array}{c}
\func{Im}\phi _{L}^{0} \\ 
\func{Im}\phi _{R}^{0}%
\end{array}%
\right) , \\
\tan 2\theta _{A} &=&\tan 2\theta _{H}  \notag \\
&=&\frac{\lambda _{7}v_{L}v_{R}}{\frac{1}{2}\lambda _{33}v_{R}^{2}+\frac{1}{2%
}\left( \lambda _{9}+\lambda _{11}\right) v_{L}^{2}+\mu _{3}^{2}-\frac{1}{2}%
\lambda _{32}v_{L}^{2}-\frac{1}{2}\left( \lambda _{10}+\lambda _{12}\right)
v_{R}^{2}-\mu _{4}^{2}}.  \notag
\end{eqnarray}

The squared mass matrix for the electrically charged scalar fields in the
basis of $\left( \chi _{L}^{\pm },\chi _{R}^{\pm },\phi _{L}^{\pm },\phi
_{R}^{\pm }\right) $ takes the form: 
\begin{equation}
\mathcal{M}_{\func{charged}}^{2}=\left( 
\begin{array}{cccc}
0 & 0 & 0 & 0 \\ 
0 & 0 & 0 & 0 \\ 
0 & 0 & \frac{1}{2}\lambda _{33}v_{R}^{2}+\frac{1}{2}\lambda
_{9}v_{L}^{2}+\mu _{3}^{2} & \frac{1}{2}\lambda _{8}v_{L}v_{R} \\ 
0 & 0 & \frac{1}{2}\lambda _{8}v_{L}v_{R} & \frac{1}{2}\lambda
_{32}v_{L}^{2}+\frac{1}{2}\lambda _{10}v_{R}^{2}+\mu _{4}^{2} \\ 
&  &  & 
\end{array}%
\right)
\end{equation}%
Then the electrically charged scalar mass spectrum is composed of the
massless scalar eigenstates $\chi _{L}^{\pm }$ and $\chi _{R}^{\pm }$, as
well as the massive scalars $H_{1}^{\pm }$ and $H_{2}^{\pm }$. The
electrically charged massless scalars $\chi _{L}^{\pm }$ and $\chi _{R}^{\pm
}$ are the Goldstone bosons associated with the longitudinal components of
the $W^{\pm }$ and $W^{\prime \pm }$ gauge bosons, respectively.
Furthermore, the massive scalars $H_{1}^{\pm }$ and $H_{2}^{\pm }$ are
defined as follows: 
\begin{eqnarray}
\left( 
\begin{array}{c}
H_{1}^{\pm } \\ 
H_{2}^{\pm }%
\end{array}%
\right) &=&\left( 
\begin{array}{cc}
\cos \theta & \sin \theta \\ 
-\sin \theta & \cos \theta%
\end{array}%
\right) \left( 
\begin{array}{c}
\func{Re}\phi _{L}^{\pm } \\ 
\func{Re}\phi _{R}^{\pm }%
\end{array}%
\right) ,\hspace{1cm}\hspace{1cm}  \notag \\
\tan 2\theta &=&\frac{\lambda _{8}v_{L}v_{R}}{\frac{1}{2}\lambda
_{33}v_{R}^{2}+\frac{1}{2}\lambda _{9}v_{L}^{2}+\mu _{3}^{2}-\frac{1}{2}%
\lambda _{10}v_{R}^{2}-\frac{1}{2}\lambda _{32}v_{L}^{2}-\mu _{4}^{2}}.
\end{eqnarray}


\section{Fermion mass matrices}

\label{fermionmasses} 

From the Yukawa interactions, we find that the mass matrices for charged
fermions in the basis $(\overline{u}_{L_{1}},\overline{u}_{L_{2}},\overline{u%
}_{L_{3}},\overline{T}_{L_{1}},\overline{T}_{L_{2}})$-$%
(u_{R_{1}},u_{R_{2}},u_{3R},T_{R_{1}},T_{R_{2}})$, $(\overline{d}_{L_{1}},%
\overline{d}_{L_{2}},\overline{d}_{L_{3}},\overline{B}_{L})$-$%
(d_{R_{1}},d_{R_{2}},d_{3R},B_{R})$ and $(\overline{l}_{L_{1}},\overline{l}%
_{L_{2}},\overline{l}_{L_{3}},\overline{E}_{L})$ -$%
(l_{R_{1}},l_{R_{2}},l_{3R},E_{R})$ are respectively given:

\begin{eqnarray}
M_{U} &=&\left( 
\begin{array}{cc}
\Delta _{U} & A_{U} \\ 
B_{U} & M_{T}%
\end{array}%
\right) ,\hspace{1cm}\hspace{1cm}A_{U}=\left( 
\begin{array}{cc}
0 & x_{1}^{\left( T\right) } \\ 
0 & x_{2}^{\left( T\right) } \\ 
x_{3}^{\left( T\right) } & 0%
\end{array}%
\right) \frac{v_{_{L}}}{\sqrt{2}},  \notag \\
B_{U} &=&\left( 
\begin{array}{ccc}
z_{11}^{\left( T\right) }, & z_{12}^{\left( T\right) } & z_{13}^{\left(
T\right) } \\ 
z_{21}^{\left( T\right) }, & z_{22}^{\left( T\right) } & z_{23}^{\left(
T\right) }%
\end{array}%
\right) \frac{v_{R}}{\sqrt{2}},\hspace{1cm}\hspace{1cm}M_{T}=\left( 
\begin{array}{cc}
m_{T} & 0 \\ 
0 & y_{T}\frac{v_{\sigma }}{\sqrt{2}}%
\end{array}%
\right) ,  \label{MU}
\end{eqnarray}%
\begin{eqnarray}
M_{D} &=&\left( 
\begin{array}{cc}
\Delta _{D} & A_{D} \\ 
B_{D} & m_{B}%
\end{array}%
\right) ,\hspace{1cm}\hspace{1cm}A_{D}=\left( 
\begin{array}{c}
0 \\ 
0 \\ 
x_{3}^{\left( B\right) }%
\end{array}%
\right) \frac{v_{L}}{\sqrt{2}},  \notag \\
B_{D} &=&\left( 
\begin{array}{ccc}
z_{1}^{\left( B\right) }, & z_{2}^{\left( B\right) }, & z_{3}^{\left(
B\right) }%
\end{array}%
\right) \frac{v_{R}}{\sqrt{2}},  \label{MD}
\end{eqnarray}%
\begin{eqnarray}
M_{E} &=&\left( 
\begin{array}{cc}
\Delta _{E} & A_{E} \\ 
B_{E} & m_{E}%
\end{array}%
\right) ,\hspace{1cm}\hspace{1cm}A_{E}=\left( 
\begin{array}{c}
x_{1}^{\left( E\right) } \\ 
x_{2}^{\left( E\right) } \\ 
x_{3}^{\left( E\right) }%
\end{array}%
\right) \frac{v_{L}}{\sqrt{2}},  \notag \\
B_{E} &=&\left( 
\begin{array}{ccc}
z_{1}^{\left( E\right) }, & z_{2}^{\left( E\right) }, & z_{3}^{\left(
E\right) }%
\end{array}%
\right) \frac{v_{R}}{\sqrt{2}},\hspace{1cm}\hspace{1cm}m_{E}=y_{E}\frac{%
v_{\sigma }}{\sqrt{2}}.  \label{ME}
\end{eqnarray}%
Furthermore, due to the preserved matter parity symmetry arising from the
spontaneous breaking of the $U\left( 1\right) _{X}$ global symmetry, the
exotic up-type quarks $T^{\prime }$ and $B_{n}^{\prime }$\ ($n=1,2$)\ do not
mix with the remaining up-type quark fields. For the same reason, the
charged exotic leptons $E_{n}^{\prime }$ ($n=1,2$), do not mix with the
remaining charged leptons. As seen from Eqs.~(\ref{MU}), (\ref{MD}) and (\ref%
{ME}), the exotic heavy vector-like fermions $T_{n}$ ($n=1,2$), $B$ and $E$
mix with the SM fermions. The masses of these vector-like fermions are much
larger than the electroweak symmetry breaking scale, since the gauge singlet
scalars $\eta $ and $\rho $ are assumed to acquire VEVs around the scale of
breaking of the LR symmetry, taken to be much larger than the Fermi scale.
Consequently, the charged exotic vector-like fermions $T_{n}$ ($n=1,2$), $B$
and $E$ induce a seesaw mechanism that gives rise to tree level masses to
the third family of SM charged fermions, as well as to the charm quark. The
remaining charged exotic vector-like fermions $T^{\prime }$, $B_{n}^{\prime
} $\ and $E_{n}^{\prime }$ ($n=1,2$)\ mediate a one-loop level radiative
seesaw mechanism that generates the masses of the light up, down and strange
quarks as well as of the electron and muon masses. Thus, the SM charged
fermion mass matrices take the form: 
\begin{eqnarray}
\widetilde{M}_{U} &=&\Delta _{U}-A_{U}M_{T}^{-1}B_{U}, \\
\widetilde{M}_{D} &=&\Delta _{D}-A_{D}m_{B}^{-1}B_{D}, \\
\widetilde{M}_{E} &=&\Delta _{E}-A_{E}m_{E}^{-1}B_{E},
\end{eqnarray}%
Where $\Delta _{U}$, $\Delta _{D}$ and $\Delta _{E}$ are the one-loop
contributions to the SM charged fermion mass matrices which are given by: 
\begin{eqnarray}
\Delta _{U} &=&\frac{W_{Q}F_{U}R_{Q}}{16\pi ^{2}}+\frac{X_{Q}G_{U}Z_{Q}}{%
16\pi ^{2}},\hspace{0.7cm}F_{U}=m_{T^{\prime }}K_{U},\hspace{0.7cm}\left(
W_{Q}\right) _{i}=w_{i}^{\left( T^{\prime }\right) }\delta _{in},\hspace{%
0.7cm}\left( R_{Q}\right) _{i}=r_{i}^{\left( T^{\prime }\right) },\hspace{%
0.7cm}n=1,2,  \notag \\
K_{U} &=&\left[ f\left( m_{H_{1}}^{2},m_{T^{\prime }}^{2}\right) -f\left(
m_{H_{2}}^{2},m_{T^{\prime }}^{2}\right) \right] \sin 2\theta _{H}+\left[
f\left( m_{A_{1}}^{2},m_{T^{\prime }}^{2}\right) -f\left(
m_{A_{2}}^{2},m_{T^{\prime }}^{2}\right) \right] \sin 2\theta _{A},\hspace{%
0.7cm}i=1,2,3,  \notag \\
&=&2\left[ f\left( m_{H_{1}}^{2},m_{T^{\prime }}^{2}\right) -f\left(
m_{H_{2}}^{2},m_{T^{\prime }}^{2}\right) \right] \sin 2\theta _{H},\hspace{%
1cm}  \notag \\
G_{U} &=&\left( 
\begin{array}{cc}
m_{B_{1}^{\prime }}\widetilde{K}_{U}^{\left( 1\right) } & 0 \\ 
0 & m_{B_{2}^{\prime }}\widetilde{K}_{U}^{\left( 2\right) }%
\end{array}%
\right) ,\hspace{1cm}\widetilde{K}_{U}^{\left( n\right) }=\left[ f\left(
m_{H_{1}^{\pm }}^{2},m_{B_{n}^{\prime }}^{2}\right) -f\left( m_{H_{2}^{\pm
}}^{2},m_{B_{n}^{\prime }}^{2}\right) \right] \sin 2\theta ,\hspace{1cm} 
\notag \\
\Delta _{D} &=&\frac{X_{Q}G_{D}Z_{Q}}{16\pi ^{2}}+\frac{W_{Q}F_{D}R_{Q}}{%
16\pi ^{2}},\hspace{0.7cm}F_{D}=m_{T^{\prime }}K_{D},\hspace{0.7cm}\left(
R_{U}\right) _{i}=r_{i}^{\left( T^{\prime }\right) },\hspace{0.7cm}\left(
X_{Q}\right) _{in}=w_{in}^{\left( B^{\prime }\right) }\delta _{im},\hspace{%
0.7cm}\left( Z_{Q}\right) _{j}=r_{nj}^{\left( B^{\prime }\right) },  \notag
\\
K_{D} &=&\left[ f\left( m_{H_{1}^{\pm }}^{2},m_{T^{\prime }}^{2}\right)
-f\left( m_{H_{2}^{\pm }}^{2},m_{T^{\prime }}^{2}\right) \right] \sin
2\theta ,\hspace{1cm}n=1,2,\hspace{1cm}i=1,2,3,  \notag \\
G_{D} &=&\left( 
\begin{array}{cc}
m_{B_{1}^{\prime }}\widetilde{K}_{D}^{\left( 1\right) } & 0 \\ 
0 & m_{B_{2}^{\prime }}\widetilde{K}_{D}^{\left( 2\right) }%
\end{array}%
\right) ,\hspace{1cm}\widetilde{K}_{D}^{\left( n\right) }=2\left[ f\left(
m_{H_{1}}^{2},m_{B_{n}^{\prime }}^{2}\right) -f\left(
m_{H_{2}}^{2},m_{B_{n}^{\prime }}^{2}\right) \right] \sin 2\theta _{H}, 
\notag \\
\Delta _{E} &=&\frac{W_{E}F_{E}R_{E}}{16\pi ^{2}},\hspace{1cm}F_{E}=\left( 
\begin{array}{cc}
m_{E_{1}^{\prime }}\widetilde{K}_{E}^{\left( 1\right) } & 0 \\ 
0 & m_{E_{2}^{\prime }}\widetilde{K}_{E}^{\left( 2\right) }%
\end{array}%
\right) ,  \notag \\
\widetilde{K}_{E}^{\left( n\right) } &=&\left[ f\left(
m_{H_{1}}^{2},m_{E_{n}^{\prime }}^{2}\right) -f\left(
m_{H_{2}}^{2},m_{E_{n}^{\prime }}^{2}\right) \right] \sin 2\theta _{H}+\left[
f\left( m_{A_{1}}^{2},m_{E_{n}^{\prime }}^{2}\right) -f\left(
m_{A_{2}}^{2},m_{E_{n}^{\prime }}^{2}\right) \right] \sin 2\theta _{A} 
\notag \\
&=&2\left[ f\left( m_{H_{1}}^{2},m_{E_{n}^{\prime }}^{2}\right) -f\left(
m_{H_{2}}^{2},m_{E_{n}^{\prime }}^{2}\right) \right] \sin 2\theta _{H} 
\notag \\
\left( W_{E}\right) _{in} &=&w_{in}^{\left( E^{\prime }\right) },\hspace{1cm}%
\left( R_{E}\right) _{nj}=r_{nj}^{\left( E^{\prime }\right) },\hspace{1cm}%
\left( A_{E}\right) _{i}=x_{i}^{\left( E\right) },\hspace{1cm}\left(
B_{E}\right) _{j}=z_{j}^{\left( E\right) },  \label{Delta}
\end{eqnarray}%
where $f\left( m_{1},m_{2}\right) $ is given by: 
\begin{equation}
f\left( m_{1},m_{2}\right) =\frac{m_{1}^{2}}{m_{1}^{2}-m_{2}^{2}}\ln \left( 
\frac{m_{1}^{2}}{m_{2}^{2}}\right) .
\end{equation}

For $\theta \ll \theta _{H},\theta _{A}$, the SM charged fermion mass
matrices can be parametrized as follows: 
\begin{eqnarray}
\widetilde{M}_{U} &\simeq &\widetilde{A}_{U}J_{U}^{-1}\widetilde{B}_{U}^{T},%
\hspace{0.7cm}\hspace{0.7cm}J_{U}=\left( 
\begin{array}{ccc}
\frac{1}{16\pi ^{2}}m_{T^{\prime }}K_{U} & 0 & 0 \\ 
0 & m_{T_{1}} & 0 \\ 
0 & 0 & m_{T_{2}}%
\end{array}%
\right) ,  \label{Mu} \\
\widetilde{M}_{D} &\simeq &\widetilde{A}_{D}J_{D}^{-1}\widetilde{B}_{D}^{T},%
\hspace{0.7cm}\hspace{0.7cm}J_{D}=\left( 
\begin{array}{ccc}
\frac{1}{16\pi ^{2}}m_{B_{1}^{\prime }}\widetilde{K}_{D}^{\left( 1\right) }
& 0 & 0 \\ 
0 & \frac{1}{16\pi ^{2}}m_{B_{2}^{\prime }}\widetilde{K}_{D}^{\left(
2\right) } & 0 \\ 
0 & 0 & m_{B}%
\end{array}%
\right) ,  \label{Md} \\
\widetilde{M}_{E} &=&\widetilde{A}_{E}J_{E}^{-1}\widetilde{B}_{E}^{T},%
\hspace{0.7cm}\hspace{0.7cm}J_{E}=\left( 
\begin{array}{ccc}
\frac{1}{16\pi ^{2}}m_{E_{1}^{\prime }}\widetilde{K}_{E}^{\left( 1\right) }
& 0 & 0 \\ 
0 & \frac{1}{16\pi ^{2}}m_{E_{2}^{\prime }}\widetilde{K}_{E}^{\left(
2\right) } & 0 \\ 
0 & 0 & m_{E}%
\end{array}%
\right) \,,  \label{Ml}
\end{eqnarray}%
where 
\begin{eqnarray}
\widetilde{A}_{U} &=&V_{L}^{\left( U\right) }M_{u}^{\frac{1}{2}}J_{U}^{\frac{%
1}{2}},\hspace{0.7cm}\hspace{0.7cm}\widetilde{B}_{U}=V_{R}^{\left( U\right)
}M_{u}^{\frac{1}{2}}J_{U}^{\frac{1}{2}},\hspace{0.7cm}\hspace{0.7cm}%
M_{u}=\left( 
\begin{array}{ccc}
m_{u} & 0 & 0 \\ 
0 & m_{c} & 0 \\ 
0 & 0 & m_{t}%
\end{array}%
\right) , \\
\widetilde{A}_{D} &=&V_{L}^{\left( D\right) }M_{d}^{\frac{1}{2}}J_{D}^{\frac{%
1}{2}},\hspace{0.7cm}\hspace{0.7cm}\widetilde{B}_{D}=V_{R}^{\left( D\right)
}M_{d}^{\frac{1}{2}}J_{D}^{\frac{1}{2}},\hspace{0.7cm}\hspace{0.7cm}%
M_{d}=\left( 
\begin{array}{ccc}
m_{d} & 0 & 0 \\ 
0 & m_{s} & 0 \\ 
0 & 0 & m_{b}%
\end{array}%
\right) , \\
\widetilde{A}_{E} &=&V_{L}^{\left( E\right) }M_{l}^{\frac{1}{2}}J_{E}^{\frac{%
1}{2}},\hspace{0.7cm}\hspace{0.7cm}\widetilde{B}_{E}=V_{R}^{\left( E\right)
}M_{l}^{\frac{1}{2}}J_{E}^{\frac{1}{2}},\hspace{0.7cm}\hspace{0.7cm}%
M_{l}=\left( 
\begin{array}{ccc}
m_{e} & 0 & 0 \\ 
0 & m_{\mu } & 0 \\ 
0 & 0 & m_{\tau }%
\end{array}%
\right)
\end{eqnarray}

The SM charged fermion mass matrices can be rewritten, in the
scenario $\theta \ll \theta _{H},\theta _{A}$, as follows: 
\begin{eqnarray}
\widetilde{M}_{U} &\simeq &-\left( 
\begin{array}{ccc}
x_{1}^{\left( T\right) }\frac{z_{21}^{\left( T\right) }}{m_{T_{2}}} & 
x_{1}^{\left( T\right) }\frac{z_{22}^{\left( T\right) }}{m_{T_{2}}} & 
x_{1}^{\left( T\right) }\frac{z_{23}^{\left( T\right) }}{m_{T_{2}}} \\ 
x_{2}^{\left( T\right) }\frac{z_{21}^{\left( T\right) }}{m_{T_{2}}} & 
x_{2}^{\left( T\right) }\frac{z_{22}^{\left( T\right) }}{m_{T_{2}}} & 
x_{2}^{\left( T\right) }\frac{z_{23}^{\left( T\right) }}{m_{T_{2}}} \\ 
x_{3}^{\left( T\right) }\frac{z_{11}^{\left( T\right) }}{m_{T_{1}}} & 
x_{3}^{\left( T\right) }\frac{z_{12}^{\left( T\right) }}{m_{T_{1}}} & 
x_{3}^{\left( T\right) }\frac{z_{13}^{\left( T\right) }}{m_{T_{1}}}%
\end{array}%
\right) \frac{v_{L}v_{R}}{2}\allowbreak  \label{MuSM} \\
&&+\left( 
\begin{array}{ccc}
w_{1}^{\left( T^{\prime }\right) }r_{1}^{\left( T^{\prime }\right) } & 
w_{1}^{\left( T^{\prime }\right) }r_{2}^{\left( T^{\prime }\right) } & 
w_{1}^{\left( T^{\prime }\right) }r_{3}^{\left( T^{\prime }\right) } \\ 
w_{2}^{\left( T^{\prime }\right) }r_{1}^{\left( T^{\prime }\right) } & 
w_{2}^{\left( T^{\prime }\right) }r_{2}^{\left( T^{\prime }\right) } & 
w_{2}^{\left( T^{\prime }\right) }r_{3}^{\left( T^{\prime }\right) } \\ 
0 & 0 & 0%
\end{array}%
\right) \frac{m_{T^{\prime }}}{8\pi ^{2}}\left[ f\left(
m_{H_{1}}^{2},m_{T^{\prime }}^{2}\right) -f\left( m_{H_{2}}^{2},m_{T^{\prime
}}^{2}\right) \right] \sin 2\theta _{H},  \notag
\end{eqnarray}
\begin{eqnarray}
\widetilde{M}_{D} &\simeq &-\left( 
\begin{array}{ccc}
0 & 0 & 0 \\ 
0 & 0 & 0 \\ 
x_{3}^{\left( B\right) }z_{1}^{\left( B\right) } & x_{3}^{\left( B\right)
}z_{2}^{\left( B\right) } & x_{3}^{\left( B\right) }z_{3}^{\left( B\right) }%
\end{array}%
\right) \frac{v_{L}v_{R}}{2m_{B}}  \label{MdSM} \\
&&+\left( 
\begin{array}{ccc}
w_{11}^{\left( B^{\prime }\right) }r_{11}^{\left( B^{\prime }\right) } & 
w_{11}^{\left( B^{\prime }\right) }r_{12}^{\left( B^{\prime }\right) } & 
w_{11}^{\left( B^{\prime }\right) }r_{13}^{\left( B^{\prime }\right) } \\ 
w_{21}^{\left( B^{\prime }\right) }r_{11}^{\left( B^{\prime }\right) } & 
w_{21}^{\left( B^{\prime }\right) }r_{12}^{\left( B^{\prime }\right) } & 
w_{21}^{\left( B^{\prime }\right) }r_{13}^{\left( B^{\prime }\right) } \\ 
0 & 0 & 0%
\end{array}%
\right) \frac{m_{B_{1}^{\prime }}}{8\pi ^{2}}\left[ f\left(
m_{H_{1}}^{2},m_{B_{1}^{\prime }}^{2}\right) -f\left(
m_{H_{2}}^{2},m_{B_{1}^{\prime }}^{2}\right) \right] \sin 2\theta _{H} 
\notag \\
&&+\left( 
\begin{array}{ccc}
w_{12}^{\left( B^{\prime }\right) }r_{21}^{\left( B^{\prime }\right) } & 
w_{12}^{\left( B^{\prime }\right) }r_{22}^{\left( B^{\prime }\right) } & 
w_{12}^{\left( B^{\prime }\right) }r_{23}^{\left( B^{\prime }\right) } \\ 
w_{22}^{\left( B^{\prime }\right) }r_{21}^{\left( B^{\prime }\right) } & 
w_{22}^{\left( B^{\prime }\right) }r_{22}^{\left( B^{\prime }\right) } & 
w_{22}^{\left( B^{\prime }\right) }r_{23}^{\left( B^{\prime }\right) } \\ 
0 & 0 & 0%
\end{array}%
\right) \frac{m_{B_{2}^{\prime }}}{8\pi ^{2}}\left[ f\left(
m_{H_{1}}^{2},m_{B_{2}^{\prime }}^{2}\right) -f\left(
m_{H_{2}}^{2},m_{B_{2}^{\prime }}^{2}\right) \right] \sin 2\theta _{H} 
\notag
\end{eqnarray}
\begin{eqnarray}
\widetilde{M}_{E} &\simeq &-\left( 
\begin{array}{ccc}
x_{1}^{\left( E\right) }z_{1}^{\left( E\right) } & x_{1}^{\left( E\right)
}z_{2}^{\left( E\right) } & x_{1}^{\left( E\right) }z_{3}^{\left( E\right) }
\\ 
x_{2}^{\left( E\right) }z_{1}^{\left( E\right) } & x_{2}^{\left( E\right)
}z_{2}^{\left( E\right) } & x_{2}^{\left( E\right) }z_{3}^{\left( E\right) }
\\ 
x_{3}^{\left( E\right) }z_{1}^{\left( E\right) } & x_{3}^{\left( E\right)
}z_{2}^{\left( E\right) } & x_{3}^{\left( E\right) }z_{3}^{\left( E\right) }%
\end{array}%
\right) \frac{v_{L}v_{R}}{2m_{E}}  \label{MlSM} \\
&&+\left( 
\begin{array}{ccc}
w_{11}^{\left( E^{\prime }\right) }r_{11}^{\left( E^{\prime }\right) } & 
w_{11}^{\left( E^{\prime }\right) }r_{12}^{\left( E^{\prime }\right) } & 
w_{11}^{\left( E^{\prime }\right) }r_{13}^{\left( E^{\prime }\right) } \\ 
w_{21}^{\left( E^{\prime }\right) }r_{11}^{\left( E^{\prime }\right) } & 
w_{21}^{\left( E^{\prime }\right) }r_{12}^{\left( E^{\prime }\right) } & 
w_{21}^{\left( E^{\prime }\right) }r_{13}^{\left( E^{\prime }\right) } \\ 
w_{31}^{\left( E^{\prime }\right) }r_{11}^{\left( E^{\prime }\right) } & 
w_{31}^{\left( E^{\prime }\right) }r_{12}^{\left( E^{\prime }\right) } & 
w_{31}^{\left( E^{\prime }\right) }r_{13}^{\left( E^{\prime }\right) }%
\end{array}%
\right) \frac{m_{E_{1}^{\prime }}}{8\pi ^{2}}\left[ f\left(
m_{H_{1}}^{2},m_{E_{1}^{\prime }}^{2}\right) -f\left(
m_{H_{2}}^{2},m_{E_{1}^{\prime }}^{2}\right) \right] \sin 2\theta _{H} 
\notag \\
&&+\left( 
\begin{array}{ccc}
w_{12}^{\left( E^{\prime }\right) }r_{21}^{\left( E^{\prime }\right) } & 
w_{12}^{\left( E^{\prime }\right) }r_{22}^{\left( E^{\prime }\right) } & 
w_{12}^{\left( E^{\prime }\right) }r_{23}^{\left( E^{\prime }\right) } \\ 
w_{22}^{\left( E^{\prime }\right) }r_{21}^{\left( E^{\prime }\right) } & 
w_{22}^{\left( E^{\prime }\right) }r_{22}^{\left( E^{\prime }\right) } & 
w_{22}^{\left( E^{\prime }\right) }r_{23}^{\left( E^{\prime }\right) } \\ 
w_{32}^{\left( E^{\prime }\right) }r_{21}^{\left( E^{\prime }\right) } & 
w_{32}^{\left( E^{\prime }\right) }r_{22}^{\left( E^{\prime }\right) } & 
w_{32}^{\left( E^{\prime }\right) }r_{23}^{\left( E^{\prime }\right) }%
\end{array}%
\right) \frac{m_{E_{2}^{\prime }}}{8\pi ^{2}}\left[ f\left(
m_{H_{1}}^{2},m_{E_{2}^{\prime }}^{2}\right) -f\left(
m_{H_{2}}^{2},m_{E_{2}^{\prime }}^{2}\right) \right] \sin 2\theta _{H} 
\notag
\end{eqnarray}
In what follows we show that our model can successfully reproduce the
following experimental values of the quark masses \cite{Xing:2020ijf}, the
CKM parameters \cite{ParticleDataGroup:2022pth} and the charged lepton
masses \cite{ParticleDataGroup:2022pth}: 
\begin{eqnarray}
&&m_{u}(MeV)=1.24\pm 0.22,\hspace{3mm}m_{d}(MeV)=2.69\pm 0.19,\hspace{3mm}%
m_{s}(MeV)=53.5\pm 4.6,  \notag \\
&&m_{c}(GeV)=0.63\pm 0.02,\hspace{3mm}m_{t}(GeV)=172.9\pm 0.4,\hspace{3mm}%
m_{b}(GeV)=2.86\pm 0.03,\hspace{3mm}  \notag \\
&&\sin \theta _{12}=0.2245\pm 0.00044,\hspace{3mm}\sin \theta
_{23}=0.0421\pm 0.00076,\hspace{3mm}\sin \theta _{13}=0.00365\pm 0.00012, 
\notag \\
&&J=\left( 3.18\pm 0.15\right) \times 10^{-5}\,,
\label{eq:Qsector-observables} \\
&&m_{e}(MeV)=0.4883266\pm 0.0000017,\ \ \ \ m_{\mu }(MeV)=102.87267\pm
0.00021,\ \ \ \ m_{\tau }(MeV)=1747.43\pm 0.12,  \notag
\end{eqnarray}%
where $J$ is the Jarlskog parameter. By solving the eigenvalue problem for
the SM charged fermion mass matrices, we find a solution for the parameters
that reproduces the values in Eq.~(\ref{eq:Qsector-observables}). It is
given by 
\begin{eqnarray}
\theta _{H} &\simeq &{10}^{-2}\mbox{rad},\hspace{1cm}m_{H_{1}}=m_{A_{1}}%
\simeq 1.4\mbox{TeV},\hspace{1cm}m_{H_{2}}=m_{A_{2}}\simeq 1.6\mbox{TeV},%
\hspace{1cm}v_{R}\simeq 10\mbox{TeV},  \notag \\
m_{T_{1}} &\simeq &10\mbox{TeV},\hspace{0.9cm}m_{T_{2}}\simeq 20\mbox{TeV},%
\hspace{0.9cm}m_{T^{\prime }}=m_{B_{1}^{\prime }}=m_{B_{2}^{\prime }}\simeq 4%
\mbox{TeV},\hspace{0.9cm}m_{B}\simeq 10\mbox{TeV}  \notag \\
x_{1}^{\left( T\right) } &\simeq &-0.0518,\hspace{0.45cm}x_{2}^{\left(
T\right) }\simeq -0.1,\hspace{0.45cm}x_{3}^{\left( T\right) }\simeq -3.0,%
\hspace{0.45cm}x_{3}^{\left( B\right) }\simeq -0.01,\hspace{0.45cm}%
z_{1}^{\left( B\right) }\simeq -0.182,\hspace{0.45cm}z_{2}^{\left( B\right)
}\simeq 1.646,\hspace{0.45cm}z_{3}^{\left( B\right) }\simeq 1.629,  \notag \\
z_{11}^{\left( T\right) } &\simeq &0.0566,\hspace{0.7cm}z_{12}^{\left(
T\right) }\simeq 0.0480,\hspace{0.7cm}z_{13}^{\left( T\right) }\simeq 0.463,%
\hspace{0.7cm}z_{21}^{\left( T\right) }\simeq 0.00869,\hspace{0.7cm}%
z_{22}^{\left( T\right) }\simeq 0.0869,\hspace{0.7cm}z_{23}^{\left( T\right)
}\simeq 0.00869,  \notag \\
r_{1}^{\left( T^{\prime }\right) } &\simeq &0.716,\hspace{1cm}r_{2}^{\left(
T^{\prime }\right) }\simeq 0.522,\hspace{1cm}r_{3}^{\left( T^{\prime
}\right) }\simeq 0.148,\hspace{1cm}w_{1}^{\left( T^{\prime }\right) }\simeq
0.948,\hspace{1cm}w_{2}^{\left( T^{\prime }\right) }\simeq 1.763,\hspace{1cm}
\notag \\
w_{11}^{\left( B^{\prime }\right) } &\simeq &1.011-1.051i,\hspace{1cm}%
w_{12}^{\left( B^{\prime }\right) }\simeq 0.0459+0.353i,\hspace{1cm}%
w_{21}^{\left( B^{\prime }\right) }\simeq 0.750+0.608i,\hspace{1cm}%
w_{22}^{\left( B^{\prime }\right) }\simeq 1.338-0.658i,  \notag \\
r_{11}^{\left( B^{\prime }\right) } &\simeq &-0.184-0.103i,\hspace{1cm}%
r_{12}^{\left( B^{\prime }\right) }\simeq 0.767+0.313i,\hspace{1cm}%
r_{13}^{\left( B^{\prime }\right) }\simeq 0.292+0.0391i,  \label{bfpfermions}
\\
r_{21}^{\left( B^{\prime }\right) } &\simeq &-0.221+0.028i,\hspace{1cm}%
r_{22}^{\left( B^{\prime }\right) }\simeq 1.063+0.0427i,\hspace{1cm}%
r_{23}^{\left( B^{\prime }\right) }\simeq 0.431+0.10i,  \notag \\
x_{1}^{\left( E\right) } &\simeq &0.0394,\hspace{0.7cm}x_{2}^{\left(
E\right) }\simeq -0.0814,\hspace{0.7cm}x_{3}^{\left( E\right) }\simeq 0.0426,%
\hspace{0.7cm}z_{1}^{\left( E\right) }\simeq -0.01,\hspace{0.7cm}%
z_{2}^{\left( E\right) }\simeq 0.0184,\hspace{0.7cm}z_{3}^{\left( E\right)
}\simeq -0.0139,  \notag \\
w_{11}^{\left( E^{\prime }\right) } &\simeq &-0.0315,\hspace{0.65cm}%
w_{21}^{\left( E^{\prime }\right) }\simeq -0.0272,\hspace{0.65cm}%
w_{31}^{\left( E^{\prime }\right) }\simeq -0.0229,\hspace{0.65cm}%
w_{21}^{\left( E^{\prime }\right) }\simeq 0.630,\hspace{0.65cm}  \notag \\
w_{22}^{\left( E^{\prime }\right) } &\simeq &-0.0909,\hspace{0.65cm}%
w_{32}^{\left( E^{\prime }\right) }\simeq -0.757,\hspace{0.9cm}m_{E}\simeq
1.7\mbox{TeV},\hspace{0.9cm}m_{E_{1}^{\prime }}\simeq 3.8\mbox{TeV},\hspace{%
0.9cm}m_{E_{2}^{\prime }}\simeq 3.3\mbox{TeV},  \notag \\
r_{11}^{\left( E^{\prime }\right) } &\simeq &-0.0290,\hspace{0.6cm}%
r_{12}^{\left( E^{\prime }\right) }\simeq -0.0888,\hspace{0.6cm}%
r_{13}^{\left( E^{\prime }\right) }\simeq -0.1,\hspace{0.6cm}r_{21}^{\left(
E^{\prime }\right) }\simeq 1.268,\hspace{0.6cm}r_{22}^{\left( E^{\prime
}\right) }\simeq 0.188,\hspace{0.6cm}r_{23}^{\left( E^{\prime }\right)
}\simeq -0.536,  \notag
\end{eqnarray}%
The above values successfully reproduce the SM charged fermion masses and
CKM parameters. The Yukawa couplings of the charged fermion sector feature a
moderate hierachy, since their order of magnitude is located in the range $%
\left[ 10^{-2},1\right] $. On the other hand, the electrically charged
fermionic seesaw mediators have $\mathcal{O}\left( 1\right) $ TeV and $%
\mathcal{O}\left( 10\right) $ TeV mases. Despite the aforementioned moderate
hierarchy in the Yukawa couplings, this situation is much better than in the
SM, where a hierarchy of about 5 orders of magnitude is present in the
charged fermion sector. Regarding the charged lepton sector, in our
numerical analysis we consider charged exotic leptons with masses of a few
TeVs, which allows us to successfully reproduce the muon and electron $\left(
g-2\right) $ anomalies. Furthermore, it is worth mentioning that the
effective Yukawa couplings are proportional to a product of two other
dimensionless couplings, so a moderate hierarchy in those couplings can
yield a quadratically larger hierarchy in the effective couplings, thus
allowing to explain the SM charged fermion mass and quark mixing pattern.\
Small quark mixing angles are attributed to the hierarchy in the rows of the
SM quark mass matrices, whose entries are proportional to the product of two
dimensionless Yukawa couplings, as shown in Eqs. (\ref{MuSM}), (\ref{MdSM})
and (\ref{MlSM}). Having all charged fermion Yukawa couplings of the same
order of magnitude would require the implementation of the sequential loop
suppression mechanism \cite{CarcamoHernandez:2016pdu}  where the first family of SM charged fermions will get
masses at two loop level. This will imply a non minimal extension of the
scalar and fermion sectors of the model and such implementation is beyond
the scope of this work and will be done elsewhere. On the other hand, it is
worth mentioning that the Yukawa interactions of the $126$ GeV SM like Higgs
boson (which is identified with the CP even neutral part of the $SU\left(
2\right) _{L}$ scalar doublet $\chi _{L}$) with SM fermion-antifermion pairs
are dynamically generated below the left-right symmetry breaking scale,
after integrating out the charged vector like seesaw mediators. This can be
seen from the Feynman diagrams of Figure \ref{Diagramschargedfermions},
where the external scalar lines are associated with the $\chi _{L}$ and $%
\chi _{R}$ scalar fields, whereas the external fermionic lines are mostly
composed of the SM charged fermions (after rotating to the physical basis).
Given that there is only one non vevless $SU\left( 2\right) _{L}$ scalar
doublet $\chi _{L}$ in the scalar sector of the model and there is no
bidoublet scalar, our model is free from tree level flavor changing neutral
currents and the couplings of the $126$ GeV SM like Higgs boson with SM
particles are very close to the SM expectation, thus implying the alignment
limit is naturally fulfilled in our model.

Concerning the neutrino sector, we find that the neutrino Yukawa
interactions give rise to the following neutrino mass terms: 
\begin{equation}
-\mathcal{L}_{mass}^{\left( \nu \right) }=\frac{1}{2}\left( 
\begin{array}{ccc}
\overline{\nu _{L}^{C}} & \overline{\nu _{R}} & \overline{N_{R}}%
\end{array}%
\right) M_{\nu }\left( 
\begin{array}{c}
\nu _{L} \\ 
\nu _{R}^{C} \\ 
N_{R}^{C}%
\end{array}%
\right) +\dsum\limits_{n=1}^{2}\left( m_{\Omega }\right) _{n}\overline{%
\Omega }_{R_{n}}\Omega _{R_{n}}^{C}+H.c,  \label{Lnu}
\end{equation}%
where the neutrino mass matrix reads: 
\begin{equation}
M_{\nu }=\left( 
\begin{array}{ccc}
0_{3\times 3} & m_{\nu D} & 0_{3\times 3} \\ 
m_{\nu D}^{T} & 0_{2\times 2} & M \\ 
0_{3\times 3} & M^{T} & \mu%
\end{array}%
\right) ,  \label{Mnu}
\end{equation}%
and the submatrices are given by: 
\begin{eqnarray}
\left( m_{\nu D}\right) _{ij} &=&\dsum\limits_{k=1}^{2}\frac{w_{ik}^{\left(
E^{\prime }\right) }r_{kj}^{\left( E^{\prime }\right) }m_{E_{k}^{\prime }}}{%
16\pi ^{2}}\left[ f\left( m_{H_{1}^{\pm }}^{2},m_{E_{k}^{\prime
}}^{2}\right) -f\left( m_{H_{2}^{\pm }}^{2},m_{E_{k}^{\prime }}^{2}\right) %
\right] \sin 2\theta ,\hspace{0.7cm}\hspace{0.7cm}  \notag \\
M_{ij} &=&x_{ij}^{\left( N\right) }\frac{v_{R}}{\sqrt{2}},\hspace{0.7cm}%
\hspace{0.7cm}i,j,n,k=1,2,3,\hspace{0.7cm}\hspace{0.7cm}r=1,2, \\
\mu _{nk} &=&\dsum\limits_{r=1}^{2}\frac{x_{R_{n}}^{\left( \Omega \right)
}x_{kr}^{\left( \Omega \right) }m_{\Omega _{r}}}{16\pi ^{2}}\left[ \frac{%
m_{\varphi _{R}}^{2}}{m_{\varphi _{R}}^{2}-m_{\Omega _{r}}^{2}}\ln \left( 
\frac{m_{\varphi _{R}}^{2}}{m_{\Omega _{r}}^{2}}\right) -\frac{m_{\varphi
_{I}}^{2}}{m_{\varphi _{I}}^{2}-m_{\Omega _{r}}^{2}}\ln \left( \frac{%
m_{\varphi _{I}}^{2}}{m_{\Omega _{r}}^{2}}\right) \right] ,\hspace{0.7cm}%
\hspace{0.7cm}.
\end{eqnarray}%
The $\mu $ block is generated at one-loop level due to the exchange of $%
\Omega _{rR}$ ($r=1,2$) and $\varphi $ in the internal lines, as shown in
Figure~\ref{Diagramsneutrinos}.

The light active masses arise from an inverse seesaw mechanism, and the
physical neutrino mass matrices are: 
\begin{eqnarray}
\widetilde{\mathbf{M}}_{\nu } &=&m_{\nu D}\left( M^{T}\right) ^{-1}\mu
M^{-1}m_{\nu D}^{T},\hspace{0.7cm}  \label{M1nu} \\
\mathbf{M}_{\nu }^{\left( 1\right) } &=&-\frac{1}{2}\left( M+M^{T}\right) +%
\frac{1}{2}\mu ,\hspace{0.7cm} \\
\mathbf{M}_{\nu }^{\left( 2\right) } &=&\frac{1}{2}\left( M+M^{T}\right) +%
\frac{1}{2}\mu .
\end{eqnarray}%
where $\widetilde{\mathbf{M}}_{\nu }$ corresponds to the mass matrix for
light active neutrinos $\nu _{a}$ ($a=1,2,3$), whereas $\mathbf{M}_{\nu
}^{\left( 1\right) }$ and $\mathbf{M}_{\nu }^{\left( 2\right) }$ are the
mass matrices for sterile neutrinos ($N_{a}^{-},N_{a}^{+}$), which are
superpositions of mostly $\nu _{aR}$ and $N_{aR}$, as $N_{a}^{\pm }\sim 
\frac{1}{\sqrt{2}}\left( \nu _{aR}\mp N_{aR}\right) $. In the limit $\mu
\rightarrow 0$, which corresponds to unbroken lepton number, the light
active neutrinos become massless. The smallness of the $\mu $ parameter is
responsible for a small mass splitting between the three pairs of sterile
neutrinos, thus implying that the sterile neutrinos form pseudo-Dirac pairs.
Notice that the physical neutrino spectrum is composed of three light active
neutrinos and six exotic neutrinos. These exotic neutrinos are pseudo-Dirac,
with masses $\sim \pm \frac{1}{2}\left( M+M^{T}\right) $ and a small
splitting $\mu $.


\section{Oblique $T$, $S$ and $U$ parameters}

\label{TnS} 

The extra scalars affect the oblique corrections of the SM, whose values are
measured in high precision experiments. Consequently, they act as a further
constraint on the validity of any New Physics model. The oblique corrections
are parametrized in terms of the three well-known quantities $T$, $S$ and $U$%
. In this section, we calculate one-loop contributions to the oblique
parameters $T $, $S$\ and $U$, defined as \cite%
{Peskin:1991sw,Altarelli:1990zd,Barbieri:2004qk} 
\begin{eqnarray}
T &=&\frac{\Pi _{33}\left( q^{2}\right) -\Pi _{11}\left( q^{2}\right) }{%
\alpha _{EM}(M_{Z})M_{W}^{2}}\biggl|_{q^{2}=0},\ \ \ \ \ \ \ \ \ \ \ S=\frac{%
2\sin 2{\theta }_{W}}{\alpha _{EM}(M_{Z})}\frac{d\Pi _{30}\left(
q^{2}\right) }{dq^{2}}\biggl|_{q^{2}=0},  \label{T-S-definition} \\
U &=&\frac{4\sin ^{2}\theta _{W}}{\alpha _{EM}(M_{Z})}\left( \frac{d\Pi
_{33}\left( q^{2}\right) }{dq^{2}}-\frac{d\Pi _{11}\left( q^{2}\right) }{%
dq^{2}}\right) \biggl|_{q^{2}=0}
\end{eqnarray}%
where $\Pi _{11}\left( 0\right) $, $\Pi _{33}\left( 0\right) $, and $\Pi
_{30}\left( q^{2}\right) $ are the vacuum polarization amplitudes with $%
\{W_{\mu }^{1},W_{\mu }^{1}\}$, $\{W_{\mu }^{3},W_{\mu }^{3}\}$ and $%
\{W_{\mu }^{3},B_{\mu }\}$ external gauge bosons, respectively, and $q$ is
their momentum. We note that in the definitions of the $T$, $S$ and $U$
parameters, the New Physics is assumed to be heavy compared to $M_{W}$ and $%
M_{Z}$.

\begin{figure}[H]
\begin{subfigure}{0.48\textwidth}
\includegraphics[keepaspectratio,width=\textwidth]{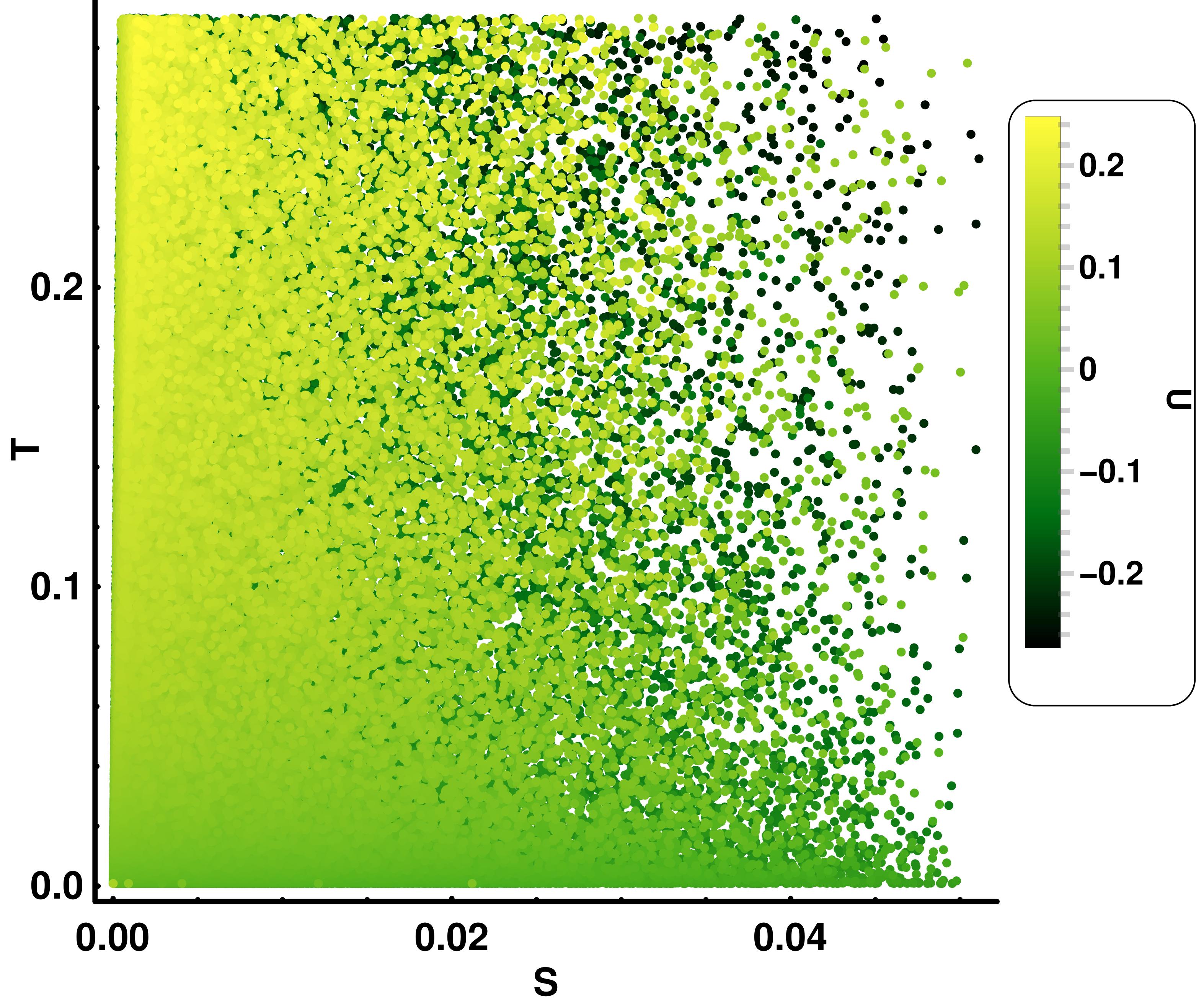}
\caption{Correlation between the oblique $S$ and $T$ parameters.}
\label{TvsS}
\end{subfigure}
\begin{subfigure}{0.48\textwidth}
\includegraphics[keepaspectratio,width=\textwidth]{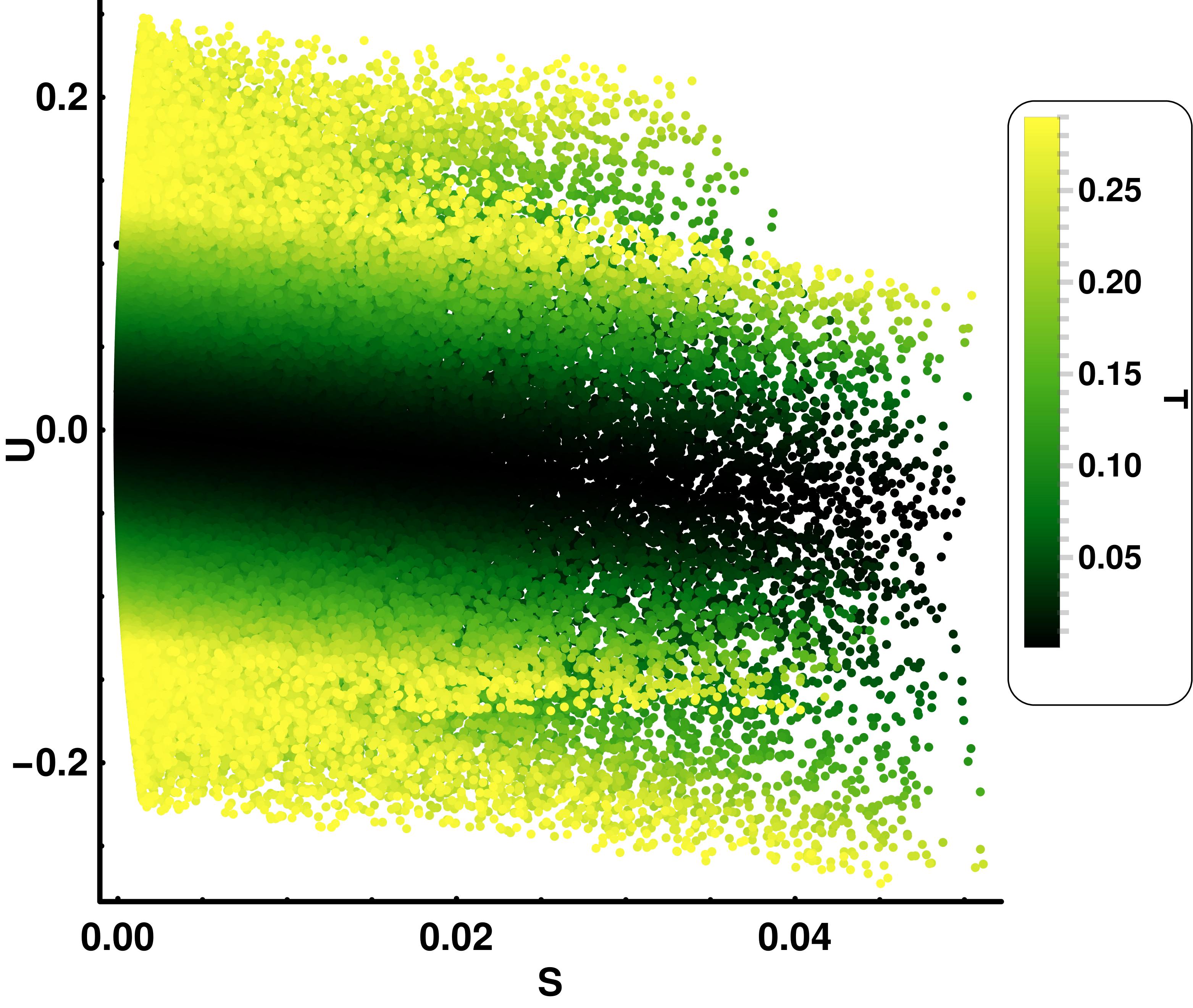}
\caption{Correlation between the oblique $S$ and $U$ parameters.}
\label{UvsS}
\end{subfigure}
\begin{subfigure}{0.98\textwidth}
\centering
\includegraphics[keepaspectratio,width=0.50\textwidth]{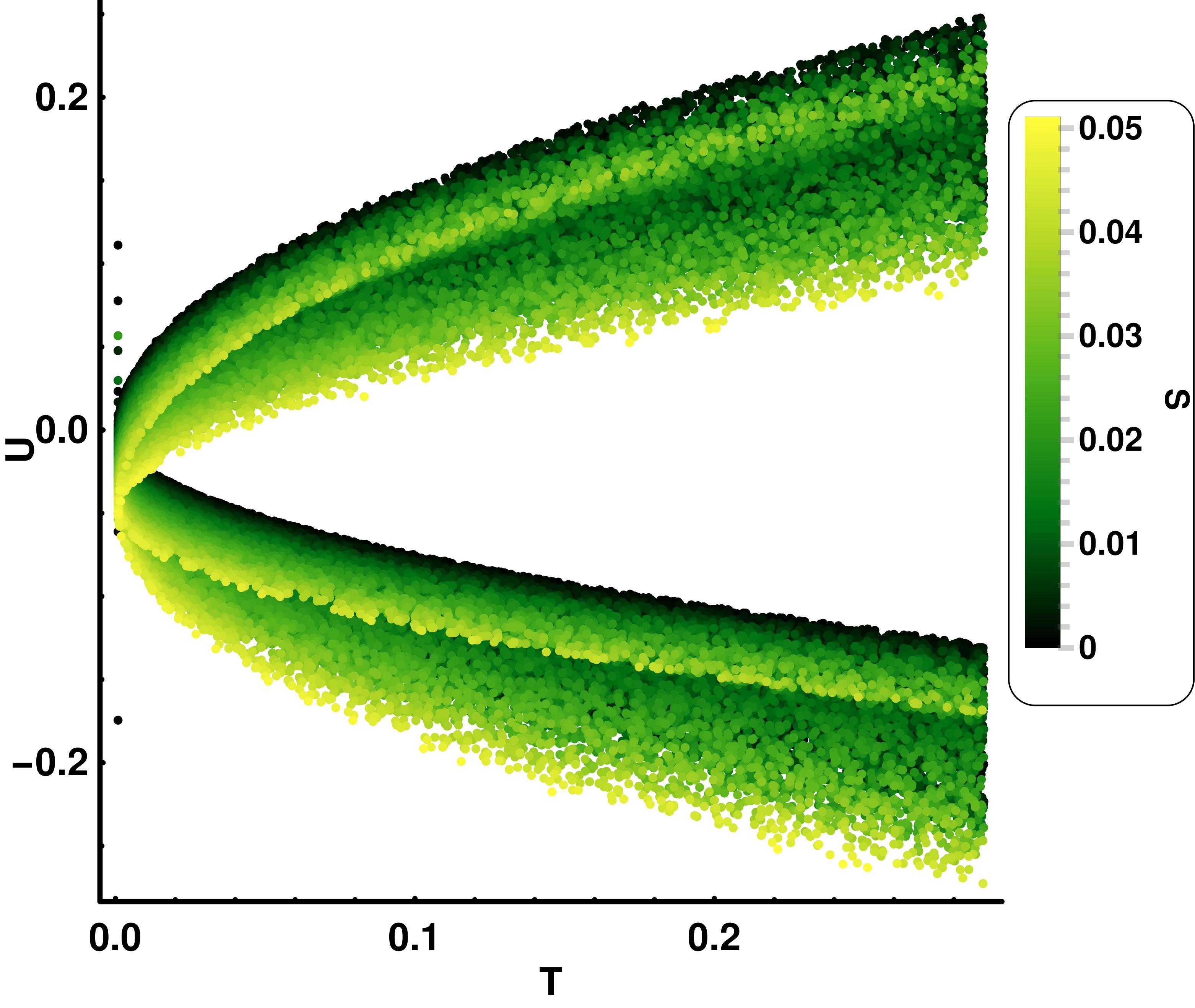}
\caption{Correlation between the oblique $U$ and $T$ parameters.}
\label{UvsT}
\end{subfigure}
\caption{Oblique parameters scanned}
\end{figure}

The contributions arising from New Physics to the $T$, $S$ and $U$
parameters are:%
\begin{eqnarray}
T &\simeq &\frac{v_{L}^{2}}{\alpha _{EM}(M_{Z})v_{R}^{2}}\left( \frac{%
g_{B-L}^{2}}{g_{2}^{2}+g_{B-L}^{2}}\right) ^{2}  \notag \\
&+&\frac{1}{16\pi ^{2}v^{2}\alpha _{EM}(M_{Z})}\left\{ \sum_{k=1}^{2}\left(
\left( R_{C}\right) _{1k}\right) ^{2}m_{H_{k}^{\pm
}}^{2}+\sum_{i=1}^{2}\sum_{j=1}^{2}\left( \left( R_{H}\right) _{1i}\right)
^{2}\left( \left( R_{A}\right) _{1j}\right) ^{2}F\left(
m_{H_{i}^{0}}^{2},m_{A_{j}^{0}}^{2}\right) \right.  \notag \\
&&-\left. \sum_{i=1}^{2}\sum_{k=1}^{2}\left( \left( R_{H}\right)
_{1i}\right) ^{2}\left( \left( R_{C}\right) _{1k}\right) ^{2}F\left(
m_{H_{i}^{0}}^{2},m_{H_{k}^{\pm }}^{2}\right)
-\sum_{i=1}^{2}\sum_{k=1}^{2}\left( \left( R_{A}\right) _{1i}\right)
^{2}\left( \left( R_{C}\right) _{1k}\right) ^{2}F\left(
m_{A_{i}^{0}}^{2},m_{H_{k}^{\pm }}^{2}\right) \right\}
\end{eqnarray}%
\begin{equation}
S\simeq \sum_{i=1}^{2}\sum_{j=1}^{2}\sum_{k=1}^{2}\frac{\left( \left(
R_{H}\right) _{1i}\right) ^{2}\left( \left( R_{A}\right) _{1j}\right) ^{2}}{%
12\pi }K\left( m_{H_{i}^{0}}^{2},m_{A_{i}^{0}}^{2},m_{H_{k}^{\pm
}}^{2}\right) ,
\end{equation}%
\begin{eqnarray}
U &\simeq &-S+\sum_{i=1}^{2}\sum_{k=1}^{2}\left( \left( R_{A}\right)
_{1i}\right) ^{2}\left( \left( R_{C}\right) _{1k}\right) ^{2}K_{2}\left(
m_{A_{i}^{0}}^{2},m_{H_{k}^{\pm }}^{2}\right)  \notag \\
&&+\sum_{i=1}^{2}\sum_{k=1}^{2}\left( \left( R_{H}\right) _{1i}\right)
^{2}\left( \left( R_{C}\right) _{1k}\right) ^{2}K_{2}\left(
m_{H_{i}^{0}}^{2},m_{H_{k}^{\pm }}^{2}\right) ,
\end{eqnarray}
where we introduced the functions \cite{CarcamoHernandez:2015smi} 
\begin{equation}
F\left( m_{1}^{2},m_{2}^{2}\right) =\frac{m_{1}^{2}m_{2}^{2}}{%
m_{1}^{2}-m_{2}^{2}}\ln \left( \frac{m_{1}^{2}}{m_{2}^{2}}\right) ,\hspace{%
1.5cm}\hspace{1.5cm}\lim_{m_{2}\rightarrow m_{1}}F\left(
m_{1}^{2},m_{2}^{2}\right) =m_{1}^{2}.
\end{equation}%
\begin{eqnarray}
K\left( m_{1}^{2},m_{2}^{2},m_{3}^{2}\right) &=&\frac{1}{\left(
m_{2}^{2}-m_{1}^{2}\right) {}^{3}}\left\{ m_{1}^{4}\left(
3m_{2}^{2}-m_{1}^{2}\right) \ln \left( \frac{m_{1}^{2}}{m_{3}^{2}}\right)
-m_{2}^{4}\left( 3m_{1}^{2}-m_{2}^{2}\right) \ln \left( \frac{m_{2}^{2}}{%
m_{3}^{2}}\right) \right.  \notag \\
&&-\left. \frac{1}{6}\left[ 27m_{1}^{2}m_{2}^{2}\left(
m_{1}^{2}-m_{2}^{2}\right) +5\left( m_{2}^{6}-m_{1}^{6}\right) \right]
\right\} ,
\end{eqnarray}%
with the properties 
\begin{eqnarray}
\lim_{m_{1}\rightarrow m_{2}}K(m_{1}^{2},m_{2}^{2},m_{3}^{2})
&=&K_{1}(m_{2}^{2},m_{3}^{2})=\ln \left( \frac{m_{2}^{2}}{m_{3}^{2}}\right) ,
\notag \\
\lim_{m_{2}\rightarrow m_{3}}K(m_{1}^{2},m_{2}^{2},m_{3}^{2})
&=&K_{2}(m_{1}^{2},m_{3}^{2})=\frac{%
-5m_{1}^{6}+27m_{1}^{4}m_{3}^{2}-27m_{1}^{2}m_{3}^{4}+6\left(
m_{1}^{6}-3m_{1}^{4}m_{3}^{2}\right) \ln \left( \frac{m_{1}^{2}}{m_{3}^{2}}%
\right) +5m_{3}^{6}}{6\left( m_{1}^{2}-m_{3}^{2}\right) ^{3}},  \notag \\
\lim_{m_{1}\rightarrow m_{3}}K(m_{1}^{2},m_{2}^{2},m_{3}^{2})
&=&K_{2}(m_{2}^{2},m_{3}^{2}).
\end{eqnarray}
Besides that, for the sake of simplicity we have set $g_{L}=g_{R}=g_{2}$, as
done in \cite{Adam:2019oes}, which yields: 
\begin{equation}
\frac{1}{g^{2}}+\frac{1}{g^{\prime 2}}=\frac{1}{g_{L}^{2}}+\frac{1}{g_{R}^{2}%
}+\frac{1}{g_{B-L}^{2}}=\frac{2}{g_{2}^{2}}+\frac{1}{g_{B-L}^{2}}.
\end{equation}%
Above, $R_{H}$, $R_{A}$ and $R_{C}$\ are the rotation matrices diagonalizing
the squared mass matrices for the inert CP-even, CP-odd and electrically
charged scalars, respectively, according to the following relations: 
\begin{eqnarray}
R_{H,A,C}^{T}M_{H,A,C}^{2}R_{H,A,C} &=&\left( M_{H,A,C}^{2}\right) _{diag},%
\hspace{0.7cm}\hspace{0.7cm}\left( 
\begin{array}{c}
\func{Re}\phi _{L}^{0} \\ 
\func{Re}\phi _{R}^{0}%
\end{array}%
\right) =R_{H}\left( 
\begin{array}{c}
H_{1} \\ 
H_{2}%
\end{array}%
\right) =\left( 
\begin{array}{cc}
\cos \theta _{H} & -\sin \theta _{H} \\ 
\sin \theta _{H} & \cos \theta _{H}%
\end{array}%
\right) \left( 
\begin{array}{c}
H_{1} \\ 
H_{2}%
\end{array}%
\right) ,  \notag \\
\left( 
\begin{array}{c}
\func{Im}\phi _{L}^{0} \\ 
\func{Im}\phi _{R}^{0}%
\end{array}%
\right) &=&R_{A}\left( 
\begin{array}{c}
A_{1} \\ 
A_{2}%
\end{array}%
\right) =\left( 
\begin{array}{cc}
\cos \theta _{A} & -\sin \theta _{A} \\ 
\sin \theta _{A} & \cos \theta _{A}%
\end{array}%
\right) \left( 
\begin{array}{c}
A_{1} \\ 
A_{2}%
\end{array}%
\right) ,  \notag \\
\left( 
\begin{array}{c}
\func{Re}\phi _{L}^{\pm } \\ 
\func{Re}\phi _{R}^{\pm }%
\end{array}%
\right) &=&R_{C}\left( 
\begin{array}{c}
H_{1}^{\pm } \\ 
H_{2}^{\pm }%
\end{array}%
\right) =\left( 
\begin{array}{cc}
\cos \theta & -\sin \theta \\ 
\sin \theta & \cos \theta%
\end{array}%
\right) \left( 
\begin{array}{c}
H_{1}^{\pm } \\ 
H_{2}^{\pm }%
\end{array}%
\right) \,.
\end{eqnarray}%
On the other hand, the experimental values of $T$, $S$ and $U$ are
constrained to be in the ranges \cite{Lu:2022bgw}: 
\begin{equation}
T=-0.01\pm 0.10,\ \ \ \ \ \ \ \ \ \ \ S=0.03\pm 0.12,\ \ \ \ \ \ \ \ \ \ \
U=0.02\pm 0.11 \,.
\end{equation}%
The correlations of the oblique $T$ and $U$ parameters with the $S$
parameters are shown in Figures~\ref{TvsS} and \ref{UvsS}, respectively.

It is worth mentioning that the analysis of the scalar sector yields the
prediction $m_{H_{i}^{0}}=m_{A_{i}^{0}}$ ($i=1,2$) and $\theta _{H}=\theta
_{A}$. In our numerical analysis we have varied the neutral scalar masses $%
m_{H_{1}^{0}}$ , $m_{H_{2}^{0}}$ , the charged scalar masses $m_{H_{1}^{\pm
}}$, $m_{H_{2}^{\pm }}$\ and the scalar mixing angles $\theta _{H}$, $\theta 
$ in the ranges ${1}${\ TeV $\leqslant m_{H_{i}^{0}},m_{H_{i}^{\pm }}{%
\leqslant 3}$ TeV} ($i=1,2$), ${0.9\times 10}^{-2}${\ }rad $\leqslant \theta
_{H}{\leqslant }1{{.1\times 10}^{-2}}${\ }rad and ${0.9\times 10}^{-3}${\ }%
rad $\leqslant \theta {\leqslant }1{{.1\times 10}^{-3}}${\ }rad,
respectively. As shown in Figures \ref{TvsS} and \ref{UvsS} our model is
consistent with the constraints arising from the experimental measurements
of the oblique $T$, $S$ and $U$ parameters. Furthermore, we have numerically
checked that our model can also successfully accommodate the W mass anomaly
for an appropriate region of parameter space. 


\section{Muon and electron anomalous magnetic moments}

\label{sec:Mu-amm} 

In this section, we will analyze the implications of our model in the muon
and electron anomalous magnetic moments. The contributions to the muon and
electron anomalous magnetic moments take the form: 
\begin{eqnarray}
\Delta a_{\mu } &=&\dsum\limits_{k=1}^{2}\frac{\func{Re}\left( \beta
_{2k}\gamma _{k2}^{\ast }\right) m_{\mu }^{2}}{8\pi ^{2}}\left[ \left(
R_{H}\right) _{11}\left( R_{H}\right) _{12}I_{S}^{\left( \mu \right) }\left(
m_{E_{k}^{\prime }},m_{H_{1}}\right) +\left( R_{H}\right) _{21}\left(
R_{H}\right) _{22}I_{S}^{\left( \mu \right) }\left( m_{E_{k}^{\prime
}},m_{H_{2}}\right) \right]  \notag \\
&&+\dsum\limits_{k=1}^{2}\frac{\func{Re}\left( \beta _{2k}\gamma _{k2}^{\ast
}\right) m_{\mu }^{2}}{8\pi ^{2}}\left[ \left( R_{A}\right) _{11}\left(
R_{A}\right) _{12}I_{P}^{\left( \mu \right) }\left( m_{E_{k}^{\prime
}},m_{A_{1}}\right) +\left( R_{A}\right) _{21}\left( R_{A}\right)
_{22}I_{P}^{\left( \mu \right) }\left( m_{E_{k}^{\prime }},m_{A_{2}}\right) %
\right]  \notag \\
&&+\frac{m_{\mu }^{2}\func{Re}\left( \kappa _{2}\vartheta _{2}^{\ast
}\right) }{8\pi ^{2}}\left[ I_{S}^{\left( \mu \right) }\left(
m_{E},m_{h}\right) -I_{S}^{\left( \mu \right) }\left( m_{E},m_{H_{3}}\right) %
\right] \sin \theta _{\chi }\cos \theta _{\chi }  \notag \\
&=&\dsum\limits_{k=1}^{2}\frac{\func{Re}\left( \beta _{2k}\gamma _{k2}^{\ast
}\right) m_{\mu }^{2}}{8\pi ^{2}}\left[ I_{S}^{\left( \mu \right) }\left(
m_{E_{k}^{\prime }},m_{H_{1}}\right) -I_{P}^{\left( \mu \right) }\left(
m_{E_{k}^{\prime }},m_{A_{1}}\right) -I_{S}^{\left( \mu \right) }\left(
m_{E_{k}^{\prime }},m_{H_{2}}\right) +I_{P}^{\left( \mu \right) }\left(
m_{E_{k}^{\prime }},m_{A_{2}}\right) \right] \sin \theta _{H}\cos \theta _{H}
\notag \\
&&+\frac{m_{\mu }^{2}\func{Re}\left( \kappa _{2}\vartheta _{2}^{\ast
}\right) }{8\pi ^{2}}\left[ I_{S}^{\left( \mu \right) }\left(
m_{E},m_{h}\right) -I_{S}^{\left( \mu \right) }\left( m_{E},m_{H_{3}}\right) %
\right] \sin \theta _{\chi }\cos \theta _{\chi }
\end{eqnarray}
\begin{eqnarray}
\Delta a_{e} &=&\dsum\limits_{k=1}^{2}\frac{\func{Re}\left( \beta
_{1k}\gamma _{k1}^{\ast }\right) m_{e}^{2}}{8\pi ^{2}}\left[ \left(
R_{H}\right) _{11}\left( R_{H}\right) _{12}I_{S}^{\left( e\right) }\left(
m_{E_{k}^{\prime }},m_{H_{1}}\right) +\left( R_{H}\right) _{21}\left(
R_{H}\right) _{22}I_{S}^{\left( e\right) }\left( m_{E_{k}^{\prime
}},m_{H_{2}}\right) \right]  \notag \\
&&+\dsum\limits_{k=1}^{2}\frac{\func{Re}\left( \beta _{1k}\gamma _{k1}^{\ast
}\right) m_{e}^{2}}{8\pi ^{2}}\left[ \left( R_{A}\right) _{11}\left(
R_{A}\right) _{12}I_{P}^{\left( e\right) }\left( m_{E_{k}^{\prime
}},m_{A_{1}}\right) +\left( R_{A}\right) _{21}\left( R_{A}\right)
_{22}I_{P}^{\left( e\right) }\left( m_{E_{k}^{\prime }},m_{A_{2}}\right) %
\right]  \notag \\
&&+\frac{m_{e}^{2}\func{Re}\left( \kappa _{1}\vartheta _{1}^{\ast }\right) }{%
8\pi ^{2}}\left[ I_{S}^{\left( e\right) }\left( m_{E},m_{h}\right)
-I_{S}^{\left( e\right) }\left( m_{E},m_{H_{3}}\right) \right] \sin \theta
_{\chi }\cos \theta _{\chi }  \notag \\
&=&\dsum\limits_{k=1}^{2}\frac{\func{Re}\left( \beta _{1k}\gamma _{k1}^{\ast
}\right) m_{e}^{2}}{8\pi ^{2}}\left[ I_{S}^{\left( e\right) }\left(
m_{E_{k}^{\prime }},m_{H_{1}}\right) -I_{P}^{\left( e\right) }\left(
m_{E_{k}^{\prime }},m_{A_{1}}\right) -I_{S}^{\left( e\right) }\left(
m_{E_{k}^{\prime }},m_{H_{2}}\right) +I_{P}^{\left( e\right) }\left(
m_{E_{k}^{\prime }},m_{A_{2}}\right) \right] \sin \theta _{H}\cos \theta _{H}
\notag \\
&&+\frac{m_{e}^{2}\func{Re}\left( \kappa _{1}\vartheta _{1}^{\ast }\right) }{%
8\pi ^{2}}\left[ I_{S}^{\left( e\right) }\left( m_{E},m_{h}\right)
-I_{S}^{\left( e\right) }\left( m_{E},m_{H_{3}}\right) \right] \sin \theta
_{\chi }\cos \theta _{\chi } \,.
\end{eqnarray}
Furthermore, the loop function $I_{S\left( P\right) }^{\left( e,\mu \right)
}\left( m_{E},m\right)$ has the form \cite%
{Diaz:2002uk,Jegerlehner:2009ry,Kelso:2014qka,Lindner:2016bgg,Kowalska:2017iqv}%
: 
\begin{equation}
I_{S\left( P\right) }^{\left( e,\mu \right) }\left( m_{E},m_{S}\right)
=\int_{0}^{1}\frac{x^{2}\left( 1-x\pm \frac{m_{E}}{m_{e,\mu }}\right) }{%
m_{\mu }^{2}x^{2}+\left( m_{E}^{2}-m_{e,\mu }^{2}\right) x+m_{S,P}^{2}\left(
1-x\right) }dx
\end{equation}
and the dimensionless parameters $\beta _{1k}$, $\beta _{2k}$, $\gamma _{k1}$%
, $\gamma _{k2}$, $\kappa _{1}$, $\kappa _{2}$, $\vartheta _{1}$, $\vartheta
_{2}$ are given by: 
\begin{eqnarray}
\beta _{1k} &=&\dsum\limits_{i=1}^{3}w_{ik}^{\left( E^{\prime }\right)
}\left( V_{lL}^{\dagger }\right) _{1i},\hspace{0.7cm}\hspace{0.7cm}\gamma
_{k1}=\dsum\limits_{j=1}^{3}r_{kj}^{\left( E^{\prime }\right) }\left(
V_{lR}\right) _{j1}, \\
\beta _{2k} &=&\dsum\limits_{i=1}^{3}w_{ik}^{\left( E^{\prime }\right)
}\left( V_{lL}^{\dagger }\right) _{2i},\hspace{0.7cm}\hspace{0.7cm}\gamma
_{k2}=\dsum\limits_{j=1}^{3}r_{kj}^{\left( E^{\prime }\right) }\left(
V_{lR}\right) _{j2}, \\
\kappa _{1} &=&\dsum\limits_{i=1}^{3}x_{i}^{\left( E\right) }\left(
V_{lL}^{\dagger }\right) _{1i},\hspace{0.7cm}\hspace{0.7cm}\vartheta
_{1}=\dsum\limits_{j=1}^{3}z_{j}^{\left( E\right) }\left( V_{lR}\right)
_{j1}, \\
\kappa _{2} &=&\dsum\limits_{i=1}^{3}x_{i}^{\left( E\right) }\left(
V_{lL}^{\dagger }\right) _{2i},\hspace{0.7cm}\hspace{0.7cm}\vartheta
_{2}=\dsum\limits_{j=1}^{3}z_{j}^{\left( E\right) }\left( V_{lR}\right)
_{j2},
\end{eqnarray}
where $V_{lL}$ and $V_{lR}$ are the rotation matrices that diagonalize $%
\widetilde{M}_{E}$ according to the relation: 
\begin{equation}
V_{lL}^{\dagger }\widetilde{M}_{E}V_{lR}=\mathrm{diag}\left( m_{e},m_{\mu
},m_{\tau }\right)
\end{equation}
\begin{figure}[tbp]
\centering
\includegraphics[width=14cm, height=10cm]{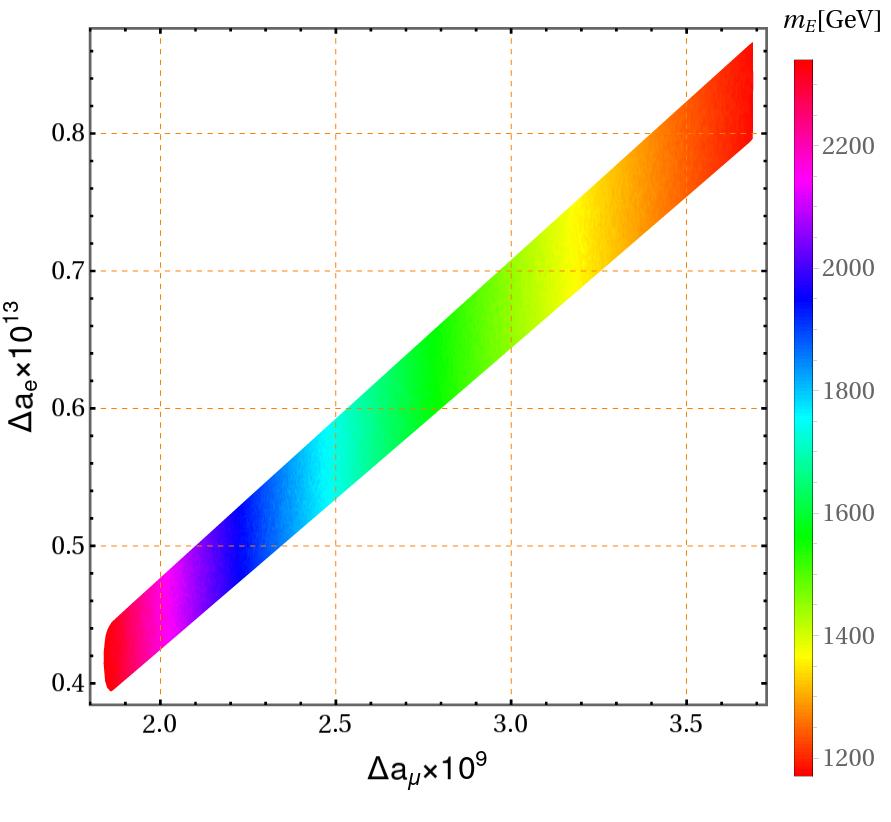}
\caption{Correlation between the electron and muon anomalous magnetic
moments.}
\label{gminus2}
\end{figure}
\begin{figure}[tbp]
\centering
\includegraphics[width=9cm, height=10cm]{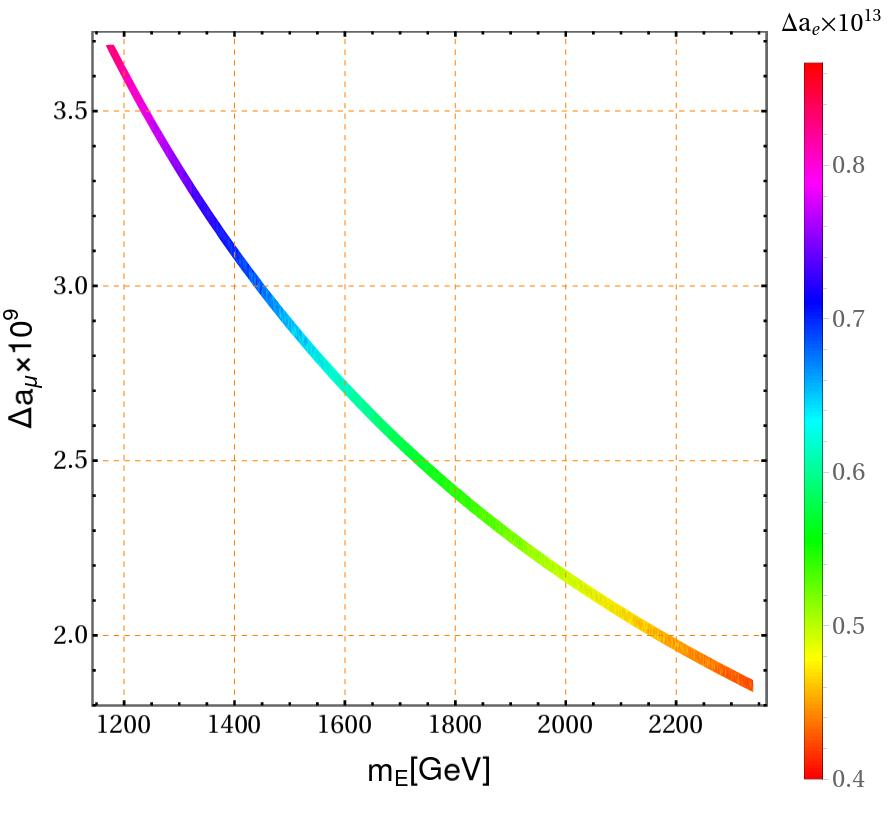}%
\includegraphics[width=9cm, height=10cm]{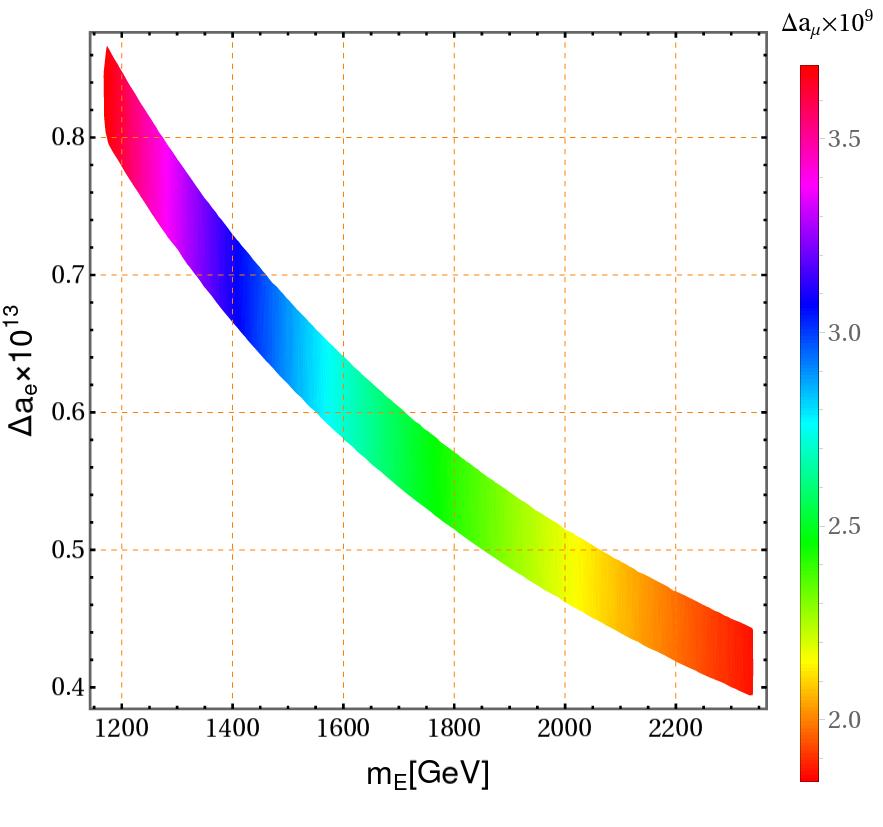}
\caption{Correlation of the muon and electron anomalous magnetic moments
with the charged exotic lepton mass $m_{E}$.}
\label{gminus2b}
\end{figure}

Considering that the muon and electron anomalous magnetic moments are
constrained to be in the ranges \cite{Abi:2021gix,Morel:2020dww}, 
\begin{eqnarray}
\left( \Delta a_{\mu }\right) _{\exp } &=&\left( 2.51\pm 0.59\right) \times
10^{-9}  \notag \\
(\Delta a_{e})_{\text{exp}} &=&(4.8\pm 3.0)\times 10^{-13},
\end{eqnarray}%
we plot in Figure~\ref{gminus2} the correlation between the electron and
muon anomalous magnetic moments, and in Figure~\ref{gminus2b} we show the
correlations of the muon and electron anomalous magnetic moments with the
charged exotic lepton mass $m_{E}$. In our numerical analysis we have fixed $%
m_{H_{1}^{0}}=1.4$ TeV, $m_{H_{2}^{0}}=1.6$ TeV, $m_{H_{3}^{0}}=14$ TeV and
we have varied the masses of the charged exotic leptons in the ranges ${1}${%
\ TeV $\leqslant m_{E}{\leqslant }100$ TeV and }${1}$ TeV $\leqslant
m_{E_{k}^{\prime }}{\leqslant }10$ TeV ($k=1.2$), and we have used the
extended Casas-Ibarra parametrization \cite{Casas:2001sr,Ibarra:2003up} for
the SM charged lepton mass matrix given in Eq.~(\ref{Ml}), to guarantee that
the obtained points of the model parameter space consistent with the muon
and electron anomalous magnetic moments are also in excellent agreement with
the experimental values of the SM charged lepton masses. It is
worth mentioning that the parameter space considered in our analysis is
consistent with the measured values of the SM charged fermion masses and
fermionic mixing parameters. As indicated by Figures \ref{gminus2} and \ref%
{gminus2b}, our model can successfully accommodate the experimental values
of the muon and electron anomalous magnetic moments.


\section{The $95~\text{GeV}$ diphoton excess}

\label{sec:diphoton-excess} 

In this section we discuss the implication of our model for a possible
interpretation of the $95~\text{GeV}$ diphoton excess recently observed by
the CMS collaboration. In what follows we discuss the possibility that the
excess of events in the diphoton final state at the invariant mass of about $%
95~$GeV be due to the real component $\sigma _{R}$ of the scalar singlet $%
\sigma $ of the model under consideration, whose mass is assumed to be equal
to $95~$GeV. The EW scalar singlet $\sigma _{R}$ is mainly produced via a
gluon fusion mechanism, involving the heavy exotic $T_{2}$, $T^{\prime }$
and $B_{n}^{\prime }$ ($n=1,2$)\ quarks running in the internal lines of the
triangular loop. The diphoton decay of the scalar singlet $\sigma _{R}$ is
mediated by triangular loops involving the virtual exchange of vector-like
quarks, charged vector-like leptons $E$ and $E^{\prime }$ and electrically
charged scalars. Consequently, the cross section for the production of the
diphoton scalar resonance at the LHC takes the form:%
\begin{equation}
\sigma _{total}\left( pp\rightarrow \sigma _{R}\rightarrow \gamma \gamma
\right) =\frac{\pi ^{2}}{8}\frac{1}{m_{\sigma _{R}}\Gamma _{\sigma _{R}}}%
\Gamma (\sigma _{R}\rightarrow \gamma \gamma )\frac{1}{s}\int_{\frac{%
m_{\sigma _{R}}^{2}}{s}}^{1}\frac{dx}{x}f_{g}(x)f_{g}\left( \frac{m_{\sigma
_{R}}^{2}}{sx}\right) \Gamma (\sigma _{R}\rightarrow gg),  \label{sigma}
\end{equation}%
where $m_{\sigma _{R}}\simeq 95\text{GeV}$ is the resonance mass, $\Gamma
_{\sigma _{R}}$ its total decay width, $f_{g}(x)$ is the gluon distribution
function, and $\sqrt{s}=13$ TeV is the LHC center of mass energy.

The $95~\text{GeV}$ diphoton excess can be interpreted as a scalar resonance
with a signal strength given by \cite{CMS-PAS-HIG-20-002,Biekotter:2023jld}: 
\begin{equation}
\mu _{\gamma \gamma }^{\left( \exp \right) }=\frac{\sigma _{\exp }\left(
pp\rightarrow \sigma _{R}\rightarrow \gamma \gamma \right) }{\sigma
_{SM}\left( pp\rightarrow h\rightarrow \gamma \gamma \right) }=0.35\pm 0.12
\,,  \label{mu95GeV}
\end{equation}
where $\sigma _{SM}$ corresponds to the total cross section for a
hypothetical SM Higgs boson at the same mass.

The corresponding decay widths of the resonance into photon and gluon pairs
are respectively given by: 
\begin{equation}
\Gamma (\sigma _{R}\rightarrow gg)=\frac{K^{gg}\alpha _{s}^{2}m_{\sigma }^{3}%
}{256\pi ^{3}}\left\vert \frac{y_{T}}{m_{T_{2}}}F(x_{T_{2}})+\frac{%
y_{T^{\prime }}}{m_{T^{\prime }}}F(x_{T^{\prime }})+\sum_{n=1}^{2}\frac{%
y_{B_{n}^{\prime }}}{m_{B_{n}^{\prime }}}F(x_{B_{n}^{\prime }})\right\vert
^{2},  \label{Gammadigluon}
\end{equation}%
\begin{eqnarray}
\Gamma (\sigma _{R}\rightarrow \gamma \gamma ) &=&\frac{\alpha ^{2}m_{\chi
}^{3}}{512\pi ^{3}}\left\vert \frac{N_{c}Q_{T_{2}}^{2}y_{T}}{m_{T_{2}}}%
F_{1/2}(\zeta _{T_{2}})+\frac{N_{c}Q_{T^{\prime }}^{2}y_{T^{\prime }}}{%
m_{T^{\prime }}}F_{1/2}(\zeta _{T^{\prime }})+\sum_{n=1}^{2}\frac{%
N_{c}Q_{B_{n}^{\prime }}^{2}y_{B_{n}^{\prime }}}{m_{B_{n}^{\prime }}}%
F_{1/2}(\zeta _{B_{n}^{\prime }})\right.  \notag \\
&&+\left. \frac{Q_{E}^{2}y_{E}}{m_{E}}F_{1/2}(\zeta
_{E})+\dsum\limits_{n=1}^{2}\frac{Q_{E_{n}^{\prime }}^{2}y_{E_{n}^{\prime }}%
}{m_{E_{n}^{\prime }}}F_{1/2}(\zeta _{E^{\prime }})+\sum_{n=1}^{2}\frac{%
C_{\sigma H_{n}^{\pm }H_{n}^{\mp }}}{\sqrt{2}m_{H_{n}^{\pm }}^{2}}%
F_{0}(\zeta _{H_{n}^{\pm }})\right\vert ^{2}\,,  \label{Gammadiphoton}
\end{eqnarray}%
where $K^{gg}\sim 1.5$ is a QCD loop enhancement factor that accounts for
the higher order QCD corrections, $\zeta _{i}=4M_{i}^{2}/m_{\sigma }^{2}$,
with $M_{i}=m_{T_{2}},m_{T^{\prime }},m_{B_{n}^{\prime }},m_{E},m_{E^{\prime
}},m_{H_{n}^{\pm }}^{2}$ ($n=1,2$) and the loop functions $F_{1/2}(\zeta )$
and $F_{0}(\zeta )$ are given by:%
\begin{equation}
F_{1/2}(\zeta )=-2\zeta \left( 1+(1-\zeta )f(\zeta )\right) ,\,\hspace{0.7cm}%
\hspace{0.7cm}F_{0}(\zeta )=\left( 1-\zeta f(\zeta )\right) \zeta ,
\label{F}
\end{equation}%
where 
\begin{equation}
f(\zeta )=\left\{ 
\begin{array}{lcc}
\left\vert \arcsin \sqrt{1/\zeta }\right\vert ^{2} & \text{for} & \zeta \geq
1 \\ 
&  &  \\ 
-\frac{1}{4}\left( \ln \left( \frac{1+\sqrt{1-\zeta }}{1-\sqrt{1-\zeta }}%
\right) -i\pi \right) ^{2} & \text{for} & \zeta <1 \\ 
&  & 
\end{array}%
\right.  \label{f}
\end{equation}%
For the sake of simplicity we consider a benchmark scenario characterized by
the absence of mixings between the $95$ GeV resonance $\sigma _{R}$ and the
remaining scalar fields. In addition, we assume that the $95$ GeV resonance $%
\sigma _{R}$ is lighter than the remaining scalar fields. Consequently,
under the aforementioned assumptions, the $95$ GeV resonance $\sigma _{R}$
does not feature tree-level decays into scalar pairs and thus it
predominantly decays into photon and gluon pairs. It is worth mentioning
that the $95$ GeV resonance $\sigma _{R}$ can also decay into SM
fermion-antifermion pairs (except for the top-antitop quark pair), however,
those decays are strongly suppressed by the fourth power of the very small
mixing angles between the SM charged fermions lighter than the top quark and
the heavy vector-like fermions. Here, we consider the 
benchmark scenario 
specified in Eq. (\ref{bfpfermions}). The parameters relevant for this part of our analysis are
\begin{eqnarray}
m_{T_{2}} &\simeq &20\mbox{TeV},\hspace{0.9cm}m_{T^{\prime
}}=m_{B_{1}^{\prime }}=m_{B_{2}^{\prime }}\simeq 4\mbox{TeV},\,\hspace{0.7cm}%
\hspace{0.7cm}v_{\sigma }=10\mbox{TeV}  \notag \\
m_{E} &\simeq &1.7\mbox{TeV},\hspace{0.9cm}m_{E_{1}^{\prime }}\simeq 3.8%
\mbox{TeV},\hspace{0.9cm}m_{E_{2}^{\prime }}\simeq 3.3\mbox{TeV},  \notag \\
m_{H_{1}^{\pm }} &=&m_{H^{\pm }},\hspace{0.9cm}m_{H_{2}^{\pm }}=m_{H^{\pm
}}+\delta ,\,\hspace{0.9cm}\delta =0.1\mbox{TeV},\,\hspace{0.9cm}C_{\sigma
H_{1}^{\pm }H_{1}^{\mp }}=C_{\sigma H_{2}^{\pm }H_{2}^{\mp }}=10\,\mbox{TeV},
\label{benchmarkdiphoton}
\end{eqnarray}
which is consistent with the measured values of the SM
charged fermion masses and fermionic mixing parameters, as well as with the
constraints arising from the $g-2$ muon and electron anomalies, and the
experimental measurements of the oblique $T$, $S$ and $U$ parameters, for
charged scalar mass $m_{H^{\pm }}$ in the range $1.3$ TeV $\leqslant
m_{H^{\pm }}{\leqslant 1.5}$ TeV.

\begin{figure}[t]
\centering
\includegraphics[width=10cm, height=8cm]{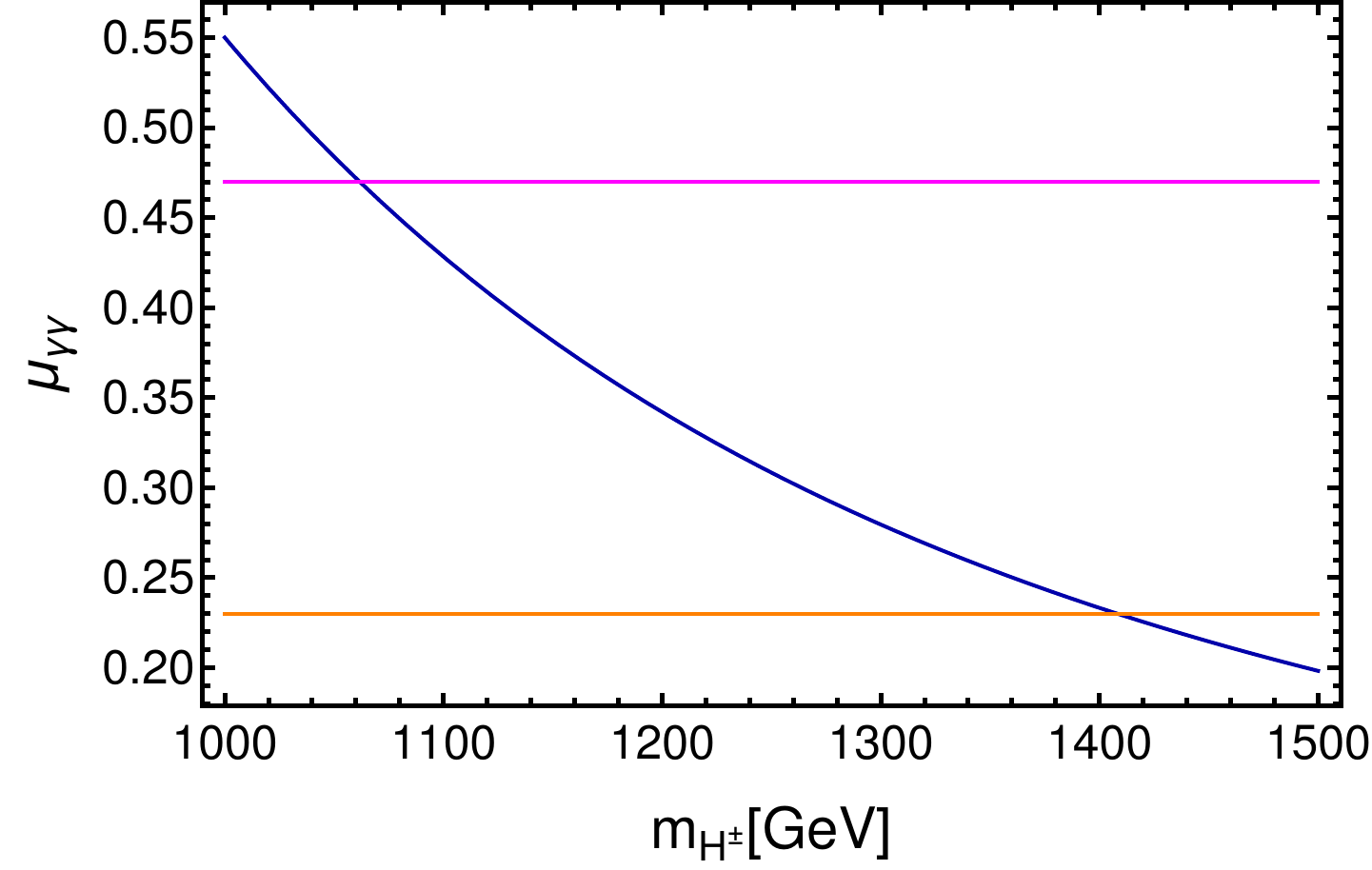}
\caption{Diphoton signal strength for a hypothetical $95~$GeV scalar
resonance as a function of the charged scalar mass $m_{H^{\pm }}$. The magenta and orange horizontal lines correspond to the upper and lower experimental limits within the $1\protect\sigma $ range,
respectively.}
\label{Fig95GeV}
\end{figure}

Figure \ref{Fig95GeV} displays the diphoton signal strength for a
hypothetical $95~$GeV scalar resonance as a function of the charged scalar
mass $m_{H^{\pm }}$. The magenta and orange horizontal lines correspond to
the upper and lower experimental limits within the $1\sigma $ range,
respectively. For the computation of the total cross section, the MSTW
next-to-leading-order (NLO) gluon distribution functions \cite{Martin:2009iq}%
, evaluated at the factorisation scale $\mu =m_{\sigma _{R}}$, have been
used. As seen from Fig.~\ref{Fig95GeV}, our model successfully accommodates
the $95\text{GeV}$ diphoton excess. 


\section{Charged lepton flavor violation}

\label{LFV} 

In this section we will discuss the implications of the model for charged
lepton flavor violation. As mentioned in the previous section, the sterile
neutrino spectrum of the model is composed of six nearly degenerate heavy
neutrinos. These sterile neutrinos, as well as the light active neutrinos
together with the SM $W$ gauge and heavy $W^{\prime }$ gauge bosons, induce
the $l_{i}\rightarrow l_{j}\gamma $ decay at one loop level, whose branching
ratio is given by \cite{Langacker:1988up,Lavoura:2003xp,Hue:2017lak}: 
\begin{equation*}
Br\left( l_{i}\rightarrow l_{j}\gamma \right) =\frac{\alpha
_{W}^{3}s_{W}^{2}m_{l_{i}}^{5}}{256\pi ^{2}m_{W}^{4}\Gamma _{i}}\left\vert
G_{ij}\right\vert ^{2}\,,
\end{equation*}%
where 
\begin{eqnarray}
G_{ij} &\simeq &\sum_{k=1}^{3}\left( \left[ \left( 1_{3\times 3}-RR^{\dagger
}\right) U_{\nu }\right] ^{\ast }\right) _{ik}\left( \left( 1_{3\times
3}-RR^{\dagger }\right) U_{\nu }\right) _{jk}G\left( \frac{m_{\nu _{k}}^{2}}{%
m_{W}^{2}}\right) +2\sum_{k=1}^{3}\left( R^{\ast }\right) _{ik}\left(
R\right) _{jk}G\left( \frac{m_{N_{k}}^{2}}{m_{W}^{2}}\right)  \notag \\
&&+2\frac{m_{W}^{2}}{m_{W^{\prime }}^{2}}\sum_{k=1}^{3}\left( \left[ \left(
1_{3\times 3}-\frac{1}{2}RR^{\dagger }\right) U_{\nu }\right] ^{\ast
}\right) _{ik}\left( \left( 1_{3\times 3}-\frac{1}{2}RR^{\dagger }\right)
U_{\nu }\right) _{jk}G\left( \frac{m_{N_{k}}^{2}}{m_{W^{\prime }}^{2}}\right)
\notag \\
&&+2\frac{m_{W}^{2}}{m_{W^{\prime }}^{2}}\sum_{k=1}^{3}\left( R^{\ast
}\right) _{ik}\left( R\right) _{jk}G\left( \frac{m_{\nu _{k}}^{2}}{%
m_{W^{\prime }}^{2}}\right)  \notag \\
G\left( x\right) &=&\frac{10-43x+78x^{2}-49x^{3}+18x^{3}\ln x+4x^{4}}{%
12\left( 1-x\right) ^{4}}.  \label{Brmutoegamma2}
\end{eqnarray}%
Here, $U_{\nu }$ is the PMNS leptonic mixing matrix, $\Gamma _{\mu }=3\times
10^{-19}$ GeV is the total muon decay width and $R$ are described by \cite%
{Catano:2012kw}: 
\begin{equation}
R=\frac{1}{\sqrt{2}}m_{D}^{\ast }M^{-1}.
\end{equation}%
In our analysis we consider the simplified scenario of degenerate heavy
neutrinos with a common mass $m_{N}$ and we also set $g_{R}=g$. 
\begin{figure}[tbp]
\centering
\includegraphics[width=10cm, height=9cm]{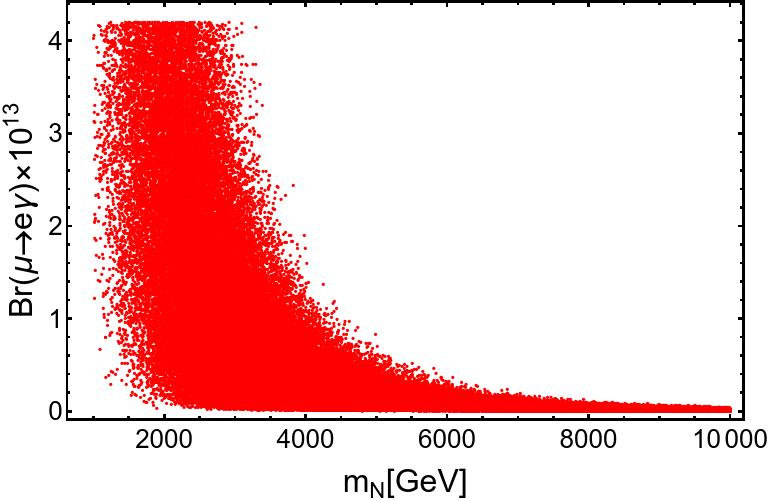}
\caption{Correlation between the branching ratio for the $\protect\mu %
\rightarrow e\protect\gamma $ decay and the mass $m_{N}$ of the sterile
neutrinos.}
\label{mutoegammas}
\end{figure}
Figure \ref{mutoegammas} shows the correlation between the branching ratio
for the $\mu \rightarrow e\gamma $ decay and the mass $m_{N}$ of the sterile
neutrinos. As indicated by Figure \ref{mutoegammas}, the obtained values for
the $\mu \rightarrow e\gamma $ decay branching ratio are below its
experimental upper bound of $4.2\times 10^{-13}$ and are within the reach of
future experimental sensitivity, in the allowed model parameter space. In
the region of parameter space consistent with $\mu \rightarrow e\gamma $
decay rate constraints, we obtain values for the branching ratios for the $%
\tau \rightarrow \mu \gamma $ and $\tau \rightarrow e\gamma $ decays, below
their corresponding upper experimental bounds of $4.4\times 10^{-8}$ and $%
3.3\times 10^{-8}$, respectively. Consequently, our model is compatible with
the current charged lepton flavor violating decay constraints.

To close this section we proceed to discuss the implications of
the proposed model in electron-muon conversion. It is worth mentioning that
the Effective Lagrangian approach of Ref. \cite{Kuno:1999jp}, used for
describing lepton flavor violating (LFV) processes in the regime of low
momentum, where the off-shell contributions from photon exchange are
negligible with respect to the real photon emission contributions, imply
that the dipole operators shown in Ref.~\cite{Kuno:1999jp} will dominate the
Lepton Flavor Violating (LFV) transitions $\mu \rightarrow 3e$, $\mu {\text{%
Al}}\rightarrow e{\text{Al and }}\mu {\text{Ti}}\rightarrow e{\text{T}}$,
thus giving rise to the following relations \cite%
{Kuno:1999jp,Lindner:2016bgg}: 
\begin{equation}
{\text{Br}}\left( \mu \rightarrow 3e\right) \simeq \frac{1}{160}{\text{Br}}%
\left( \mu \rightarrow e\gamma \right) ,\hspace{1cm}{\text{CR}}\left( \mu {%
\text{Ti}}\rightarrow e{\text{Ti}}\right) \simeq \frac{1}{200}{\text{Br}}%
\left( \mu \rightarrow e\gamma \right) ,\hspace{1cm}{\text{CR}}\left( \mu {%
\text{Al}}\rightarrow e{\text{Al}}\right) \simeq \frac{1}{350}{\text{Br}}%
\left( \mu \rightarrow e\gamma \right)  \label{eq:CR-BR}
\end{equation}%
where the $\mu ^{-}-e^{-}$ conversion ratio is defined~\cite{Lindner:2016bgg}
as follows: 
\begin{equation}
{\text{CR}}\left( \mu -e\right) =\frac{\Gamma \left( \mu ^{-}+{\text{Nucleus}%
}\left( A,Z\right) \rightarrow e^{-}+{\text{Nucleus}}\left( A,Z\right)
\right) }{\Gamma \left( \mu ^{-}+{\text{Nucleus}}\left( A,Z\right)
\rightarrow \nu _{\mu }+{\text{Nucleus}}\left( A,Z-1\right) \right) }
\label{eq:Conversion-Rate}
\end{equation}%
Consequently, for our model we expect that the resulting rates for the LFV
transitions $\mu \rightarrow 3e$, $\mu {\text{Al}}\rightarrow e{\text{Al and 
}}\mu {\text{Ti}}\rightarrow e{\text{T}}$ will be of the order of $10^{-15}$%
, i.e, two orders of magnitude lower than the obtained rates for the $\mu
\rightarrow e\gamma $ decay, thus implying that in our model the
corresponding values are below the current experimental bounds of about $%
10^{-12}$ for these lepton flavor violating transitions.


\section{Dark matter candidate}

\label{sec:DM} 

As we described in section~\ref{Sect:model}, in this model there is a residual matter parity symmetry (\ref{eq:M-P}), surviving the spontaneous breaking  of the global $U(1)_X$ group.
%
This symmetry is responsible for stabilizing the dark matter particle candidate in the our model. 
In this case, the fields that compose the dark sector have no electric
charge and are odd under the matter parity symmetry. Following these
criteria, one can identify the neutral components of the VEV-less scalars $%
\phi_L$, $\phi_R$, and $\varphi$, and the Majorana fermions $N_{R_i}$ $%
(i=1,2,3)$ as part of the dark sector. All these fields are involved in the
fermion mass generation at the loop-level, see Figs.~\ref%
{Diagramschargedfermions} and \ref{Diagramsneutrinos}. Notice that the
Majorana fermions $N_{R_i}$ mix the with active neutrinos, which will cause
their decay. For this reason, the $N_{R_i}$ are discarded as dark matter
candidates. Therefore, the dark matter candidate in this model has a scalar
nature and would be the lightest neutral component of the scalars $%
(\phi_L^0,\phi_R^0,\varphi)$. Here, $(\phi_L^0,\phi_R^0)$ generate the
corrections to the quark and charged lepton mass matrices, Fig.~\ref%
{Diagramschargedfermions}, and $\varphi$ generates the $\mu$ term in Eq.~(%
\ref{Mnu}) needed to get the neutrino masses, see Fig.~\ref%
{Diagramsneutrinos}.

This model has enough freedom to consider different simplified situations in
the parameter space. One of these limits is to assume that the CP-even (odd)
component of the singlet scalar $\varphi $ is the DM candidate. One can see
from Eqs.~(\ref{M2CPeven}) and (\ref{M2CPodd}) that this field does not mix
with any other dark scalar. In this case, the mass hierarchy between dark
scalars, $m_{\varphi }<{m_{\phi _{R}^{0}},m_{\phi _{L}^{0}}}$, is satisfied.
Notice that in such a benchmark the dark matter relic abundance will be
dominated by the DM annihilation into scalars, similar to Refs.~\cite%
{Burgess:2000yq,Casas:2017jjg}. The DM annihilation channel (s-channel with
Higgs exchange) into SM particles is expected to appear by effective
couplings that mix SM and heavy fermions. The DM annihilation processes into
heavy fermions might be possible, if allowed by kinematics.

In what follows we analyze the implications of the model for the
dark matter relic density as well as in the direct detection for dark
matter. In order to simplify our analysis we consider a benchmark scenario
where the scalar DM candidate $\varphi _{I}$ annihilates into a pair of SM
particles as well as into a pair of the heavy CP even state $H$ arising from
the $SU\left( 2\right) _{R}$ scalar doublet $\chi _{R}$. In this benchmark
scenario we take all dark sector scalar couplings excepting $\lambda _{24}$
and $\lambda _{25}$ to be small. Then, in the previously described
simplified benmark scenario, the scalar dark matter candidate mainly
annihilates into $WW$, $ZZ$, $t\overline{t}$, $b\overline{b}$ and $hh$, via
a Higgs portal scalar interaction $\lambda _{24}(\varphi ^{\dagger }\varphi
)(\chi _{L}^{\dagger }\chi _{L})$, as well into $HH$ thanks to the quartic
scalar interaction $\lambda _{25}(\varphi ^{\dagger }\varphi )(\chi
_{R}^{\dagger }\chi _{R})$.

The corresponding annihilation cross sections are given by \cite{Bhattacharya:2016ysw}: 
\begin{eqnarray}
v_{rel}\sigma \left( \varphi_{I}\varphi_{I}\rightarrow WW\right) &=&\frac{%
\lambda _{24}^{2}}{8\pi }\frac{s\left( 1+\frac{12m_{W}^{4}}{s^{2}}-\frac{%
4m_{W}^{2}}{s}\right) }{\left( s-m_{h}^{2}\right) ^{2}+m_{h}^{2}\Gamma
_{h}^{2}}\sqrt{1-\frac{4m_{W}^{2}}{s}},  \notag \\
v_{rel}\sigma \left( \varphi_{I}\varphi_{I}\rightarrow ZZ\right) &=&\frac{%
\lambda _{24}^{2}}{16\pi }\frac{s\left( 1+\frac{12m_{Z}^{4}}{s^{2}}-\frac{%
4m_{Z}^{2}}{s}\right) }{\left( s-m_{h}^{2}\right) ^{2}+m_{h}^{2}\Gamma
_{h}^{2}}\sqrt{1-\frac{4m_{Z}^{2}}{s}},  \notag \\
v_{rel}\sigma \left( \varphi_{I}\varphi_{I}\rightarrow q\overline{q}\right) &=&%
\frac{N_{c}\lambda _{24}^{2}m_{q}^{2}}{4\pi }\frac{\sqrt{\left( 1-\frac{%
4m_{f}^{2}}{s}\right) ^{3}}}{\left( s-m_{h}^{2}\right) ^{2}+m_{h}^{2}\Gamma
_{h}^{2}},  \notag \\
v_{rel}\sigma \left( \varphi_{I}\varphi_{I}\rightarrow hh\right) &=&\frac{%
\lambda _{24}^{2}}{16\pi s}\left( 1+\frac{3m_{h}^{2}}{s-m_{h}^{2}}-\frac{%
4\lambda _{24}v^{2}}{s-2m_{h}^{2}}\right) ^{2}\sqrt{1-\frac{4m_{h}^{2}}{s}},
\notag \\
v_{rel}\sigma \left( \varphi_{I}\varphi_{I}\rightarrow HH\right) &=&\frac{\lambda _{25}^{2}}{16\pi s}\sqrt{1-\frac{4m_{H}^{2}}{s}},
\end{eqnarray}
where $\sqrt{s}$ is the centre-of-mass energy, $N_{c}=3$ is the color
factor, $m_{h}=125.7$ GeV and $\Gamma _{h}=4.1$ MeV are the SM Higgs boson $%
h $ mass and its total decay width, respectively; $H\simeq \func{Re}\chi
_{R}^{0}$, $\lambda _{24}$ and $\lambda _{25}$ are the quartic scalar
coupling corresponding to the interactions $\lambda _{24}(\varphi ^{\dagger
}\varphi )(\chi _{L}^{\dagger }\chi _{L})$ and $\lambda _{25}(\varphi
^{\dagger }\varphi )(\chi _{R}^{\dagger }\chi _{R})$, respectively.

\begin{figure}[tbp]
\centering
\includegraphics[width=12cm, height=10cm]{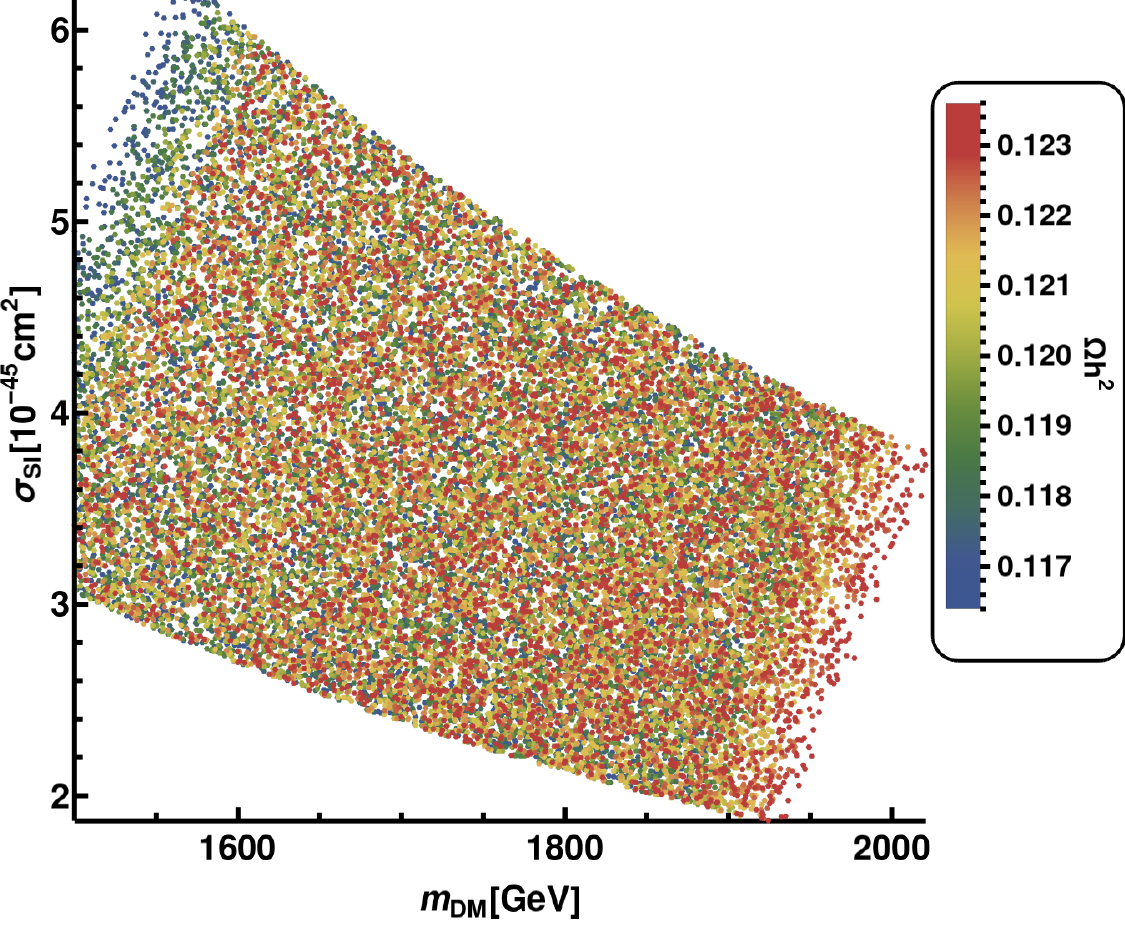}%
\caption{Correlation of the spin independent cross section $\sigma_{SI}$ with the dark matter mass $m_{DM}$.}
\label{DM}
\end{figure}

It is known that the Dark Matter relic abundance of the present Universe can
be determined as follows \cite{ParticleDataGroup:2022pth,Edsjo:1997bg}:
\begin{equation}
\Omega h^{2}=\frac{0.1\;\text{pb}}{\left\langle \sigma v\right\rangle },\,%
\hspace{1cm}\left\langle \sigma v\right\rangle =\frac{A}{n_{eq}^{2}}\,,
\end{equation}%
where $\left\langle \sigma v\right\rangle $ is the thermally averaged
annihilation cross section, $A$ is the total annihilation rate per unit
volume at temperature $T$ and $n_{eq}$ is the equilibrium value of the
particle density, which are given in \cite{Edsjo:1997bg} 
\begin{eqnarray}
A &=&\frac{T}{32\pi ^{4}}\dint\limits_{4m_{\varphi }^{2}}^{\infty
}\dsum\limits_{p=W,Z,t,b,h,H}g_{p}^{2}\frac{s\sqrt{s-4m_{\rho _{I}}^{2}}}{2}%
v_{rel}\sigma \left( \varphi _{I}\varphi _{I}\rightarrow p\overline{p}%
\right) K_{1}\left( \frac{\sqrt{s}}{T}\right) ds,  \notag \\
n_{eq} &=&\frac{T}{2\pi ^{2}}\dsum\limits_{p=W,Z,t,b,h,H}g_{p}m_{\varphi
_{I}}^{2}K_{2}\left( \frac{\varphi _{I}}{T}\right) ,
\end{eqnarray}%
with $K_{1}$ and $K_{2}$ being the modified Bessel functions of the second
kind of order 1 and 2, respectively \cite{Edsjo:1997bg}. For the relic
density calculation, we take $T=m_{\varphi _{I}}/20$ as in Ref. \cite%
{Edsjo:1997bg}, which corresponds to a typical freeze-out temperature.We
require that the obtained values for the dark matter relic abundance to be
consistent with the experimentally allowed range measured by the
Planck collaboration \cite{Planck:2018vyg}: 
\begin{equation}
\Omega _{DM}h^{2}=0.1200\pm 0.0012
\end{equation}
Besides that, direct detection constraints should also be taken into
account. They are obtained from the comparison of the spin-independent cross
section for the scattering of a dark matter off a nucleon,
\begin{equation}
\sigma _{SI}=\frac{\lambda _{24}^{2}m_{N}^{4}f^{2}}{8\pi m_{h}^{4}m_{\varphi
_{I}}^{2}}
\end{equation}
to the most recent upper bounds on $\sigma _{SI}$ arising from the Xenon1T experiment \cite{XENON:2018voc}. As shown in Ref. \cite{XENON:2018voc}, these upper bounds are of the order of $10^{-44}$ $cm^{2}$ for dark matter masses in the range $1.{5}$ TeV $\leqslant m_{DM}{\leqslant 2}$ TeV, with $m_{DM}=m_{\varphi _{I}}$. Here $m_{N}$ is the nucleon mass and $f\simeq \frac{1}{3}$ corresponds to the form factor \cite{Farina:2009ez,Giedt:2009mr}. Furthermore, we display in Figure \ref{DM} the correlation of the spin independent cross section with the dark matter mass. Here we have set $m_H=1$ TeV and we have varied the quartic scalar couplings $\lambda_{24}$ and $\lambda_{25}$ in the range $0-4\pi$. As indicated in Figure \ref{DM} our model is consistent with the constraints arising from dark matter direct detection and can successfully reproduce the experimental values of the dark matter relic density.


\section{Conclusions}

\label{conclusions} 

We have proposed a renormalizable extended Left-Right symmetric theory with
an additional global $U(1)_{X}$ symmetry, capable of explaining and
accommodating the observed SM fermion mass hierarchy and the tiny values of
the light active neutrino masses. As the same time, the model has been
demonstrated to be consistent with the current amount of dark matter relic
density observed in the Universe, the muon anomalous magnetic moment, the
oblique $T$, $S$ and $U$ parameters, as well as the $95$ GeV diphoton
excess. In the proposed model, the global $U(1)_{X}$ symmetry is
spontaneously broken down to a matter parity symmetry, thus allowing the
existence of stable scalar dark matter candidates in the particle spectrum
of our model and ensuring the radiative nature of the seesaw mechanisms that
generate the masses of the light up, down and strange quarks, as well as
those ones of the electron, muon and active neutrinos.

As an important feature of our model, the third generation of SM charged
fermions and the charm quark obtain their masses via a tree-level seesaw
mechanism triggered by their mixing with heavy charged vector-like fermions.
The masses of the light up, down and strange quarks, as well as the electron
and muon masses, are generated via a one-loop radiative seesaw mechanism
mediated by heavy charged vector-like fermions and non-SM scalars. The tiny
active neutrino masses are produced through a three-loop inverse seesaw
mechanism, with the Dirac and Majorana submatrices generated at one-loop
level. In the low-energy limit, the model features one naturally light $126$
GeV SM-like boson, strongly decoupled from the other heavy scalars, as well
as the absence of tree-level FCNC processes, rendering the model safe
against the existing flavor physics bounds.


\section*{Acknowledgments}

A.E.C.H, I.S. and S.K are supported by ANID-Chile FONDECYT 1210378, ANID-Chile FONDECYT 1230391, ANID-Chile FONDECYT 1180232, ANID-Chile FONDECYT 3150472, ANID-Chile
FONDECYT 1230160, ANID PIA/APOYO AFB220004 and Proyecto Milenio-ANID: ICN2019\_044.
The work of C.B. was supported by FONDECYT grant No. 11201240. R.P.~is
supported in part by the Swedish Research Council grant, contract number
2016-05996, as well as by the European Research Council (ERC) under the
European Union's Horizon 2020 research and innovation programme (grant
agreement No 668679). H.L. is supported by the National Science Centre
(Poland) under the research Grant No. 2017/26/E/ST2/00470. C.B. would like
to acknowledge the hospitality and support from the ICTP through the
Associates Programme (2023-2028).


\bibliographystyle{utphys}
\bibliography{BiblioLRMay2022}

\providecommand{\href}[2]{#2}\begingroup\raggedright\begin{thebibliography}{10}

\bibitem{Davidson:1987mh}
A.~Davidson and K.~C. Wali, ``{Universal Seesaw Mechanism?},''
\href{http://dx.doi.org/10.1103/PhysRevLett.59.393}{{\em Phys. Rev. Lett.}
  {\bfseries 59} (1987) 393}.

\bibitem{Ma:2009gu}
E.~Ma, ``{Radiative inverse seesaw mechanism for nonzero neutrino mass},''
  \href{http://dx.doi.org/10.1103/PhysRevD.80.013013}{{\em Phys. Rev. D}
  {\bfseries 80} (2009) 013013},
  \href{http://arxiv.org/abs/0904.4450}{{\ttfamily arXiv:0904.4450 [hep-ph]}}.

\bibitem{Law:2012mj}
S.~S.~C. Law and K.~L. McDonald, ``{Inverse seesaw and dark matter in models
  with exotic lepton triplets},''
  \href{http://dx.doi.org/10.1016/j.physletb.2012.06.044}{{\em Phys. Lett. B}
  {\bfseries 713} (2012) 490--494},
  \href{http://arxiv.org/abs/1204.2529}{{\ttfamily arXiv:1204.2529 [hep-ph]}}.

\bibitem{Ahriche:2016acx}
A.~Ahriche, S.~M. Boucenna, and S.~Nasri, ``{Dark Radiative Inverse Seesaw
  Mechanism},'' \href{http://dx.doi.org/10.1103/PhysRevD.93.075036}{{\em Phys.
  Rev. D} {\bfseries 93} no.~7, (2016) 075036},
  \href{http://arxiv.org/abs/1601.04336}{{\ttfamily arXiv:1601.04336
  [hep-ph]}}.

\bibitem{CarcamoHernandez:2017kra}
A.~E. C\'arcamo~Hern\'andez and H.~N. Long, ``{A highly predictive $A_{4}$
  flavour 3-3-1 model with radiative inverse seesaw mechanism},''
  \href{http://dx.doi.org/10.1088/1361-6471/aaace7}{{\em J. Phys. G} {\bfseries
  45} no.~4, (2018) 045001}, \href{http://arxiv.org/abs/1705.05246}{{\ttfamily
  arXiv:1705.05246 [hep-ph]}}.

\bibitem{CarcamoHernandez:2017cwi}
A.~E. C\'arcamo~Hern\'andez, S.~Kovalenko, H.~N. Long, and I.~Schmidt, ``{A
  variant of 3-3-1 model for the generation of the SM fermion mass and mixing
  pattern},'' \href{http://dx.doi.org/10.1007/JHEP07(2018)144}{{\em JHEP}
  {\bfseries 07} (2018) 144}, \href{http://arxiv.org/abs/1705.09169}{{\ttfamily
  arXiv:1705.09169 [hep-ph]}}.

\bibitem{CarcamoHernandez:2018hst}
A.~E. Cárcamo~Hernández, S.~Kovalenko, J.~W.~F. Valle, and C.~A.
  Vaquera-Araujo, ``{Neutrino predictions from a left-right symmetric flavored
  extension of the standard model},''
  \href{http://dx.doi.org/10.1007/JHEP02(2019)065}{{\em JHEP} {\bfseries 02}
  (2019) 065},
\href{http://arxiv.org/abs/1811.03018}{{\ttfamily arXiv:1811.03018 [hep-ph]}}.

\bibitem{Mandal:2019oth}
S.~Mandal, N.~Rojas, R.~Srivastava, and J.~W.~F. Valle, ``{Dark matter as the
  origin of neutrino mass in the inverse seesaw mechanism},''
  \href{http://dx.doi.org/10.1016/j.physletb.2021.136609}{{\em Phys. Lett. B}
  {\bfseries 821} (2021) 136609},
  \href{http://arxiv.org/abs/1907.07728}{{\ttfamily arXiv:1907.07728
  [hep-ph]}}.

\bibitem{CarcamoHernandez:2019lhv}
A.~E. C\'arcamo~Hern\'andez, D.~T. Huong, and H.~N. Long, ``{Minimal model for
  the fermion flavor structure, mass hierarchy, dark matter, leptogenesis, and
  the electron and muon anomalous magnetic moments},''
  \href{http://dx.doi.org/10.1103/PhysRevD.102.055002}{{\em Phys. Rev. D}
  {\bfseries 102} no.~5, (2020) 055002},
  \href{http://arxiv.org/abs/1910.12877}{{\ttfamily arXiv:1910.12877
  [hep-ph]}}.

\bibitem{Abada:2021yot}
A.~Abada, N.~Bernal, A.~E.~C. Hern\'andez, X.~Marcano, and G.~Piazza, ``{Gauged
  inverse seesaw from dark matter},''
  \href{http://dx.doi.org/10.1140/epjc/s10052-021-09535-5}{{\em Eur. Phys. J.
  C} {\bfseries 81} no.~8, (2021) 758},
  \href{http://arxiv.org/abs/2107.02803}{{\ttfamily arXiv:2107.02803
  [hep-ph]}}.

\bibitem{Hernandez:2021kju}
A.~E.~C. Hern\'andez, C.~Espinoza, J.~C. G\'omez-Izquierdo, and M.~Mondrag\'on,
  ``{Fermion masses and mixings, dark matter, leptogenesis and $g-2$ muon
  anomaly in an extended 2HDM with inverse seesaw},''
  \href{http://dx.doi.org/10.1140/epjp/s13360-022-03432-w}{{\em Eur. Phys. J.
  Plus} {\bfseries 137} no.~11, (2022) 1224},
  \href{http://arxiv.org/abs/2104.02730}{{\ttfamily arXiv:2104.02730
  [hep-ph]}}.

\bibitem{Hernandez:2021xet}
A.~E.~C. Hern\'andez, D.~T. Huong, and I.~Schmidt, ``{Universal inverse seesaw
  mechanism as a source of the SM fermion mass hierarchy},''
  \href{http://dx.doi.org/10.1140/epjc/s10052-022-10011-x}{{\em Eur. Phys. J.
  C} {\bfseries 82} no.~1, (2022) 63},
  \href{http://arxiv.org/abs/2109.12118}{{\ttfamily arXiv:2109.12118
  [hep-ph]}}.

\bibitem{Hernandez:2021uxx}
A.~E.~C. Hern\'andez and I.~Schmidt, ``{A renormalizable left-right symmetric
  model with low scale seesaw mechanisms},''
  \href{http://dx.doi.org/10.1016/j.nuclphysb.2022.115696}{{\em Nucl. Phys. B}
  {\bfseries 976} (2022) 115696},
  \href{http://arxiv.org/abs/2101.02718}{{\ttfamily arXiv:2101.02718
  [hep-ph]}}.

\bibitem{Dekens:2014ina}
W.~Dekens and D.~Boer, ``{Viability of minimal left–right models with
  discrete symmetries},''
  \href{http://dx.doi.org/10.1016/j.nuclphysb.2014.10.025}{{\em Nucl. Phys.}
  {\bfseries B889} (2014) 727--756},
\href{http://arxiv.org/abs/1409.4052}{{\ttfamily arXiv:1409.4052 [hep-ph]}}.

\bibitem{Nomura:2016run}
T.~Nomura, H.~Okada, and Y.~Orikasa, ``{Radiative neutrino mass in alternative
  left–right model},''
  \href{http://dx.doi.org/10.1140/epjc/s10052-017-4657-4}{{\em Eur. Phys. J.}
  {\bfseries C77} no.~2, (2017) 103},
\href{http://arxiv.org/abs/1602.08302}{{\ttfamily arXiv:1602.08302 [hep-ph]}}.

\bibitem{Brdar:2018sbk}
V.~Brdar and A.~Y. Smirnov, ``{Low Scale Left-Right Symmetry and Naturally
  Small Neutrino Mass},'' \href{http://dx.doi.org/10.1007/JHEP02(2019)045}{{\em
  JHEP} {\bfseries 02} (2019) 045},
\href{http://arxiv.org/abs/1809.09115}{{\ttfamily arXiv:1809.09115 [hep-ph]}}.

\bibitem{Ma:2020lnm}
E.~Ma, ``{Universal Scotogenic Fermion Masses in Left-Right Gauge Model},''
  \href{http://dx.doi.org/10.1016/j.nuclphysb.2021.115406}{{\em Nucl. Phys.}
  {\bfseries B967} (2021) 115406},
\href{http://arxiv.org/abs/2012.03128}{{\ttfamily arXiv:2012.03128 [hep-ph]}}.

\bibitem{Babu:2020bgz}
K.~S. Babu and A.~Thapa, ``{Left-Right Symmetric Model without Higgs
  Triplets},''
\href{http://arxiv.org/abs/2012.13420}{{\ttfamily arXiv:2012.13420 [hep-ph]}}.

\bibitem{ATLAS:2018cjd}
{\bfseries ATLAS} Collaboration, M.~Aaboud {\em et~al.}, ``{Search for large
  missing transverse momentum in association with one top-quark in
  proton-proton collisions at $ \sqrt{s} $ = 13 TeV with the ATLAS detector},''
  \href{http://dx.doi.org/10.1007/JHEP05(2019)041}{{\em JHEP} {\bfseries 05}
  (2019) 041}, \href{http://arxiv.org/abs/1812.09743}{{\ttfamily
  arXiv:1812.09743 [hep-ex]}}.

\bibitem{ATLAS:2018alq}
{\bfseries ATLAS} Collaboration, M.~Aaboud {\em et~al.}, ``{Search for new
  phenomena in events with same-charge leptons and $b$-jets in $pp$ collisions
  at $\sqrt{s}= 13$ TeV with the ATLAS detector},''
  \href{http://dx.doi.org/10.1007/JHEP12(2018)039}{{\em JHEP} {\bfseries 12}
  (2018) 039}, \href{http://arxiv.org/abs/1807.11883}{{\ttfamily
  arXiv:1807.11883 [hep-ex]}}.

\bibitem{ATLAS:2018uky}
{\bfseries ATLAS} Collaboration, M.~Aaboud {\em et~al.}, ``{Search for pair
  production of heavy vector-like quarks decaying into hadronic final states in
  $pp$ collisions at $\sqrt{s} = 13$ TeV with the ATLAS detector},''
  \href{http://dx.doi.org/10.1103/PhysRevD.98.092005}{{\em Phys. Rev. D}
  {\bfseries 98} no.~9, (2018) 092005},
  \href{http://arxiv.org/abs/1808.01771}{{\ttfamily arXiv:1808.01771
  [hep-ex]}}.

\bibitem{ATLAS:2017nap}
{\bfseries ATLAS} Collaboration, M.~Aaboud {\em et~al.}, ``{Search for pair
  production of heavy vector-like quarks decaying to high-p$_{T}$ W bosons and
  b quarks in the lepton-plus-jets final state in pp collisions at $
  \sqrt{s}=13 $ TeV with the ATLAS detector},''
  \href{http://dx.doi.org/10.1007/JHEP10(2017)141}{{\em JHEP} {\bfseries 10}
  (2017) 141}, \href{http://arxiv.org/abs/1707.03347}{{\ttfamily
  arXiv:1707.03347 [hep-ex]}}.

\bibitem{ATLAS:2016scx}
{\bfseries ATLAS} Collaboration, G.~Aad {\em et~al.}, ``{Search for single
  production of vector-like quarks decaying into Wb in pp collisions at
  $\sqrt{s} = 8$ TeV with the ATLAS detector},''
  \href{http://dx.doi.org/10.1140/epjc/s10052-016-4281-8}{{\em Eur. Phys. J. C}
  {\bfseries 76} no.~8, (2016) 442},
  \href{http://arxiv.org/abs/1602.05606}{{\ttfamily arXiv:1602.05606
  [hep-ex]}}.

\bibitem{ATLAS:2015ktd}
{\bfseries ATLAS} Collaboration, G.~Aad {\em et~al.}, ``{Search for production
  of vector-like quark pairs and of four top quarks in the lepton-plus-jets
  final state in $pp$ collisions at $\sqrt{s}=8$ TeV with the ATLAS
  detector},'' \href{http://dx.doi.org/10.1007/JHEP08(2015)105}{{\em JHEP}
  {\bfseries 08} (2015) 105}, \href{http://arxiv.org/abs/1505.04306}{{\ttfamily
  arXiv:1505.04306 [hep-ex]}}.

\bibitem{ATLAS:2015lpr}
{\bfseries ATLAS} Collaboration, G.~Aad {\em et~al.}, ``{Search for pair
  production of a new heavy quark that decays into a $W$ boson and a light
  quark in $pp$ collisions at $\sqrt{s} = 8$ TeV with the ATLAS detector},''
  \href{http://dx.doi.org/10.1103/PhysRevD.92.112007}{{\em Phys. Rev. D}
  {\bfseries 92} no.~11, (2015) 112007},
  \href{http://arxiv.org/abs/1509.04261}{{\ttfamily arXiv:1509.04261
  [hep-ex]}}.

\bibitem{ATLAS:2011tvb}
{\bfseries ATLAS} Collaboration, G.~Aad {\em et~al.}, ``{Search for heavy
  vector-like quarks coupling to light quarks in proton-proton collisions at
  $\sqrt{s}=7$ TeV with the ATLAS detector},''
  \href{http://dx.doi.org/10.1016/j.physletb.2012.03.082}{{\em Phys. Lett. B}
  {\bfseries 712} (2012) 22--39},
  \href{http://arxiv.org/abs/1112.5755}{{\ttfamily arXiv:1112.5755 [hep-ex]}}.

\bibitem{Freitas:2022cno}
F.~F. Freitas, J.~a. Gon\c{c}alves, A.~P. Morais, and R.~Pasechnik,
  ``{Phenomenology at the large hadron collider with deep learning: the case of
  vector-like quarks decaying to light jets},''
  \href{http://dx.doi.org/10.1140/epjc/s10052-022-10799-8}{{\em Eur. Phys. J.
  C} {\bfseries 82} no.~9, (2022) 826},
  \href{http://arxiv.org/abs/2204.12542}{{\ttfamily arXiv:2204.12542
  [hep-ph]}}.

\bibitem{Freitas:2020ttd}
F.~F. Freitas, J.~a. Gon\c{c}alves, A.~P. Morais, and R.~Pasechnik,
  ``{Phenomenology of vector-like leptons with Deep Learning at the Large
  Hadron Collider},'' \href{http://dx.doi.org/10.1007/JHEP01(2021)076}{{\em
  JHEP} {\bfseries 01} (2021) 076},
  \href{http://arxiv.org/abs/2010.01307}{{\ttfamily arXiv:2010.01307
  [hep-ph]}}.

\bibitem{Morais:2021ead}
A.~P. Morais, A.~Onofre, F.~F. Freitas, J.~a. Gon\c{c}alves, R.~Pasechnik, and
  R.~Santos, ``{Deep learning searches for vector-like leptons at the LHC and
  electron/muon colliders},''
  \href{http://dx.doi.org/10.1140/epjc/s10052-023-11314-3}{{\em Eur. Phys. J.
  C} {\bfseries 83} no.~3, (2023) 232},
  \href{http://arxiv.org/abs/2108.03926}{{\ttfamily arXiv:2108.03926
  [hep-ph]}}.

\bibitem{Xing:2020ijf}
Z.-z. Xing, ``{Flavor structures of charged fermions and massive neutrinos},''
  \href{http://dx.doi.org/10.1016/j.physrep.2020.02.001}{{\em Phys. Rept.}
  {\bfseries 854} (2020) 1--147},
  \href{http://arxiv.org/abs/1909.09610}{{\ttfamily arXiv:1909.09610
  [hep-ph]}}.

\bibitem{ParticleDataGroup:2022pth}
{\bfseries Particle Data Group} Collaboration, R.~L. Workman {\em et~al.},
  ``{Review of Particle Physics},''
  \href{http://dx.doi.org/10.1093/ptep/ptac097}{{\em PTEP} {\bfseries 2022}
  (2022) 083C01}.

\bibitem{CarcamoHernandez:2016pdu}
A.~E. C\'arcamo~Hern\'andez, S.~Kovalenko, and I.~Schmidt, ``{Radiatively
  generated hierarchy of lepton and quark masses},''
  \href{http://dx.doi.org/10.1007/JHEP02(2017)125}{{\em JHEP} {\bfseries 02}
  (2017) 125}, \href{http://arxiv.org/abs/1611.09797}{{\ttfamily
  arXiv:1611.09797 [hep-ph]}}.

\bibitem{Peskin:1991sw}
M.~E. Peskin and T.~Takeuchi, ``{Estimation of oblique electroweak
  corrections},'' \href{http://dx.doi.org/10.1103/PhysRevD.46.381}{{\em Phys.
  Rev. D} {\bfseries 46} (1992) 381--409}.

\bibitem{Altarelli:1990zd}
G.~Altarelli and R.~Barbieri, ``{Vacuum polarization effects of new physics on
  electroweak processes},''
  \href{http://dx.doi.org/10.1016/0370-2693(91)91378-9}{{\em Phys. Lett. B}
  {\bfseries 253} (1991) 161--167}.

\bibitem{Barbieri:2004qk}
R.~Barbieri, A.~Pomarol, R.~Rattazzi, and A.~Strumia, ``{Electroweak symmetry
  breaking after LEP-1 and LEP-2},''
  \href{http://dx.doi.org/10.1016/j.nuclphysb.2004.10.014}{{\em Nucl. Phys. B}
  {\bfseries 703} (2004) 127--146},
  \href{http://arxiv.org/abs/hep-ph/0405040}{{\ttfamily arXiv:hep-ph/0405040}}.

\bibitem{CarcamoHernandez:2015smi}
A.~E. C\'arcamo~Hern\'andez, S.~Kovalenko, and I.~Schmidt, ``{Precision
  measurements constraints on the number of Higgs doublets},''
  \href{http://dx.doi.org/10.1103/PhysRevD.91.095014}{{\em Phys. Rev. D}
  {\bfseries 91} (2015) 095014},
  \href{http://arxiv.org/abs/1503.03026}{{\ttfamily arXiv:1503.03026
  [hep-ph]}}.

\bibitem{Adam:2019oes}
A.~S. Adam, A.~Ferdiyan, and M.~Satriawan, ``{A New Left-Right Symmetry
  Model},'' \href{http://dx.doi.org/10.1155/2020/3090783}{{\em Adv. High Energy
  Phys.} {\bfseries 2020} (2020) 3090783},
  \href{http://arxiv.org/abs/1903.03370}{{\ttfamily arXiv:1903.03370
  [hep-ph]}}.

\bibitem{Lu:2022bgw}
C.-T. Lu, L.~Wu, Y.~Wu, and B.~Zhu, ``{Electroweak precision fit and new
  physics in light of the W boson mass},''
  \href{http://dx.doi.org/10.1103/PhysRevD.106.035034}{{\em Phys. Rev. D}
  {\bfseries 106} no.~3, (2022) 035034},
  \href{http://arxiv.org/abs/2204.03796}{{\ttfamily arXiv:2204.03796
  [hep-ph]}}.

\bibitem{Diaz:2002uk}
R.~A. Diaz, R.~Martinez, and J.~A. Rodriguez, ``{Phenomenology of lepton flavor
  violation in 2HDM(3) from (g-2)(mu) and leptonic decays},''
  \href{http://dx.doi.org/10.1103/PhysRevD.67.075011}{{\em Phys. Rev.}
  {\bfseries D67} (2003) 075011},
\href{http://arxiv.org/abs/hep-ph/0208117}{{\ttfamily arXiv:hep-ph/0208117
  [hep-ph]}}.

\bibitem{Jegerlehner:2009ry}
F.~Jegerlehner and A.~Nyffeler, ``{The Muon g-2},''
  \href{http://dx.doi.org/10.1016/j.physrep.2009.04.003}{{\em Phys. Rept.}
  {\bfseries 477} (2009) 1--110},
\href{http://arxiv.org/abs/0902.3360}{{\ttfamily arXiv:0902.3360 [hep-ph]}}.

\bibitem{Kelso:2014qka}
C.~Kelso, H.~N. Long, R.~Martinez, and F.~S. Queiroz, ``{Connection of
  $g-2_{\mu}$, electroweak, dark matter, and collider constraints on 331
  models},'' \href{http://dx.doi.org/10.1103/PhysRevD.90.113011}{{\em Phys.
  Rev.} {\bfseries D90} no.~11, (2014) 113011},
\href{http://arxiv.org/abs/1408.6203}{{\ttfamily arXiv:1408.6203 [hep-ph]}}.

\bibitem{Lindner:2016bgg}
M.~Lindner, M.~Platscher, and F.~S. Queiroz, ``{A Call for New Physics : The
  Muon Anomalous Magnetic Moment and Lepton Flavor Violation},''
  \href{http://dx.doi.org/10.1016/j.physrep.2017.12.001}{{\em Phys. Rept.}
  {\bfseries 731} (2018) 1--82},
\href{http://arxiv.org/abs/1610.06587}{{\ttfamily arXiv:1610.06587 [hep-ph]}}.

\bibitem{Kowalska:2017iqv}
K.~Kowalska and E.~M. Sessolo, ``{Expectations for the muon g-2 in simplified
  models with dark matter},''
  \href{http://dx.doi.org/10.1007/JHEP09(2017)112}{{\em JHEP} {\bfseries 09}
  (2017) 112},
\href{http://arxiv.org/abs/1707.00753}{{\ttfamily arXiv:1707.00753 [hep-ph]}}.

\bibitem{Abi:2021gix}
{\bfseries Muon g-2} Collaboration, B.~Abi {\em et~al.}, ``{Measurement of the
  Positive Muon Anomalous Magnetic Moment to 0.46 ppm},''
  \href{http://dx.doi.org/10.1103/PhysRevLett.126.141801}{{\em Phys. Rev.
  Lett.} {\bfseries 126} no.~14, (2021) 141801},
\href{http://arxiv.org/abs/2104.03281}{{\ttfamily arXiv:2104.03281 [hep-ex]}}.

\bibitem{Morel:2020dww}
L.~Morel, Z.~Yao, P.~Cladé, and S.~Guellati-Khélifa, ``{Determination of the
  fine-structure constant with an accuracy of 81 parts per trillion},''
\href{http://dx.doi.org/10.1038/s41586-020-2964-7}{{\em Nature} {\bfseries 588}
  no.~7836, (2020) 61--65}.

\bibitem{Casas:2001sr}
J.~A. Casas and A.~Ibarra, ``{Oscillating neutrinos and $\mu \to e, \gamma$},''
  \href{http://dx.doi.org/10.1016/S0550-3213(01)00475-8}{{\em Nucl. Phys. B}
  {\bfseries 618} (2001) 171--204},
  \href{http://arxiv.org/abs/hep-ph/0103065}{{\ttfamily arXiv:hep-ph/0103065}}.

\bibitem{Ibarra:2003up}
A.~Ibarra and G.~G. Ross, ``{Neutrino phenomenology: The Case of two
  right-handed neutrinos},''
  \href{http://dx.doi.org/10.1016/j.physletb.2004.04.037}{{\em Phys. Lett. B}
  {\bfseries 591} (2004) 285--296},
  \href{http://arxiv.org/abs/hep-ph/0312138}{{\ttfamily arXiv:hep-ph/0312138}}.

\bibitem{CMS-PAS-HIG-20-002}
{\bfseries CMS} Collaboration, ``{Search for a standard model-like Higgs boson
  in the mass range between 70 and 110$~\mathrm{GeV}$ in the diphoton final
  state in proton-proton collisions at $\sqrt{s}=13~\mathrm{TeV}$},'' tech.
  rep., CERN, Geneva, 2023.
\newblock \url{https://cds.cern.ch/record/2852907}.

\bibitem{Biekotter:2023jld}
T.~Biek\"otter, S.~Heinemeyer, and G.~Weiglein, ``{The CMS di-photon excess at
  95 GeV in view of the LHC Run 2 results},''
  \href{http://arxiv.org/abs/2303.12018}{{\ttfamily arXiv:2303.12018
  [hep-ph]}}.

\bibitem{Martin:2009iq}
A.~D. Martin, W.~J. Stirling, R.~S. Thorne, and G.~Watt, ``{Parton
  distributions for the LHC},''
  \href{http://dx.doi.org/10.1140/epjc/s10052-009-1072-5}{{\em Eur. Phys. J. C}
  {\bfseries 63} (2009) 189--285},
  \href{http://arxiv.org/abs/0901.0002}{{\ttfamily arXiv:0901.0002 [hep-ph]}}.

\bibitem{Langacker:1988up}
P.~Langacker and D.~London, ``{Lepton Number Violation and Massless
  Nonorthogonal Neutrinos},''
  \href{http://dx.doi.org/10.1103/PhysRevD.38.907}{{\em Phys. Rev. D}
  {\bfseries 38} (1988) 907}.

\bibitem{Lavoura:2003xp}
L.~Lavoura, ``{General formulae for $f(1) \to f(2) + \gamma$},''
  \href{http://dx.doi.org/10.1140/epjc/s2003-01212-7}{{\em Eur. Phys. J. C}
  {\bfseries 29} (2003) 191--195},
  \href{http://arxiv.org/abs/hep-ph/0302221}{{\ttfamily arXiv:hep-ph/0302221}}.

\bibitem{Hue:2017lak}
L.~T. Hue, L.~D. Ninh, T.~T. Thuc, and N.~T.~T. Dat, ``{Exact one-loop results
  for $l_i \to l_j\gamma$ in 3-3-1 models},''
  \href{http://dx.doi.org/10.1140/epjc/s10052-018-5589-3}{{\em Eur. Phys. J. C}
  {\bfseries 78} no.~2, (2018) 128},
  \href{http://arxiv.org/abs/1708.09723}{{\ttfamily arXiv:1708.09723
  [hep-ph]}}.

\bibitem{Catano:2012kw}
M.~E. Catano, R.~Martinez, and F.~Ochoa, ``{Neutrino masses in a 331 model with
  right-handed neutrinos without doubly charged Higgs bosons via inverse and
  double seesaw mechanisms},''
  \href{http://dx.doi.org/10.1103/PhysRevD.86.073015}{{\em Phys. Rev.}
  {\bfseries D86} (2012) 073015},
\href{http://arxiv.org/abs/1206.1966}{{\ttfamily arXiv:1206.1966 [hep-ph]}}.

\bibitem{Kuno:1999jp}
Y.~Kuno and Y.~Okada, ``{Muon decay and physics beyond the standard model},''
  \href{http://dx.doi.org/10.1103/RevModPhys.73.151}{{\em Rev. Mod. Phys.}
  {\bfseries 73} (2001) 151--202},
  \href{http://arxiv.org/abs/hep-ph/9909265}{{\ttfamily arXiv:hep-ph/9909265}}.

\bibitem{Burgess:2000yq}
C.~P. Burgess, M.~Pospelov, and T.~ter Veldhuis, ``{The Minimal model of
  nonbaryonic dark matter: A Singlet scalar},''
  \href{http://dx.doi.org/10.1016/S0550-3213(01)00513-2}{{\em Nucl. Phys. B}
  {\bfseries 619} (2001) 709--728},
  \href{http://arxiv.org/abs/hep-ph/0011335}{{\ttfamily arXiv:hep-ph/0011335}}.

\bibitem{Casas:2017jjg}
J.~A. Casas, D.~G. Cerde\~no, J.~M. Moreno, and J.~Quilis, ``{Reopening the
  Higgs portal for single scalar dark matter},''
  \href{http://dx.doi.org/10.1007/JHEP05(2017)036}{{\em JHEP} {\bfseries 05}
  (2017) 036}, \href{http://arxiv.org/abs/1701.08134}{{\ttfamily
  arXiv:1701.08134 [hep-ph]}}.

\bibitem{Bhattacharya:2016ysw}
S.~Bhattacharya, P.~Poulose, and P.~Ghosh, ``{Multipartite Interacting Scalar
  Dark Matter in the light of updated LUX data},''
  \href{http://dx.doi.org/10.1088/1475-7516/2017/04/043}{{\em JCAP} {\bfseries
  04} (2017) 043}, \href{http://arxiv.org/abs/1607.08461}{{\ttfamily
  arXiv:1607.08461 [hep-ph]}}.

\bibitem{Edsjo:1997bg}
J.~Edsjo and P.~Gondolo, ``{Neutralino relic density including
  coannihilations},'' \href{http://dx.doi.org/10.1103/PhysRevD.56.1879}{{\em
  Phys. Rev. D} {\bfseries 56} (1997) 1879--1894},
  \href{http://arxiv.org/abs/hep-ph/9704361}{{\ttfamily arXiv:hep-ph/9704361}}.

\bibitem{Planck:2018vyg}
{\bfseries Planck} Collaboration, N.~Aghanim {\em et~al.}, ``{Planck 2018
  results. VI. Cosmological parameters},''
  \href{http://dx.doi.org/10.1051/0004-6361/201833910}{{\em Astron. Astrophys.}
  {\bfseries 641} (2020) A6}, \href{http://arxiv.org/abs/1807.06209}{{\ttfamily
  arXiv:1807.06209 [astro-ph.CO]}}. [Erratum: Astron.Astrophys. 652, C4
  (2021)].

\bibitem{XENON:2018voc}
{\bfseries XENON} Collaboration, E.~Aprile {\em et~al.}, ``{Dark Matter Search
  Results from a One Ton-Year Exposure of XENON1T},''
  \href{http://dx.doi.org/10.1103/PhysRevLett.121.111302}{{\em Phys. Rev.
  Lett.} {\bfseries 121} no.~11, (2018) 111302},
  \href{http://arxiv.org/abs/1805.12562}{{\ttfamily arXiv:1805.12562
  [astro-ph.CO]}}.

\bibitem{Farina:2009ez}
M.~Farina, D.~Pappadopulo, and A.~Strumia, ``{CDMS stands for Constrained Dark
  Matter Singlet},''
  \href{http://dx.doi.org/10.1016/j.physletb.2010.04.025}{{\em Phys. Lett. B}
  {\bfseries 688} (2010) 329--331},
  \href{http://arxiv.org/abs/0912.5038}{{\ttfamily arXiv:0912.5038 [hep-ph]}}.

\bibitem{Giedt:2009mr}
J.~Giedt, A.~W. Thomas, and R.~D. Young, ``{Dark matter, the CMSSM and lattice
  QCD},'' \href{http://dx.doi.org/10.1103/PhysRevLett.103.201802}{{\em Phys.
  Rev. Lett.} {\bfseries 103} (2009) 201802},
  \href{http://arxiv.org/abs/0907.4177}{{\ttfamily arXiv:0907.4177 [hep-ph]}}.

\end{thebibliography}\endgroup

\end{document}